\documentstyle[12pt,epsf]{article}
\newcommand{\yourtitle}[1]{
\mbox{}\\
\vskip 4\baselineskip
{\bf\noindent #1}\\ }

\newcommand{\youraddress}[1]{

\noindent\mbox{}\hspace{1in}\parbox[t]{4.0in}{#1}\\ }

\newcommand{\yournames}[1]{
\mbox{}\\
\mbox{}\\
\noindent\mbox{}\hspace{1in}{#1}\\ }

\newcommand{\yourabstract}[1]{
\mbox{}\\
\mbox{}\\
{\bf\noindent Abstract}\\
\begin{center}
\mbox{}\parbox[t]{5.in}{#1}
\end{center} }

\newcommand{\yoursection}[1]{
\vskip 2\baselineskip
{\bf\noindent #1}\\
\mbox{}\\
\vspace{-0.19in}}

\begin{document}

\yourtitle{The $A$ Dependence of Open Charm and Bottom Production}
\yournames{R. Vogt$^{a,b}$\footnote{
This work was supported in part by the Director, Office of Energy
Research, Division of Nuclear Physics of the Office of High Energy
and Nuclear Physics of the U. S. Department of Energy under Contract
Number DE-AC03-76SF00098.}}
\youraddress{$\!\!^a$Nuclear Science Division, MS 70-319\\
        Lawrence Berkeley Laboratory, \\
        University of California, Berkeley, California 94720}
\youraddress{$\!\!^b$Physics Department, \\
            University of California, Davis, California, 95616}

\vspace{1in}

\yourabstract{We study inclusive heavy quark and exclusive heavy quark pair
production in $pp$, $pA$ and $AA$
interactions.  Intrinsic transverse momentum is
introduced in $pp$ interactions.  Nuclear effects, limited to 
$k_T$ broadening and nuclear shadowing, are introduced in $pA$ and
$AA$ interactions.  The nuclear
dependence is studied over a range of energies, both in fixed target and
collider setups.}

\yoursection{Introduction}

Previously, we calculated the heavy quark production cross section in $pp$
interactions.  In that work, we focussed on the scale dependence and fixed 
the quark mass and scale 
parameters by comparison to the total cross section data \cite{hpc}.  
In this extension to $pA$ and $AA$
interactions, we study the effects of shadowing and $k_T$ broadening on the
spectra of heavy quark production.

We first briefly update the calculation of the total cross section with new
parton distribution functions and then go
on to discuss the relevant differential distributions.

\yoursection{Total cross section}

Recall that at leading order (LO) heavy quarks are produced by $gg$ fusion and
$q \overline q$ annihilation while at next-to-leading order (NLO), $qg +
\overline q g$ scattering is also included.  At any order, the partonic 
cross section may be expressed in terms of dimensionless scaling functions
$f^{(k,l)}_{ij}$ that depend only on the variable $\eta$ \cite{KLMV}
\begin{eqnarray}
\label{scalingfunctions}
\hat \sigma_{ij}(\hat s,m_Q^2,\mu^2) = \frac{\alpha^2_s(\mu)}{m^2}
\sum\limits_{k=0}^{\infty} \,\, \left( 4 \pi \alpha_s(\mu) \right)^k
\sum\limits_{l=0}^k \,\, f^{(k,l)}_{ij}(\eta) \,\,
\ln^l\left(\frac{\mu^2}{m_Q^2}\right) \, , 
\end{eqnarray} 
where $\hat s$ is the partonic center of mass, $m_Q$ is the heavy quark mass,
$\mu$ is the scale, and $\eta = \hat s/4 m_Q^2 - 1$.  
The cross section is calculated as an expansion in powers of $\alpha_s$
with $k=0$ corresponding to the Born cross section at order ${\cal
O}(\alpha_s^2)$.  The first correction, $k=1$, corresponds to the NLO cross
section at ${\cal O}(\alpha_s^3)$.  It is only at this order and above that
the renormalization scale dependence enters the calculation
since when $k=1$
and $l=1$, the logarithm $\ln(\mu^2/m_Q^2)$ 
appears.  The dependence on the factorization scale, the argument of
$\alpha_s$, appears already at LO.  We assume that the renormalization and
factorization scales are the same.  The next-to-next-to-leading
order (NNLO) corrections to next-to-next-to-leading logarithm
have been calculated near threshold \cite{KLMV} but
the complete calculation only exists to NLO.

The total partonic cross section can also be written as 
\begin{eqnarray}
\label{totpartonic}
\hat \sigma_{ij}(\hat s,m_Q^2,\mu^2) = \int d\hat t \, d\hat u \, \hat s^2
\frac{d^2\hat \sigma_{ij}(\hat s,\hat t, \hat u)}{d \hat t d \hat u}
\end{eqnarray} 
where the double differential partonic cross section at leading order is
proportional to the Born cross section, 
\begin{eqnarray}
\label{dpartonicdtdu}
\hat s^2 
\frac{d^2\hat \sigma_{ij}^{(0)}(\hat s,\hat t, \hat u)}{d \hat t d \hat u}
= \delta(\hat s - \hat t - \hat u - 2m_Q^2) \hat \sigma_{ij}^B(\hat s, \hat t,
\hat u) \, \, .
\end{eqnarray} 
The expressions for the Born cross section can be found in {\it e.g.}\
Ref.~\cite{GOR}. 
The NLO and NNLO cross sections in the $q \overline q$ channel are also
proportional to $\hat \sigma^B_{q \overline q}$, albeit with more complicated
scaling functions.  In the $gg$ channel, the higher order cross
sections are again proportional to the Born cross section to leading logarithm
and as are the highest powers of the factorization scale dependent logarithms
\cite{KLMV}.  However, beyond leading logarithm, the scaling functions also
depend on the color state in which the $Q \overline Q$ pair was produced.
The expressions for the Born cross section in both channels may
be found in Ref.~\cite{KLMV} along with the NLO and NNLO expressions near
threshold.  

The cross sections may be calculated in either one-particle inclusive
kinematics or pair invariant mass
kinematics, depending on whether one or both heavy quarks
are detected.  Single heavy quark hadroproduction in $pp$ interactions is 
defined by 
\begin{eqnarray}
\label{eq:3}
p(P_1) + p(P_2) \longrightarrow Q(p_1) + X(p_X)\, , 
\end{eqnarray}
where $X$ denotes any
hadronic final state containing the heavy antiquark and
$Q(p_1)$ is the identified heavy quark.  
The reaction in Eq.~(\ref{eq:3}) is dominated 
by the partonic reactions
\begin{eqnarray}
\label{eq:7}
q(k_1) + \overline q(k_2) &\longrightarrow& Q(p_1) +
X[\overline Q](p_2)\, ,
\\
\label{eq:8}  
g(k_1) + g(k_2) &\longrightarrow& {\rm{Q}}(p_1) +
X[\overline Q](p_2)\, . 
\end{eqnarray}
At LO or if $X[\overline Q](p_2) \equiv \overline Q(p_2')$, the reaction is
at partonic threshold with $\overline Q$ momentum $p_2'$.  At threshold the
heavy quarks do not have to be produced at rest but are instead produced with
equal and opposite momentum, leading to trivial values of the pair transverse
momentum ($\vec p_{T_1} + \vec p_{T_2}^{\, \prime} = 0$) and azimuthal 
angle $(\phi = \pi)$.
The Mandelstam invariants on the
hadronic level are
\begin{equation}
  \label{eq:5}
s = (P_1+P_2)^2 \quad,\quad t = (P_1-p_1)^2 \quad,\quad u =
(P_2-p_1)^2 \,, 
\end{equation}
with the corresponding partonic invariants
\begin{eqnarray}
  \label{eq:9}
\hat s =(k_1+k_2)^2 \quad,\quad \hat t = (k_2-p_1)^2 \quad,\quad
\hat u = (k_1-p_1)^2 \, .
\end{eqnarray}

The pair invariant mass kinematics are 
somewhat different.  Now the pair is treated as a single unit and we have
\begin{eqnarray}
  \label{eq:14}
p(P_1) + p(P_2) &\longrightarrow& Q \overline Q(P')
+ X(P_X) \, .  
\end{eqnarray}
On the parton level, Eqs.~(\ref{eq:7}) and (\ref{eq:8}) are now
\begin{eqnarray}
\label{qq_PIM}
q(k_1) + \overline q(k_2) &\longrightarrow& Q \overline Q(p')
+ X(p_X)\, ,
\\
\label{gg_PIM}
g(k_1) + g(k_2) &\longrightarrow& Q \overline Q(p') +
X(p_X)\, . 
\end{eqnarray}
The square of the heavy quark pair mass is $p'^2 = M^2$.
In the case when $p_X \neq 0$, more invariants than the three typical
Mandelstam invariants can be defined \cite{MNR}.  At partonic threshold,
$p_X  = 0$, only the three Mandelstam invariants remain with $\hat s = M^2$
and
\begin{eqnarray}
\label{tupidef}
\hat t - m_Q^2 &=& - \frac{M^2}{2} ( 1 - \beta_M\, \cos \theta )\, , \\
\hat u - m_Q^2 &=& - \frac{M^2}{2} ( 1 + \beta_M\, \cos \theta )\, ,
\end{eqnarray}
where $\beta_M=\sqrt{1-4m_Q^2/M^2}$, and $\theta$ is the scattering
angle in the parton center of mass frame.

The total hadronic cross section is obtained by convoluting the total partonic
cross section with the parton distribution functions of the initial hadrons,
\begin{eqnarray}
\label{totalhadroncrs}
\sigma_{pp}(s,m_Q^2) = \sum_{i,j = q,{\overline q},g} \,\,
\int_{4m_Q^2/s}^{1}\,\frac{d\tau}{\tau}\, \delta(x_1 x_2 - \tau) \,
F_i^p(x_1,\mu^2) F_j^p(x_2,\mu^2) \, 
\hat \sigma_{ij}(\tau ,m_Q^2,\mu^2)\, , 
\end{eqnarray}
where the sum $i$ is over all massless partons and
$x_1$ and $x_2$ are the  hadron momentum fractions carried by 
the interacting partons.
The parton distribution functions, denoted by $F_i^p$, are evaluated at
the factorization scale, assumed to be equal to the renormalization scale
in all our calculations. 

\begin{figure}[htbp]
\setlength{\epsfxsize=\textwidth}
\setlength{\epsfysize=0.6\textheight}
\centerline{\epsffile{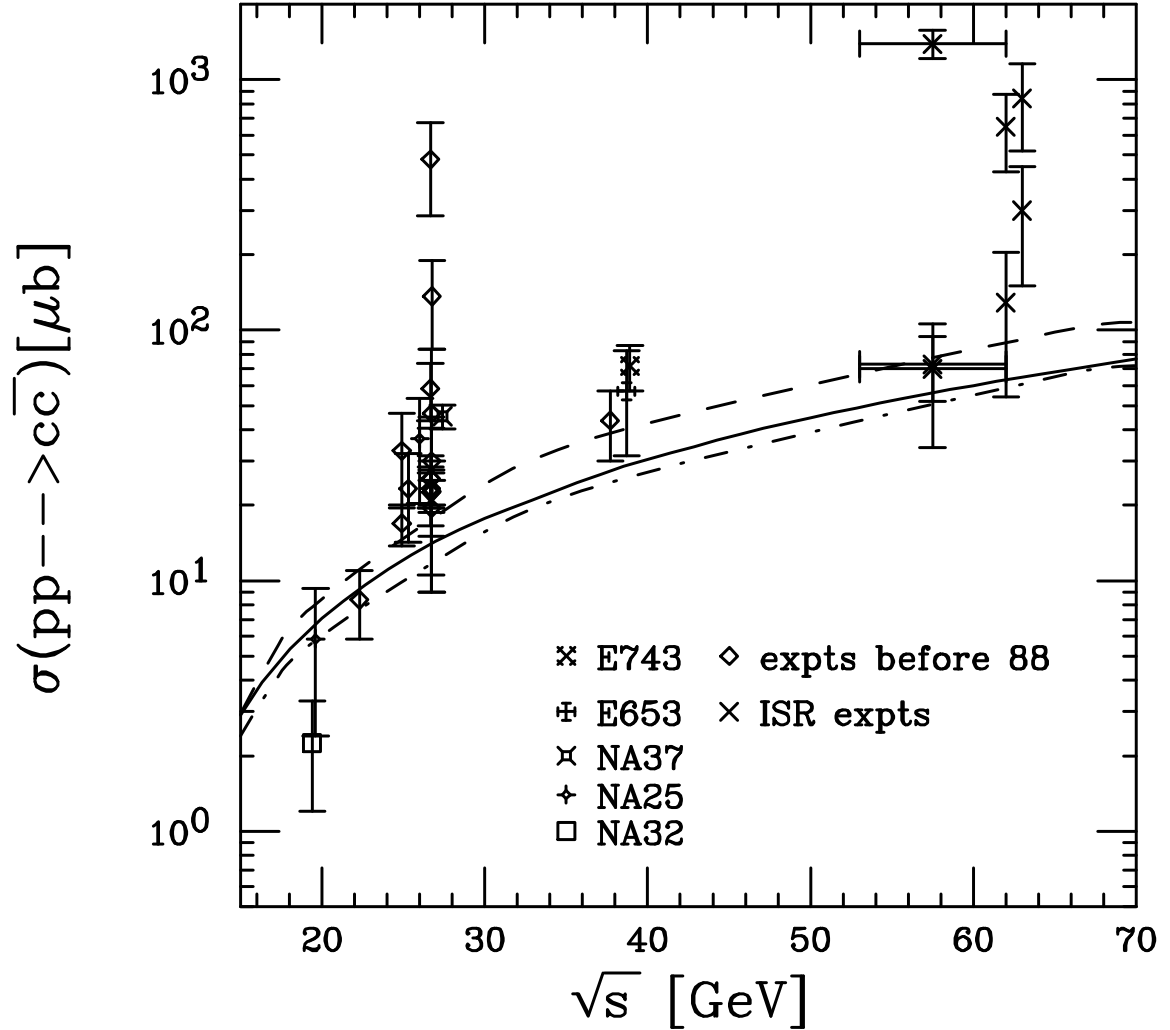}}
\caption[]{Total charm production cross sections from $pp$ and $pA$
measurements \cite{Reu,AMM,E6531,NA27,NA321} compared to calculations 
with $m_c=1.2$ GeV and $\mu = 2m_c$.  The parton distributions used are
MRS D$-'$ (solid); MRST HO (dashed) and MRST LO (dot-dashed).  Updated from
Ref.~\cite{hpc}.} 
\label{totcross}
\end{figure}

In our previous work on $pp$ interactions \cite{hpc}, we obtained our results 
with two sets of parton distribution functions that had successfully predicted
the general behavior of the low $x$ deep-inelastic scattering data 
\cite{HERA}, MRS D$-'$ \cite{mrsdm} and GRV HO \cite{grv}.  Much more low $x$ 
data has become
available since then which shows that these parton densities are too
high at low $x$ and new fits to the most recent data \cite{HERA2} 
are now available \cite{PDFLIB}.  (The low $x$ region is particularly relevant
for $Q \overline Q$ production at the LHC.)  Our latest calculations employ 
the most recent NLO parton
distribution functions from the MRS group, MRST HO (central gluon) 
\cite{mrstho}. 
This group has also made a leading order set available \cite{mrstlo}.  We
compare the MRS D$-'$, MRST HO, and MRST LO results in Fig.~\ref{totcross}
and in Tables~\ref{charmtot} and \ref{bottomtot}, all updated from our earlier
work \cite{hpc}.  Table~\ref{charmtot} has also been expanded to include the 
fixed target energies at which we compare the $pp$ and $pA$ calculations.
As before, all our numerical results are obtained using the NLO code by 
Mangano, Nason and Ridolfi (MNR) \cite{MNR}.

All the cross sections are calculated at LO and NLO
with the same mass and
scale parameters obtained from the $pp$ study---$m_c = 1.2$ GeV, 
$\mu_c = 2m_c$, $m_b = 4.75$ GeV, and $\mu_b = m_b$.  The agreement with the
total charm cross section data is at least as good with the MRST sets as with 
the MRS D$-'$ set.  In fact, the somewhat larger NLO cross section 
at fixed target energies with MRST HO tends to agree better with the data 
than the older set, in part because some of the gluon density at low $x$ has
been redistributed to higher $x$.  
However, at the $pp$ and $AA$ collider energies shown in
Tables~\ref{charmtot} and \ref{bottomtot}, the growth of the cross sections is
considerably slower with the new distribution functions than with the MRS D$-'$
set.  

The MRST LO results are for comparative purposes only.  A LO set is not 
consistent with a NLO calculation just as a NLO set is inconsistent with a
strictly LO calculation.  We note however that the tabulated results are 
all calculated with $\alpha_s$ at two
loops, regardless of whether the LO or NLO result is given, because in the MNR
code $\alpha_s$
is calculated to two loops regardless of the order of the 
chosen parton density.  Since the QCD
scale, $\Lambda$, is tuned to fit all available data in global analyses of
parton distributions, whether at LO or NLO, the number of loops in the
evaluation is important.  We remark however, that it is rather standard
procedure to evaluate all orders of the hadronic cross section with the same
NLO parton density and a two loop calculation of $\alpha_s$.  
This is done so that one can study the variation of the
partonic cross section alone while keeping the nonperturbative results fixed.
Also, beyond NLO, no standard NNLO or resummed 
evaluations of the parton densities exist, generally because the partonic
matrix elements are limited to near threshold beyond NLO.  In addition, the
data used in the global analyses of the parton densities are typically far
from threshold and of only limited use in such an evaluation.  Therefore the
uncertainties in such an analysis would be large.

We note that when a fully LO calculation of the charm cross section
is done with a one loop evaluation of
$\alpha_s$ in the global analyses of the parton densities, as with the MRST LO
set, the LO cross sections in Table~\ref{charmtot} increase by $\sim 60$\%.
This change is almost solely
due to the difference between the one and two loop evaluations of
$\alpha_s$.  The MRST LO $\Lambda$ is 0.204 GeV when $n_f=3$, leading
to $\alpha_s^{\rm 1-loop} = 0.287$ and $\alpha_s^{\rm 2-loops} = 0.220$
for $\mu = 2m_c$.
The $\Lambda$ associated with the MRST HO set is larger, $\Lambda = 0.353$ GeV
when $n_f = 3$, corresponding to $\alpha_s^{\rm 1-loop} = 0.364$ and 
$\alpha_s^{\rm 2-loops} = 0.263$ at the same scale.  Thus a fully LO
calculation with MRST LO partons would be in better agreement with the
tabulated LO results for MRST HO partons. These results show that one must
use caution when using a LO set with a NLO calculation or a NLO set for a LO
calculation. 

\begin{table}[htbp]
\begin{tabular}{|c|c|c|c|c|c|c|} \hline
& \multicolumn{2}{c|}{MRST LO} & \multicolumn{2}{c|}{MRST HO} 
& \multicolumn{2}{c|}{MRS D$-'$} \\ 
$\sqrt{s}$(GeV) & $\sigma_{c \overline c}^{\rm LO}$ ($\mu$b) &
$\sigma_{c \overline c}^{\rm NLO}$ ($\mu$b) & $\sigma_{c \overline
c}^{\rm LO}$ ($\mu$b) & $\sigma_{c \overline c}^{\rm NLO}$ ($\mu$b) &
$\sigma_{c \overline c}^{\rm LO}$ ($\mu$b)
& $\sigma_{c \overline c}^{\rm NLO}$ ($\mu$b) \\ \hline
17.3  & 1.56  & 4.16  & 1.85  & 5.54  & 1.67  & 4.62 \\
19.4  & 2.16  & 5.65  & 2.67  & 7.87  & 2.36  & 6.38 \\
29.1  & 5.84  & 14.52 & 8.04  & 22.36 & 6.42  & 16.24 \\
38.8  & 10.60 & 25.85 & 15.10 & 40.57 & 11.47 & 28.20 \\
63    & 25.37 & 60.74 & 35.79 & 91.88 & 26.88 & 64.97 \\
200   & 132.6 & 314.1 & 152.8 & 381.7 & 139.3 & 343.7 \\
500   & 374.9 & 897.0 & 356.8 & 922.8 & 449.4 & 1138 \\
5500  & 3040  & 7733  & 2002  & 5833  & 7013  & 17680 \\
14000 & 5668  & 14920 & 3332  & 10240 & 16450 & 41770 \\ \hline
\end{tabular}
\caption{Total $c \overline c$ production cross sections at fixed-target and
collider energies.}
\label{charmtot}
\end{table}

\begin{table}[htbp]
\begin{tabular}{|c|c|c|c|c|c|c|} \hline
& \multicolumn{2}{c|}{MRST LO} & \multicolumn{2}{c|}{MRST HO} 
& \multicolumn{2}{c|}{MRS D$-'$} \\ 
$\sqrt{s}$(GeV) & $\sigma_{b \overline b}^{\rm LO}$ ($\mu$b) &
$\sigma_{b \overline b}^{\rm NLO}$ ($\mu$b) & $\sigma_{b \overline
b}^{\rm LO}$ ($\mu$b) & $\sigma_{b \overline b}^{\rm NLO}$ ($\mu$b) &
$\sigma_{b \overline b}^{\rm LO}$ ($\mu$b)
& $\sigma_{b \overline b}^{\rm NLO}$ ($\mu$b) \\ \hline
63    & 0.0374 & 0.0688 & 0.0394 & 0.0763 & 0.0397 & 0.0746 \\
200   & 0.744  & 1.371  & 0.987  & 1.903  & 0.796  & 1.47 \\ 
500   & 3.825  & 7.422  & 4.809  & 9.799  & 3.847  & 7.597 \\
5500  & 82.71  & 186.3  & 72.85  & 186.3  & 98.8   & 224 \\
14000 & 211.0  & 502.2  & 164.5  & 454.1  & 296.8  & 687.5 \\ \hline
\end{tabular}
\caption{Total $b \overline b$ production cross sections at collider energies.}
\label{bottomtot}
\end{table}

\yoursection{Differential cross sections}

While the total cross section predicts the yield of heavy quark production
over all phase space,
it cannot provide much useful information on nuclear effects such a
$p_T$-broadening and shadowing.  It is also independent of the choice of
kinematic schemes.  Any $k_T$ broadening will not affect the total cross
section but will have a strong influence on the $p_T$ distributions.
Shadowing may reduce or enhance the nuclear
cross section relative to that of the proton 
but the effect may be more apparent in
some regions of phase space than others.
To obtain more information on nuclear effects, it is thus
necessary to turn to distributions.  In addition, a real detector does not 
cover all phase space.  The differential distributions can be tuned to a 
detector acceptance.

The double differential hadronic cross sections are written in the factorized
form 
\begin{eqnarray}
\label{stdfact}
s^2\, \frac{d^2\sigma_{pp}(s,t,u)}{dt du} =
\sum_{i,j=q,\overline q,g} \int_{x_1^-}^{1}\frac{dx_1}{x_1}
\int_{x_2^-}^{1}\frac{dx_2}{x_2} f_i^p(x_1,\mu^2)
f_j^p(x_2,\mu^2) {\cal J}_K(\hat s,\hat t,\hat u) \hat s^2
\frac{d^2\hat \sigma_{ij}(\hat s, \hat t, \hat u)}{d\hat t d\hat u} 
\end{eqnarray}
where $f_i^p$ is the parton density, $f_i^p(x,\mu^2) = F_i^p(x,\mu^2)/x$,
and ${\cal J}_K$ is a kinematics-dependent Jacobian.
These results are difficult to calculate analytically and expressions only
exist for $\hat 
s^2 d^2\hat \sigma_{ij}(\hat s, \hat t, \hat u)/d\hat t d\hat u$, see
{\it e.g.}\ \cite{KLMV}.  However, the MNR code calculates single and double 
differential distributions
as well as total cross sections.
In our first paper \cite{hpc}, we
neglected any fragmentation effects on the produced heavy quark and 
intrinsic transverse momentum of the initial
partons because the inclusive $D$ meson $p_T$ distributions agreed with the
bare charm quark distributions, showing no need for such effects.  However, as
pointed out in Ref.~\cite{MLM1}, 
the pair distributions, particularly the pair $p_T$
and the azimuthal angle distributions at fixed target energies are broader than
can be accounted for by the bare quark distributions.  The 
general behavior of the pair data was reproduced in Ref.~\cite{MLM1} 
by including a Peterson-type
fragmentation function \cite{Pete} along with an intrinsic $k_T$ of the initial
partons.  Indeed, the combination of fragmentation with an intrinsic average
$k_T^2$ of 1 GeV$^2$ leads to single $D$ meson $p_T$ distributions in good
agreement with the bare charm distribution.  These effects can be added by
modifying Eq.~(\ref{stdfact}) to include the additional integrals
\begin{eqnarray}
\label{extras}
\int dz_3 dz_4 d^2k_{T 1} d^2k_{T 2} \frac{D_{H/Q}(z_3,\mu^2)}{z_3}
\frac{D_{\overline H/\overline Q}(z_4,\mu^2)}{z_4} g_p(k_{T 1}) g_p(k_{T 2})
\, \, .
\end{eqnarray}

The fragmentation functions $D_{H/Q}(z)$ 
describe the hadronization of the heavy
quark $Q$ into hadron $H$.  The final-state hadron $H$ carries a fraction $z$
of the heavy quark momentum and is assumed to be collinear to the parent
heavy quark.  The Peterson fragmentation function
\cite{Pete}, fit to $D$ production data in $e^+ e^-$
collisions, 
\begin{eqnarray}
\label{Petefun}
D_{H/Q}(z) = \frac{N}{z(1 - 1/z - \epsilon_Q/(1-z))^2} \, \, , 
\end{eqnarray}
is used in the code.  The parameter $\epsilon_Q$ is 0.06 for charm and 
0.006 for bottom \cite{Chirin}.  The normalization $N$ is defined by
$\sum_H \int D_{H/Q}(z) dz = 1$. 

The Peterson function results in a nearly 30\% decrease 
in the average charm
quark momentum during hadronization.  This decrease must be compensated for in
inclusive production because the decreased momentum of the $D$ meson would no
longer agree with the single $D$ $p_T$ distributions.  The intrinsic $k_T$ of
the initial partons is such a compensation
mechanism which has been used successfully to describe the
$p_T$ distributions of other processes such as Drell-Yan \cite{Field} and hard
photon production \cite{Owens}.  Some of the low $p_T$ Drell-Yan data has
subsequently been described by resummation to all orders but the inclusion of
higher orders has not eliminated the need for this intrinsic $k_T$.  It was
found that assuming $\langle k_T^2 \rangle = 1$ GeV$^2$ in $pp$ and $\pi p$
interactions produces agreement with the inclusive data on the level of that
found with the bare quarks alone and does a better job of reconciling the
exclusive $D \overline D$ data with the calculated distributions \cite{MLM1}.

The implementation of intrinsic $k_T$ in the MNR code is not the same as in
other processes due to the nature of their calculation.  The cancellation of
divergences is handled numerically in the code.  Since adding additional
numerical integrations would slow this process, the $k_T$ kick is added in the
final, rather than the initial state. 
In Eq.~(\ref{extras}), the Gaussian function $g_p(k_T)$,
\begin{eqnarray}
g_p(k_T) = \frac{1}{\pi \langle k_T^2 \rangle_p} \exp(-k_T^2/\langle k_T^2
\rangle_p) \, \, ,
\label{intkt}
\end{eqnarray}
with $\langle k_T^2 \rangle_p = 1$ GeV$^2$ \cite{MLM1}, multiplies the parton
distribution functions, assuming the $x$ and $k_T$ dependencies completely
factorize.  If this is true, it does not matter whether the $k_T$ dependence
appears in the initial or final state modulo the caveat described below.  
In the code, the $Q \overline Q$ system
is boosted to rest from its longitudinal center of mass frame.  Intrinsic 
transverse
momenta of the incoming partons, $\vec k_{T 1}$ and $\vec k_{T 2}$, are chosen
at random with $k_{T 1}^2$ and $k_{T 2}^2$ distributed according to
Eq.~(\ref{intkt}).   A second transverse boost out of the pair rest frame
changes the initial transverse momentum of
the $Q \overline Q$ pair, $\vec p_T^{\, \prime}$ in Eqs.~(\ref{qq_PIM}) and
(\ref{gg_PIM}), to $\vec p_T^{\, \prime} 
+ \vec k_{T 1} + \vec k_{T 2}$.  The initial
$k_T$ of the partons could have alternatively been given to the entire
final-state system, as is essentially done if applied in the initial state,
instead of to the $Q \overline Q$ pair.  There is no difference if the
calculation is to LO only but at NLO an additional light parton also appears in
the final state.  In Ref.~\cite{MLM2}, it is claimed that the difference in the
two methods is rather small if $k_T^2 \leq 2$ GeV$^2$.  The effect of the
intrinsic $k_T$ decreases as the center of mass energy increases.

\yoursection{$k_T$ broadening in nuclei}

The average intrinsic $k_T$ is expected to increase in $pA$ interactions.  This
broadening is observed in Drell-Yan \cite{E7722}, $J/\psi$ \cite{NA3}, and
$\Upsilon$ \cite{E7722} production and has been used to explain high $p_T$ pion
production in nuclear interactions \cite{XNWkt}.  Since $k_T$ broadening has
not been explicitly measured in charm production, we follow the formulation of
Ref.~\cite{XNWkt}, similar to Ref.~\cite{GGetc} for $J/\psi$ and Drell-Yan
production where the $k_T$ broadening arises from multiple scattering of the
projectile partons in the target,
\begin{eqnarray}
\langle k_T^2 \rangle_A = \langle k_T^2 \rangle_p + (\langle \nu \rangle -1)
\Delta^2(\mu) \, \, .
\label{ktbroad}
\end{eqnarray}
We define the number of collisions in a proton-nucleus interaction, 
averaged over impact parameter, as
\begin{eqnarray}
\langle \nu \rangle = \sigma_{NN} \frac{\int d^2b \, T_A^2(b)}{\int d^2b \,
T_A(b)} = \frac{3}{2} \sigma_{NN} \rho_0 R_A \, \, , 
\label{avenu}
\end{eqnarray}
where $T_A(b) = \int dz \rho_A(b,z)$ is the nuclear profile function, and use
this to calculate the average broadening.  In the definition of 
$\langle \nu \rangle$, $\sigma_{NN}$ is the inelastic nucleon-nucleon
scattering cross section, $\rho_0$ is the central nuclear density, and $R_A$ is
the nuclear radius.  Our calculations are done with $A=200$, $R_A = 1.2
A^{1/3}$, and $\rho_0 = 0.16/$fm$^3$.

The strength of the nuclear broadening, $\Delta^2$, depends on $\mu$, the 
scale of the interaction \cite{XNWkt}
\begin{eqnarray}
\Delta^2(\mu) = 0.225 \frac{\ln^2(\mu/{\rm GeV})}{1 + 
\ln(\mu/{\rm GeV})} {\rm GeV}^2 \, \, .
\label{deltasq}
\end{eqnarray}
Thus $\Delta^2$ is larger
for $b \overline b$ production than $c \overline c$ production.  This
empirically reflects the larger $k_T$ broadening of the $\Upsilon$ \cite{E7722}
relative to the $J/\psi$ \cite{NA3}.  We evaluate $\Delta^2(\mu)$ at $\mu = 
2m_Q$ for both charm and bottom production.
We find $(\langle \nu \rangle - 1) \Delta^2(\mu) = 0.35$ GeV$^2$
for charm and 1.57 GeV$^2$ for bottom in $pA$ collisions at $b = 0$ and $A = 
200$. We can change the centrality by changing $\langle \nu \rangle - 1$.
Reducing $\langle \nu \rangle - 1$ by two corresponds to a more peripheral $pA$
collision while multiplying $\langle \nu \rangle - 1$ 
by two corresponds to central $AA$ collisions.
We will use the central values given above
in our $pA$ results and the larger values,
0.70 GeV$^2$ for charm and 3.14 GeV$^2$ for bottom, in our $AA$ results.  Note
that the total $\langle k_T^2 \rangle_A$ in both cases is obtained by adding
$\langle k_T^2 \rangle_p = 1$ GeV$^2$ to the above nuclear broadening.

\begin{figure}[htbp]
\setlength{\epsfxsize=\textwidth}
\setlength{\epsfysize=0.6\textheight}
\centerline{\epsffile{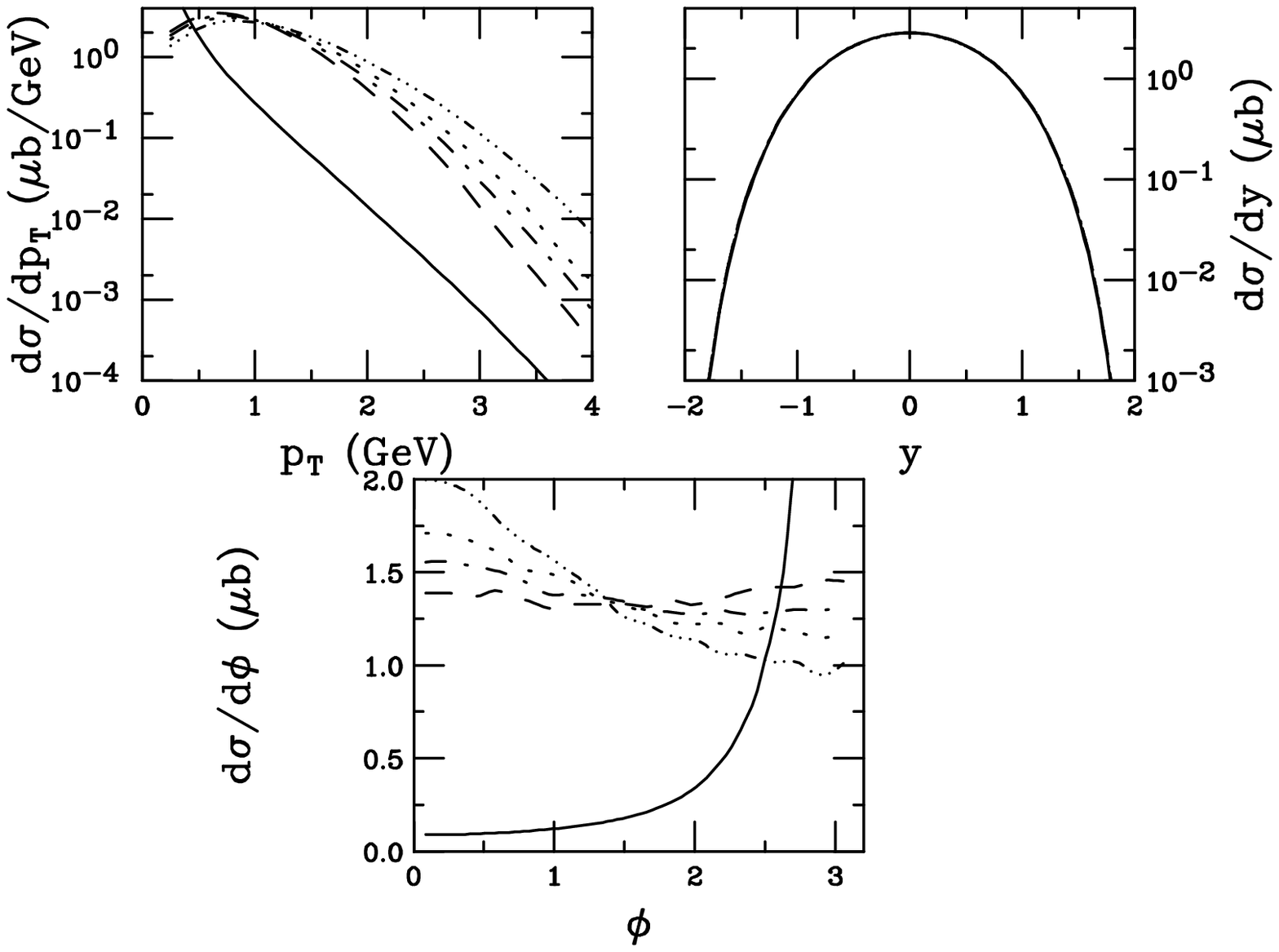}}
\caption[]{Effects of $k_T$ broadening alone on exclusive 
NLO $c \overline c$ pair production in $pA$
interactions at 158 GeV as a function of $p_T$, $y$, and $\phi$.  
The per nucleon cross section is given.
The curves are the bare quark distributions alone 
($\langle k_T^2 \rangle_A = 0$ and no fragmentation) (solid) and, including
fragmentation, $\langle k_T^2 \rangle = 1$ (dashed), 1.175
(dot-dashed), 1.35 (dotted) and 1.7 (dot-dot-dot-dashed) GeV$^2$.}
\label{158pairskt}
\end{figure}

\begin{figure}[htbp]
\setlength{\epsfxsize=\textwidth}
\setlength{\epsfysize=0.6\textheight}
\centerline{\epsffile{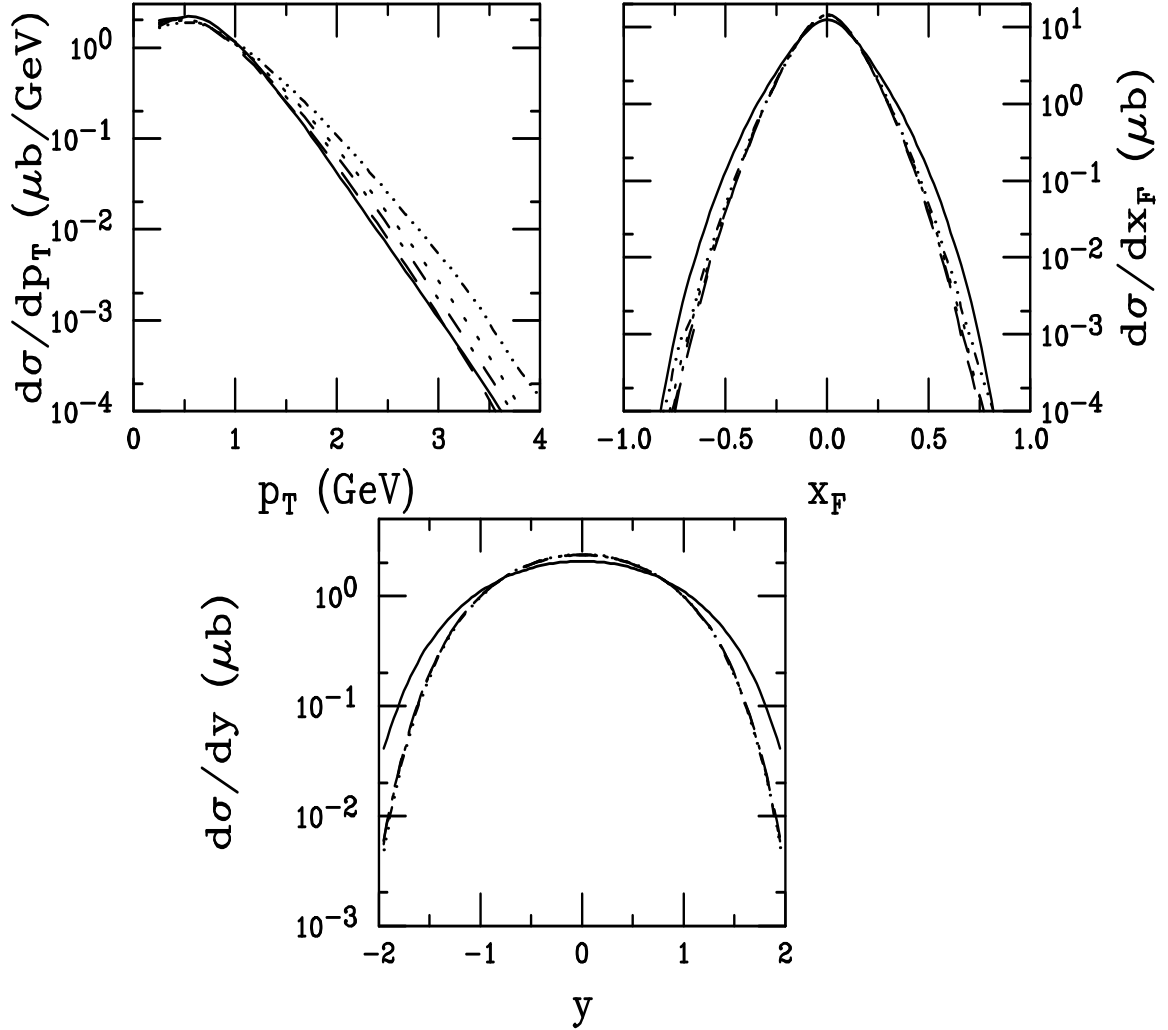}}
\caption[]{Effects of $k_T$ broadening alone on inclusive 
NLO $c$ quark production in $pA$
interactions at 158 GeV as a function of $p_T$, $x_F$, and $y$.
The per nucleon cross section is given.
The curves are the bare quark distributions alone 
($\langle k_T^2 \rangle_A = 0$ and no fragmentation) (solid) and, including
fragmentation, $\langle k_T^2 \rangle = 1$ (dashed), 1.175
(dot-dashed), 1.35 (dotted) and 1.7 (dot-dot-dot-dashed) GeV$^2$.}
\label{158singleskt}
\end{figure}

The effects of $k_T$ broadening alone on several pair and single inclusive
quantities are shown for a fixed target beam energy of 158 GeV
in Figs.~\ref{158pairskt} and \ref{158singleskt}.  The changes in the shape of
the pair quantities in
Fig.~\ref{158pairskt} are rather striking.  The bare charm pair distributions
alone are compared to results with $\langle k_T^2 \rangle = 1$, 1.175, 1.35,
and 1.7 GeV$^2$.
The differences in the pair $p_T$ and $\phi$ distributions are quite apparent.
The change in the pair $p_T$ distribution is due entirely to $k_T$ broadening
since fragmentation effects would steepen the solid curve.  The
broadening is still important at large pair $p_T$, causing more than an order
of magnitude difference in the cross section between $\langle k_T^2 \rangle =
1$ GeV$^2$ and 1.7 GeV$^2$ at $p_T = 4$ GeV.  Likewise, introducing intrinsic
$k_T$ changes the $\phi$ distribution from sharply peaked at $\phi = \pi$ for
the bare charm quarks to nearly flat when $\langle k_T^2 \rangle = 1$ GeV$^2$
to a peak at $\phi = 0$ when $\langle k_T^2 \rangle = 1.7$ GeV$^2$.  On the
contrary, the pair rapidity shows no effect of $k_T$ broadening.

The broadening effect on the single charm distributions shown
in Fig.~\ref{158singleskt} is not as strong.  
Note that all the distributions, including the
charm rapidity distribution, are affected by broadening and fragmentation.
The charm $p_T$ distributions for the bare quark and including fragmentation
with $\langle k_T^2 \rangle = 1$ GeV$^2$ are almost identical, in accord with
the observation that no intrinsic $k_T$ was needed to describe the $D$ meson
$p_T$ distributions.  The
difference between the bare quark and $\langle k_T^2 \rangle = 1.7$ GeV$^2$ 
is less than a factor of five for $p_T = 3.5$ GeV.  Fragmentation is
the most important effect on the $x_F$ and rapidity distributions since there
is little change in the distributions as $\langle k_T^2 \rangle$ is increased.
Any type of fragmentation, even that of a delta function, 
will cause the rapidity and $x_F$ distributions to be modified.
This is because fragmentation enhances these distributions in the central
region \cite{linvogt}.  If $D_{H/Q}(z)$ is a
delta function, the momentum does not change,
$p_H = p_Q$, but $E_H^2 = E_Q^2
- m_Q^2 + m_H^2$, resulting in a rapidity shift \cite{linvogt} with
\begin{eqnarray}
\label{rapfrag} 
dn \propto
dy_Q = \frac{dp_{zQ}}{E_Q} =  \frac{dp_{zH}}{E_Q} = \frac{m_{T,H} \cosh y_H
dy_H}{\sqrt{m_{T,H}^2 \cosh^2 y_H - m^2_H + m_Q^2}} \approx \frac{\cosh y_H
dy_H}{\sqrt{\cosh^2 y_H - \alpha^2}} 
\end{eqnarray}
where 
\begin{eqnarray} 
\alpha^2 = \frac{m_H^2 - m_Q^2}{m_{T,H}^2} \, \, .
\label{alffrag} 
\end{eqnarray}
For $m_c = 1.2$ GeV, $m_D = 1.87$ GeV and $m_{T,D} \approx \sqrt{2} m_D$,
$\alpha^2 = 0.25$, increasing the $D$ cross section at $y_D = 0$ by
$\approx 15$\%.  The total cross section is unchanged.
When $m_b = 4.75$ GeV, $m_B = 5.27$ GeV and $m_{T,B} =
\sqrt{2}m_B$, $\alpha^2 = 0.09$, increasing the $B$ cross section by
$\approx 5$\% at $y_B=0$.  The range of the
enhancement is $|y_H| < 2.5$, independent of energy.  
If the Peterson function is used instead, $\alpha^2$ increases to
\begin{eqnarray} 
\alpha^2 = \frac{m_H^2 - z^2 m_Q^2}{m_{T,H}^2} \,\, , \label{alfpete} 
\end{eqnarray}
increasing the $D$ cross section at $y_D=0$ by $\approx 30$\% for $\langle z
\rangle \approx 0.7$ and the $B$ cross section by $\approx 15$\% at $y_B = 0$
for $\langle z
\rangle \approx 0.85$.  These $\langle z \rangle$ values are typical for the
Peterson function with the $\epsilon_Q$ 
values given above.  The $x_F$ distribution
is also affected because $x_F = 2m_T \sinh y/\sqrt{s}$.

\yoursection{Nuclear shadowing}

Experiments \cite{Arn} have shown that the proton and neutron
structure functions are modified in a nuclear environment.  For
momentum fractions $x< 0.1$ and $0.3<x<0.7$, a depletion is observed
in the nuclear parton distributions.  The low $x$, or shadowing,
region and the larger $x$, or EMC region, is bridged by an enhancement
known as antishadowing for $0.1 < x < 0.3$.  For simplicity, the entire
characteristic modification as a function of $x$ has been referred to
as shadowing.

\begin{figure}[htbp]
\setlength{\epsfxsize=\textwidth}
\setlength{\epsfysize=0.6\textheight}
\centerline{\epsffile{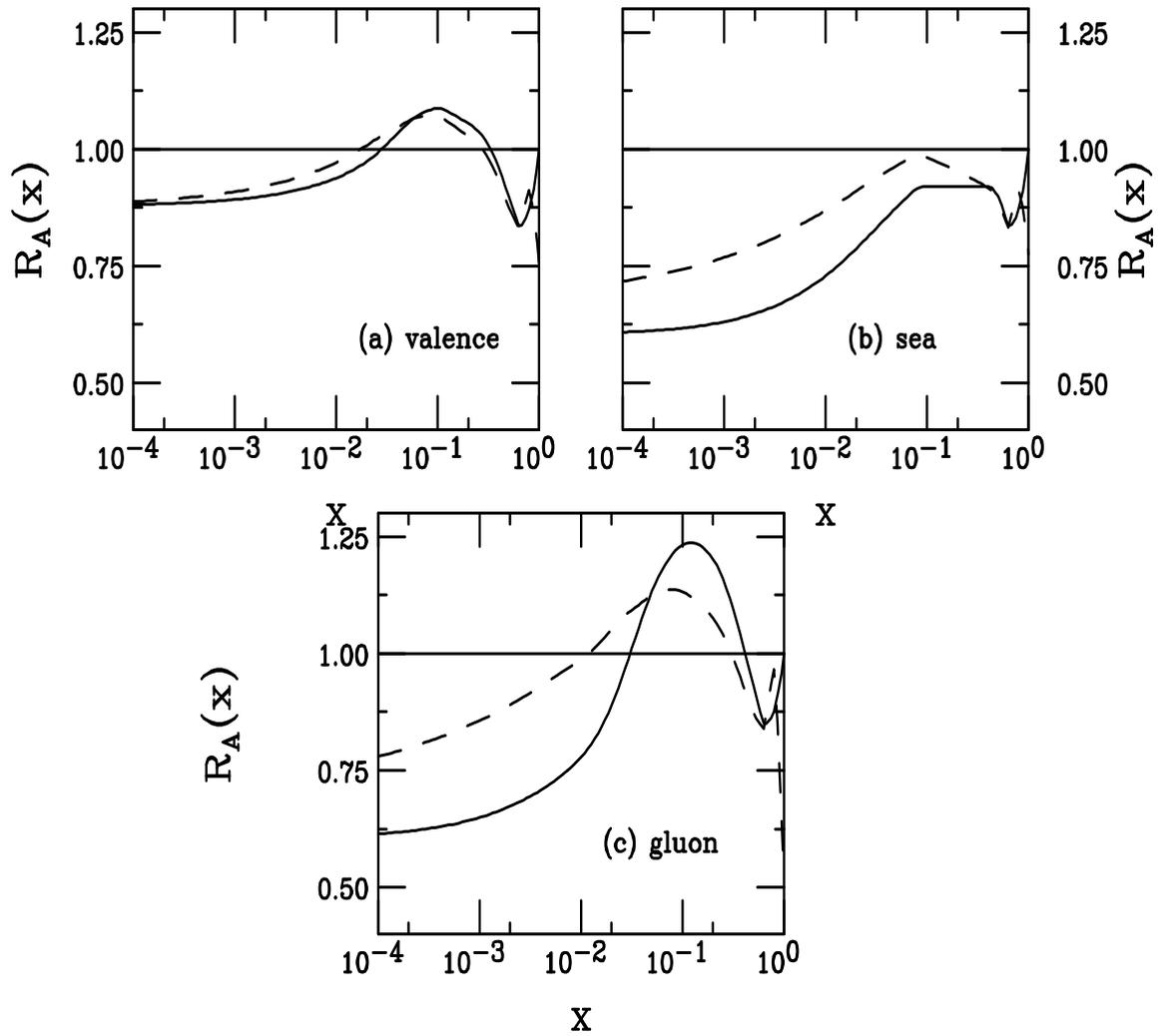}}
\caption[]{The shadowing parameterizations for $A =
200$ for (a) $u_V$ valence quarks, (b) $\overline u$ sea quarks, and (c) 
gluons.  The solid curves show the ratios at $\mu = \mu_0$ while the dashed 
curves are at $\mu=10$ GeV.}
\label{fshadow}
\end{figure}

Shadowing is an area of intense study with numerous models available
in the literature \cite{Arn}.  However, none of the models can
satisfactorily explain the behavior of the nuclear parton
distributions over the entire $x$ and $\mu^2$ range.  Therefore, we
choose to use parameterizations of shadowing based on 
nuclear DIS data.  As in DIS with
protons, the nuclear gluon distribution is not directly measured and
can only be inferred from conservation rules. 
We use the recent EKS98 parameterization
based on the GRV LO
\cite{grv} parton densities in the proton.  (See the contribution to this
volume by K.J. Eskola {\it et al.} for further details.)
Each parton type is evolved separately
above $\mu_0 = 1.5$ GeV \cite{EKRS3,EKRparam}.  The initial gluon
distribution shows significant antishadowing for $0.1<x<0.3$
while the sea quark distributions are shadowed. Figure~\ref{fshadow} compares
shows the shadowing ratios at $\mu=\mu_0$ and $\mu=10$ GeV.  The valence ratio
changes little with scale but the evolution of the gluon and sea ratios with 
scale is more significant.  The gluon ratio changes somewhat more
rapidly with scale than the sea ratio but the difference is not large.  Less
shadowing is seen in the antishadowing region at larger scales for the sea 
quark ratio while the gluon ratio decreases with scale in the same region.
The effects of shadowing alone on the $D \overline D$ and $B \overline B$ decay
contributions to the dilepton continuum have been studied in Ref.~\cite{EKV}.

Shadowing should depend on the spatial location of the parton in the nucleus.
Unfortunately, there is little data on this topic.  One relevant experiment,
Fermilab E745, studied the spatial distribution of nuclear parton densities
with $\nu N$ interactions in emulsion.  The presence of one or more
dark tracks from slow protons is used to infer a more central
interaction \cite{E745}.  For events with no dark tracks, no shadowing
is observed while for events with dark tracks, shadowing is enhanced
over other experimental measurements insensitive to the spatial location.

The spatial dependence of nuclear shadowing has been studied in charm
and bottom production \cite{us,firstprl}.  While the effect can be important 
in peripheral collisions since nucleons near the nuclear surface are expected
to behave as free particles, in central collisions where modifications are
larger than average, this increase is $\sim 1$\% over the average, negligible
for our purposes.

\yoursection{Rates}

All the results presented here are in units of cross section per nucleon pair.
The total rates in $pA$ and $AA$ interactions are obtained by extrapolation.
To calculate the rate of charm and bottom production in these interactions as a
function of impact parameter, the per nucleon $Q \overline Q$ cross sections
for $pA$ interactions given in this paper should be multiplied by the nuclear
profile function $T_A(b)$,
\begin{eqnarray}
N_A^{Q \overline Q}(b) = \sigma_{pA}^{Q \overline Q} T_A(b) \, \, .
\label{rate}
\end{eqnarray}
Likewise, the $AA$ rates are found by multiplying the $AA$ per nucleon cross
section by the nuclear overlap function, $T_{AA}(b) = \int d^2s T_A(s) T_A(|
\vec b - \vec s|)$,
\begin{eqnarray}
N_{AA}^{Q \overline Q}(b) = \sigma_{AA}^{Q \overline Q} T_{AA}(b) \, \, .
\label{rateaa}
\end{eqnarray}
Thus, integrating
over all impact parameters, the $pA$ cross section should scale as $A$
times the $pp$ cross section and the $AA$ cross section by $A^2$ times the $pp$
cross section without any nuclear effects.  Since the only nuclear effect 
discussed here that can change the total cross section is shadowing, total
charm and bottom production should have a nearly linear $A$ dependence.  The
exact $A$ dependence is energy dependent because higher energies correspond to
lower $x$ and larger shadowing effects.  One can maximize the $A$ dependence by
{\it e.g.}\ studying the rapidity distribution of low $p_T$ $Q \overline Q$
pairs. 

We note that in the case of $pA$ interactions, the impact parameter dependence
of charm and bottom production is the same as that of the geometric cross
section.  This similarity arises because the average number of collisions and 
the average number of participants is the same in $pA$ interactions.  In
nucleus-nucleus collisions, hard production is biased toward central
collisions, see Ref.~\cite{rvgeom} for more details.

\yoursection{Results}

We will show inclusive charm and bottom and exclusive $c \overline c$ and $b
\overline b$ distributions at several different energies.  First we show fixed
target charm production at 158, 450, and 800 GeV.  These energies are chosen
because Pb+Pb interactions are studied at the CERN SPS at 158 GeV/nucleon and
it would be useful to 
examine charm production in $pp$ and $p$Pb interactions at
the same energy.  However, since this would require a secondary beam, we also
show the results at the SPS primary proton beam energy of 450 GeV.  The
Fermilab fixed target program uses an 800 GeV proton 
beam which could also be employed in
studies of the charm $A$ dependence.  Indeed, a great many experiments have
studied charm production, both at CERN and at Fermilab, but unfortunately the
statistics for proton beams 
have generally been poor and the newer results have not been used to
determine the $A$ dependence except in very limited regions of phase space.  
Better statistics are available for $\pi$ beams
but the data is typically averaged over all targets to enhance statistics
rather than to study the $A$ dependence.  Since the RHIC and LHC heavy ion
programs will include $pA$ studies, we also show results at the complementary
ion-ion energy, $\sqrt{s} = 200$ GeV/nucleon and 5.5 TeV/nucleon, respectively.
Our $AA$ results are only shown for RHIC and LHC energies.

All the $pp$ calculations are shown with $\langle k_T^2 \rangle = 1$ GeV$^2$
and all the $pA$ calculations are presented with $\langle k_T^2 \rangle = 1.35$
GeV$^2$ for charm and 2.57 GeV$^2$ for bottom.  The $AA$ results are calculated
with $\langle k_T^2 \rangle = 1.7$ GeV$^2$ for charm and 4.16 GeV$^2$ for
bottom.  We note that the $\langle k_T^2 \rangle$ values used for bottom
are somewhat larger than may be suitable for the implementation in the MNR 
code.  The $pA$ and $AA$ calculations are
modified by spatially homogeneous shadowing with the EKS98 parameterization 
\cite{EKRS3,EKRparam} at $A=200$.

\begin{figure}[p]
\setlength{\epsfxsize=\textwidth}
\setlength{\epsfysize=0.6\textheight}
\centerline{\epsffile{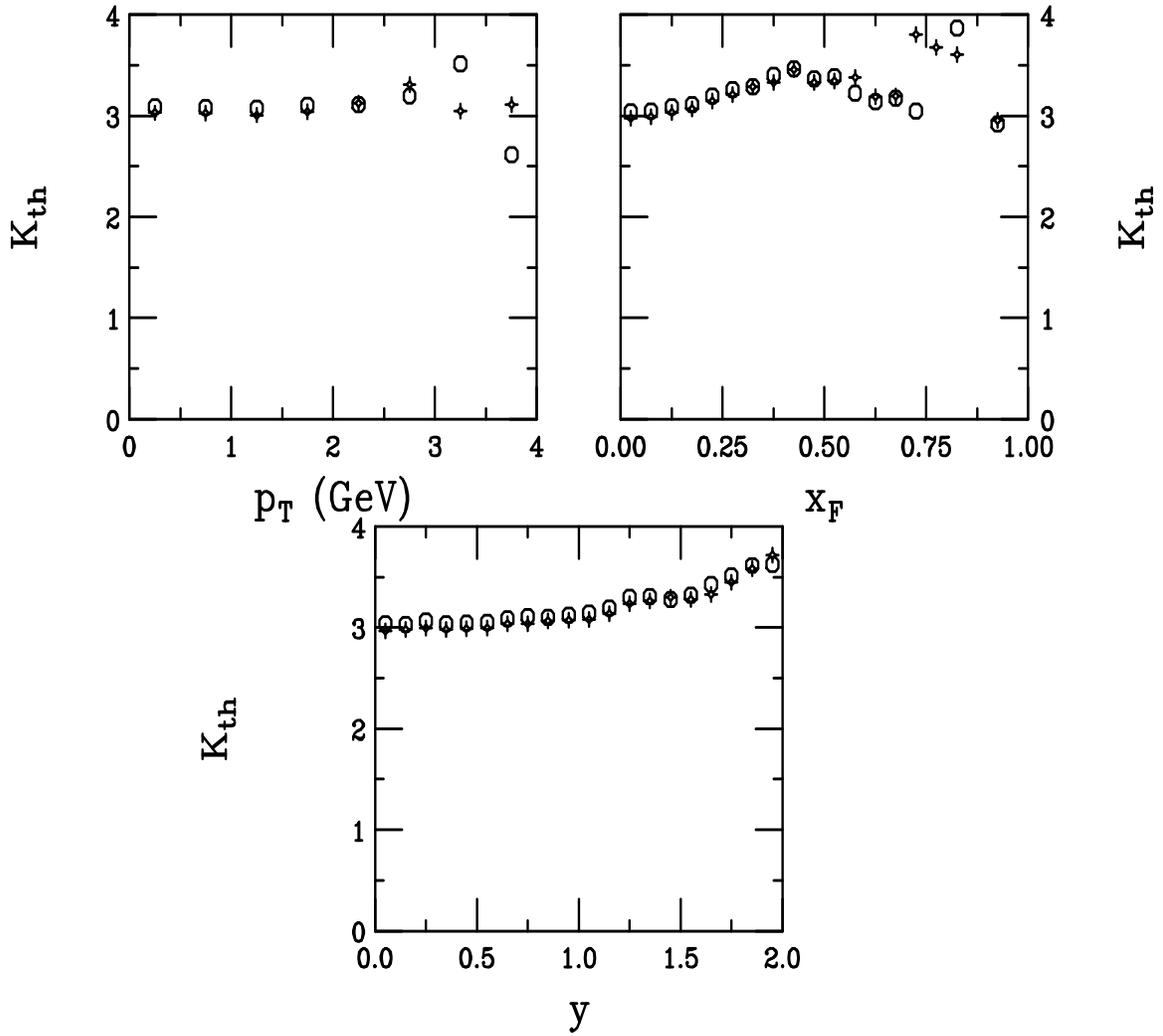}}
\caption[]{Ratio of NLO to LO $c$ quark production in $pp$ and $pA$
interactions at 158 GeV as a function of $p_T$, $x_F$, and $y$.
The crosses are the $pp$ results while the circles are the $pA$ 
results.} 
\label{k158}
\end{figure}

We first show that the theoretical $K$ factor, $K_{\rm th}$, 
the ratio of the NLO to LO cross sections,
both calculated with MRST HO in the MNR code 
is constant in $pp$ and $pA$ interactions
at 158 GeV in Fig.~\ref{k158}.  There is some variation near the edges of phase
space where the statistics become poor.  However, we have checked that neither
shadowing nor $k_T$ broadening leads to significant changes in $K_{\rm th}$.
The bare quark $K_{\rm th}$ has also been calculated at collider energies 
in Ref.~\cite{rvkfact} with the MRS D$-'$ distributions and was shown to be
constant with quark $p_T$ and $x_F$ as well as pair mass and rapidity.  
The exact value of $K_{\rm th}$ depends on the set of parton
distributions chosen and the interaction energy.

\begin{figure}[p]
\setlength{\epsfxsize=\textwidth}
\setlength{\epsfysize=0.6\textheight}
\centerline{\epsffile{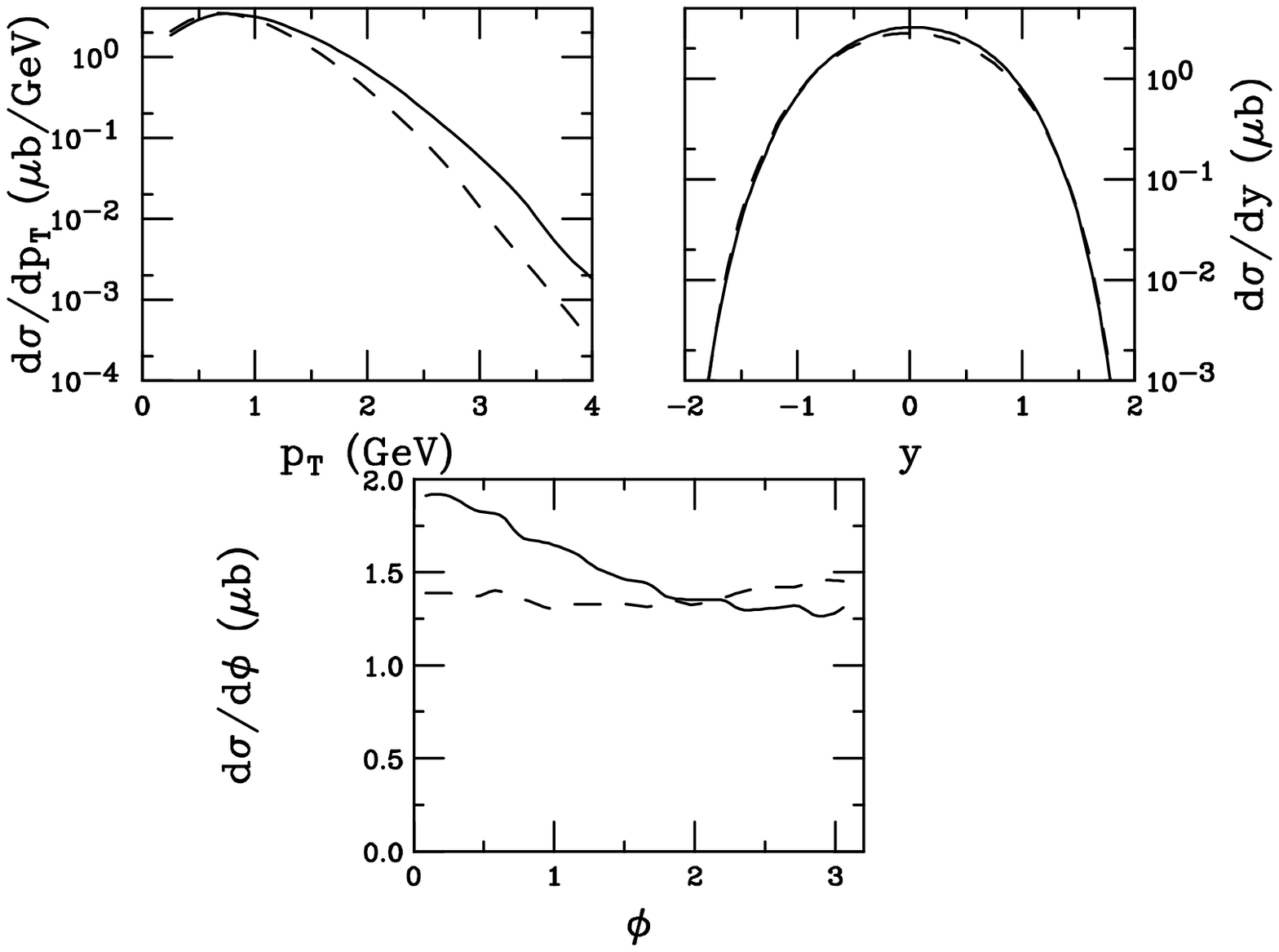}}
\caption[]{Exclusive NLO $c \overline c$ pair production in $pp$ and $pA$
interactions at 158 GeV as a function of $p_T$, $y$, and $\phi$.
The per nucleon cross section is given for $pA$ interactions.
The dashed curves are the $pp$ results while the solid curves are the $pA$ 
results.} 
\label{158pairs}
\end{figure}

\begin{figure}[p]
\setlength{\epsfxsize=\textwidth}
\setlength{\epsfysize=0.6\textheight}
\centerline{\epsffile{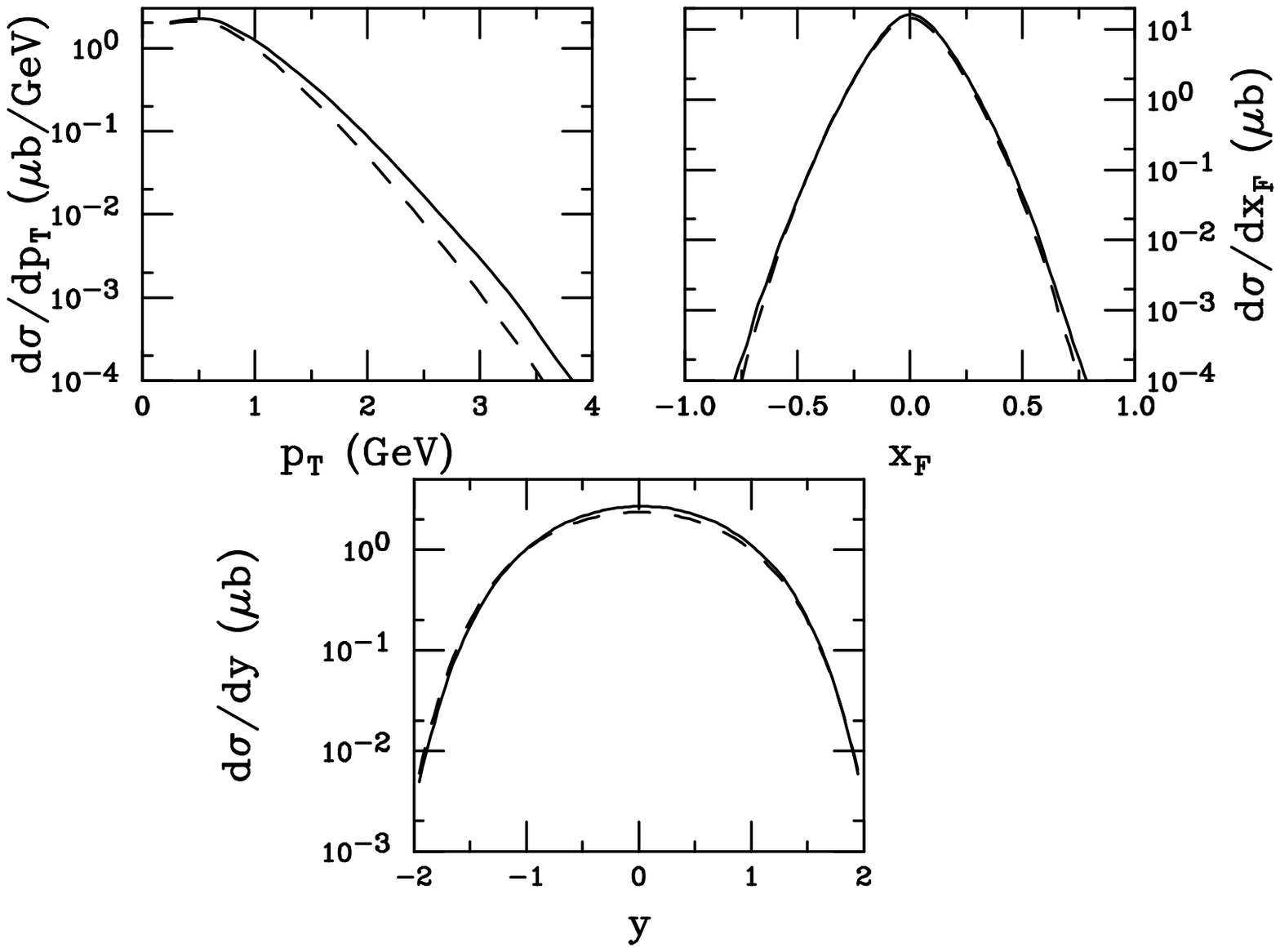}}
\caption[]{Inclusive NLO $c$ quark production in $pp$ and $pA$
interactions at 158 GeV as a function of $p_T$, $x_F$, and $y$.
The $p_T$ distributions are integrated over $x_F > 0$ only.
The per nucleon cross section is given for $pA$ interactions.
The dashed curves are the $pp$ results while the solid curves are the $pA$ 
results.} 
\label{158singles}
\end{figure}

\begin{figure}[p]
\setlength{\epsfxsize=\textwidth}
\setlength{\epsfysize=0.6\textheight}
\centerline{\epsffile{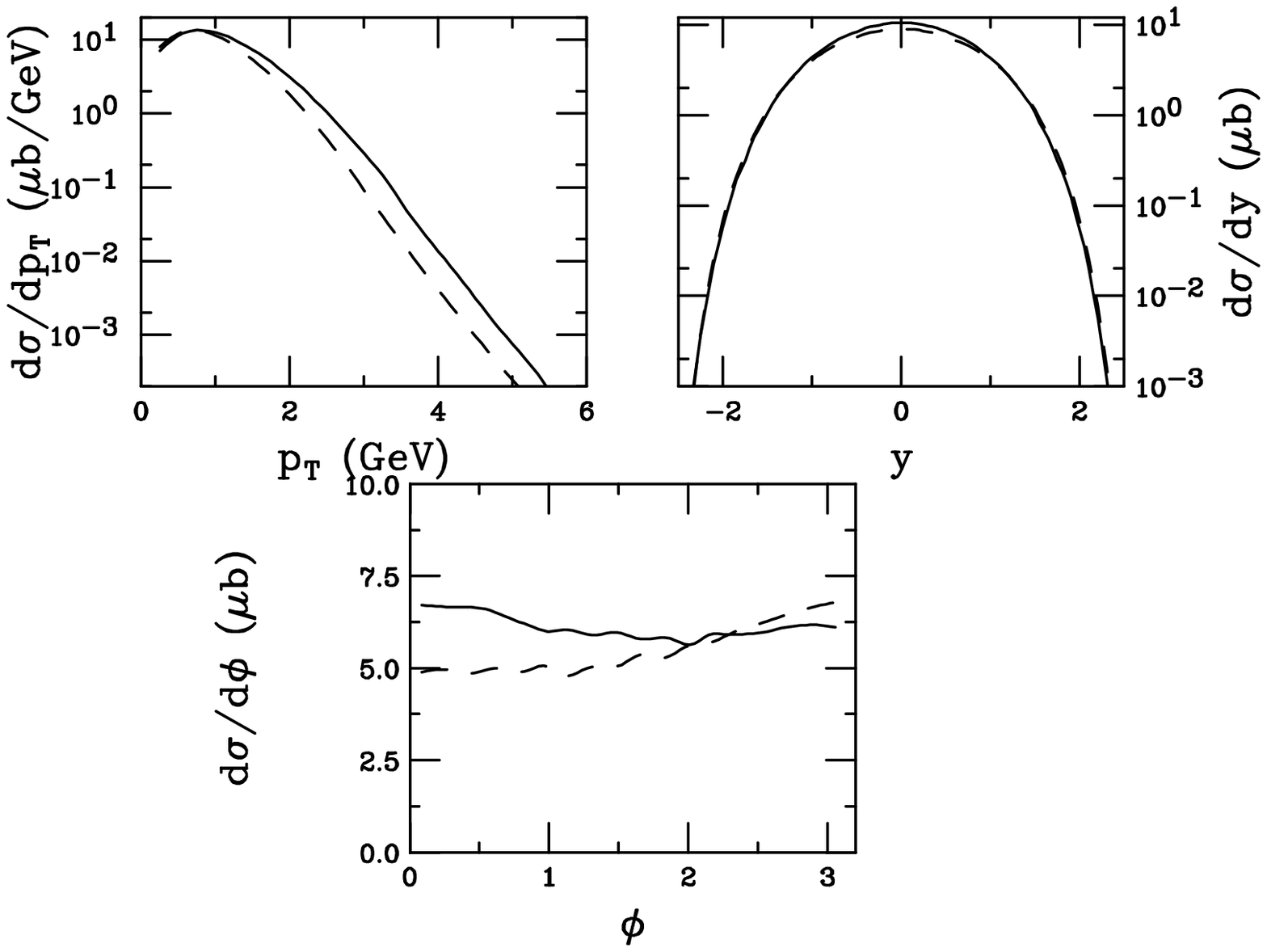}}
\caption[]{Exclusive NLO $c \overline c$ pair production in $pp$ and $pA$
interactions at 450 GeV as a function of $p_T$, $y$, and $\phi$.
The per nucleon cross section is given for $pA$ interactions.
The dashed curves are the $pp$ results while the solid curves are the $pA$ 
results.} 
\label{450pairs}
\end{figure}

\begin{figure}[p]
\setlength{\epsfxsize=\textwidth}
\setlength{\epsfysize=0.6\textheight}
\centerline{\epsffile{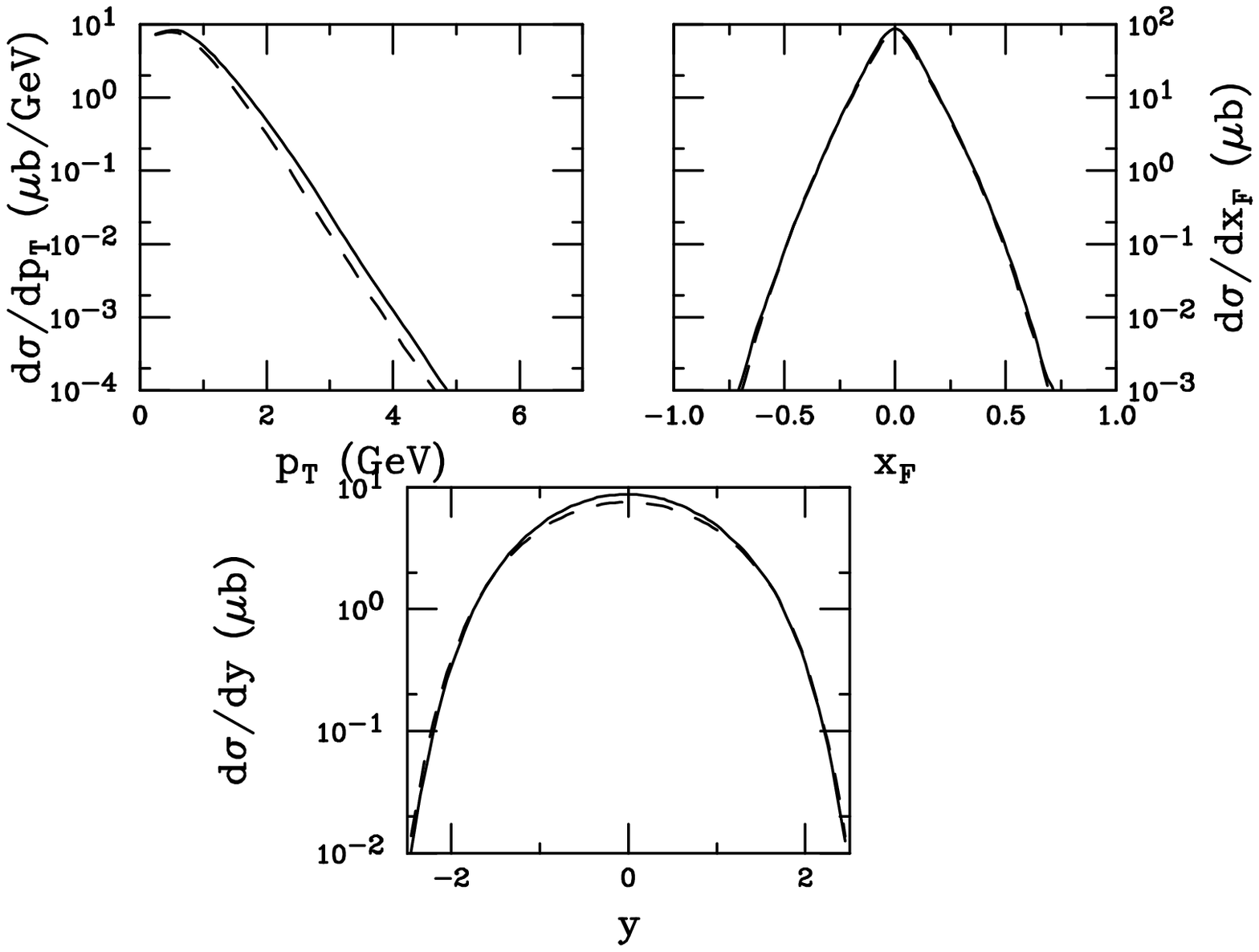}}
\caption[]{Inclusive NLO $c$ quark production in $pp$ and $pA$
interactions at 450 GeV as a function of $p_T$, $x_F$, and $y$.
The $p_T$ distributions are integrated over $x_F > 0$ only.
The per nucleon cross section is given for $pA$ interactions.
The dashed curves are the $pp$ results while the solid curves are the $pA$ 
results.} 
\label{450singles}
\end{figure}

\begin{figure}[p]
\setlength{\epsfxsize=\textwidth}
\setlength{\epsfysize=0.6\textheight}
\centerline{\epsffile{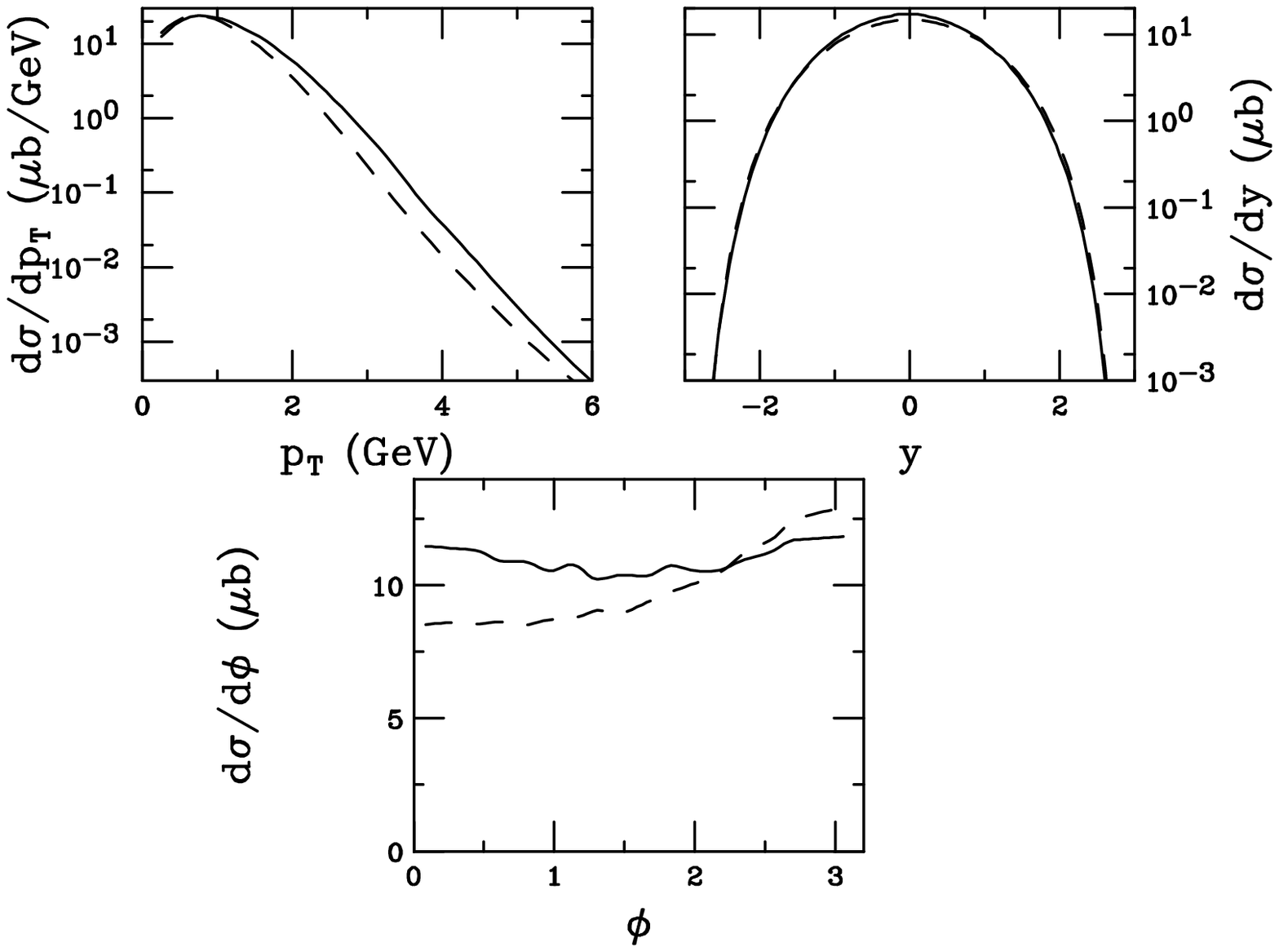}}
\caption[]{Exclusive NLO $c \overline c$ pair production in $pp$ and $pA$
interactions at 800 GeV as a function of $p_T$, $y$, and $\phi$.
The per nucleon cross section is given for $pA$ interactions.
The dashed curves are the $pp$ results while the solid curves are the $pA$ 
results.} 
\label{800pairs}
\end{figure}

\begin{figure}[p]
\setlength{\epsfxsize=\textwidth}
\setlength{\epsfysize=0.6\textheight}
\centerline{\epsffile{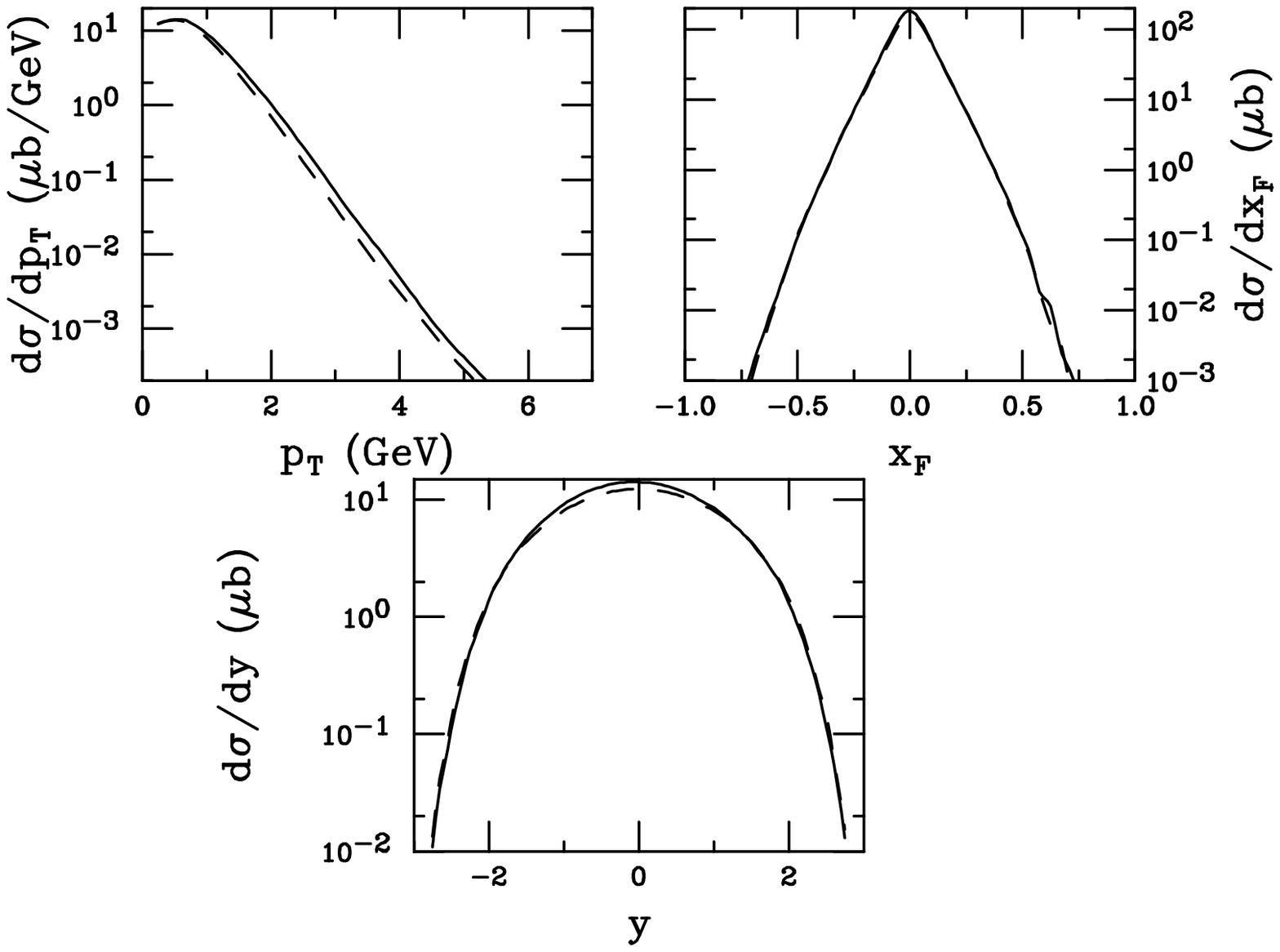}}
\caption[]{Inclusive NLO $c$ quark production in $pp$ and $pA$
interactions at 800 GeV as a function of $p_T$, $x_F$, and $y$.
The $p_T$ distributions are integrated over $x_F > 0$ only.
The per nucleon cross section is given for $pA$ interactions.
The dashed curves are the $pp$ results while the solid curves are the $pA$ 
results.} 
\label{800singles}
\end{figure}

\begin{figure}[p]
\setlength{\epsfxsize=\textwidth}
\setlength{\epsfysize=0.6\textheight}
\centerline{\epsffile{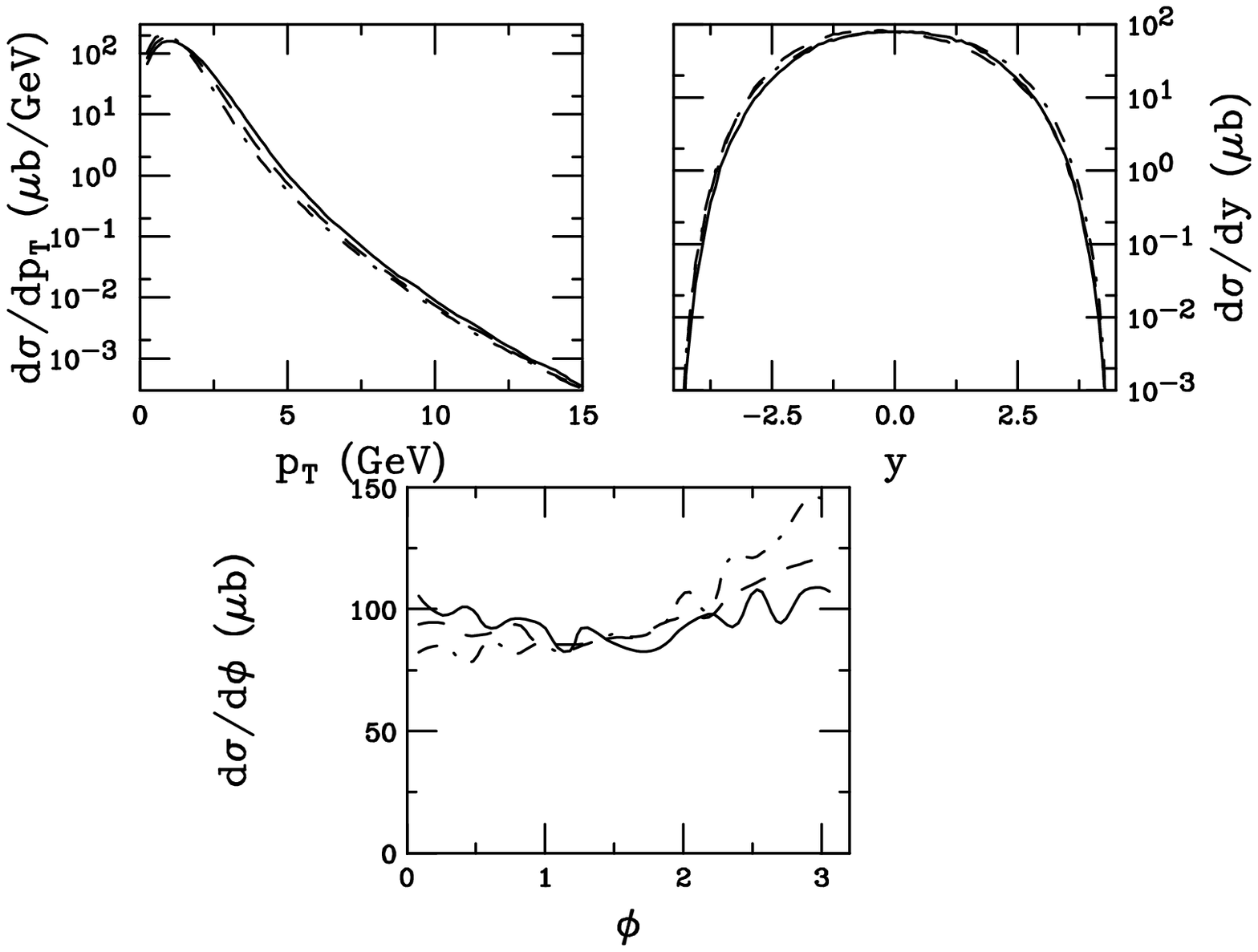}}
\caption[]{Exclusive NLO $c \overline c$ pair production in $pp$, $pA$ and $AA$
interactions at $\sqrt{s} = 200$ GeV as a function of 
$p_T$, $y$, and $\phi$.
The per nucleon cross section is given for $pA$ and $AA$ interactions.
The dot-dashed curves are the $pp$ results, the dashed curves are the $pA$
results, and the solid curves are the $AA$ 
results.} 
\label{c200pairs}
\end{figure}

\begin{figure}[p]
\setlength{\epsfxsize=\textwidth}
\setlength{\epsfysize=0.3\textheight}
\centerline{\epsffile{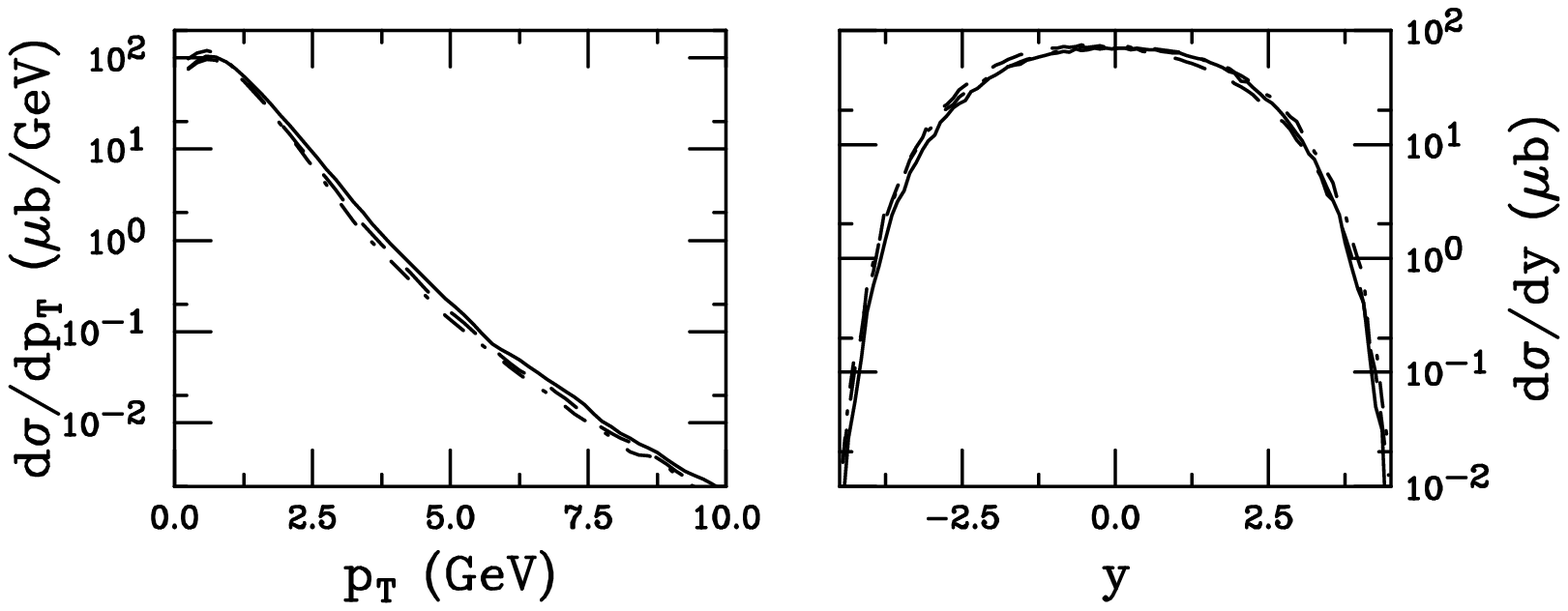}}
\caption[]{Inclusive NLO $c$ quark production in $pp$, $pA$ and $AA$ 
interactions at $\sqrt{s} = 200$ GeV 
as a function of $p_T$ and $y$.
The $p_T$ distributions are integrated over $x_F > 0$ only.
The per nucleon cross section is given for $pA$ and $AA$ interactions.
The dot-dashed curves are the $pp$ results, the dashed curves are the $pA$
results, and the solid curves are the $AA$ 
results.} 
\label{c200singles}
\end{figure}

\begin{figure}[p]
\setlength{\epsfxsize=\textwidth}
\setlength{\epsfysize=0.6\textheight}
\centerline{\epsffile{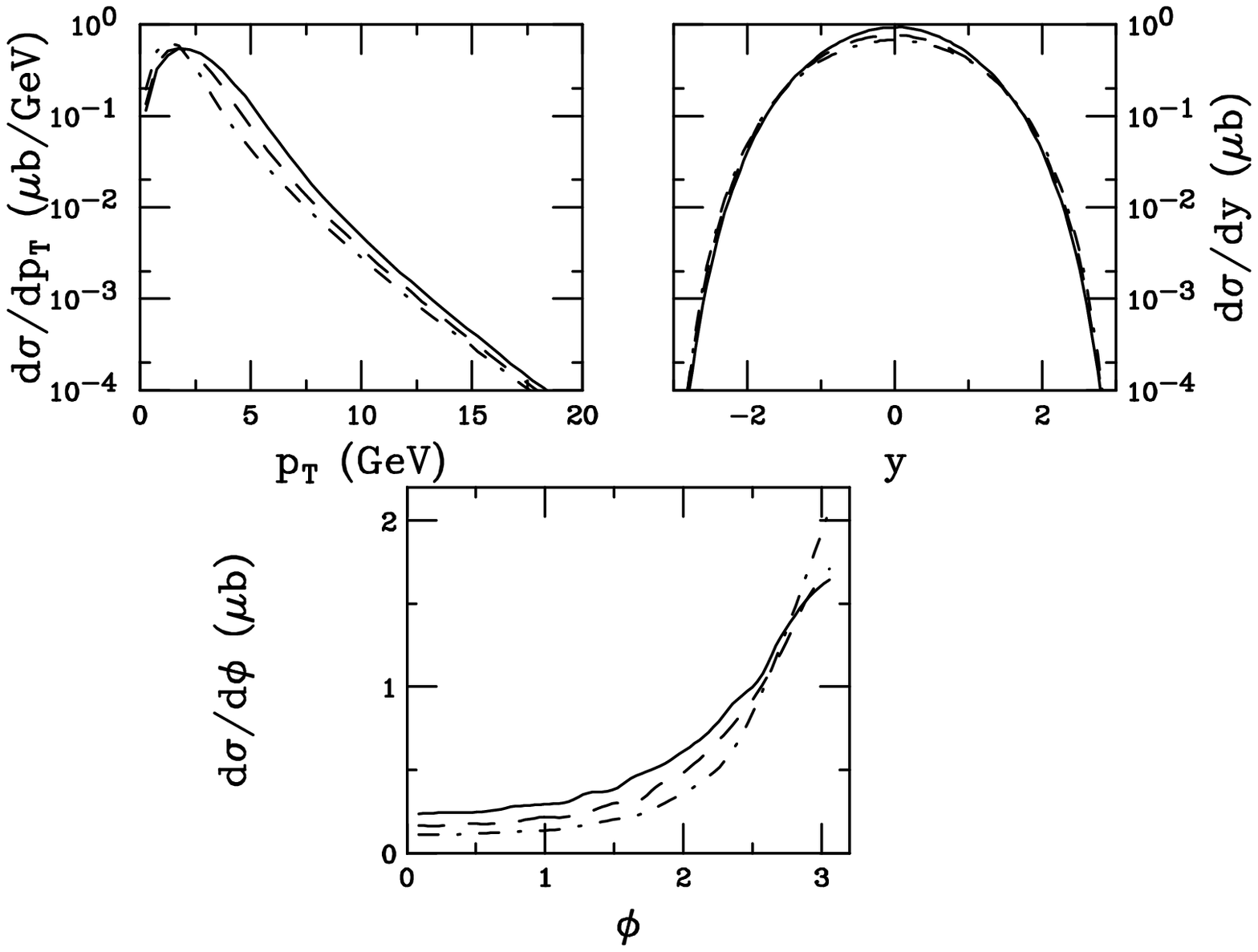}}
\caption[]{Exclusive NLO $b \overline b$ pair production in $pp$, $pA$ and $AA$
interactions at $\sqrt{s} = 200$ GeV 
as a function of $p_T$, $y$, and $\phi$.
The per nucleon cross section is given for $pA$ and $AA$ interactions.
The dot-dashed curves are the $pp$ results, the dashed curves are the $pA$
results, and the solid curves are the $AA$ 
results.} 
\label{b200pairs}
\end{figure}

\begin{figure}[p]
\setlength{\epsfxsize=\textwidth}
\setlength{\epsfysize=0.3\textheight}
\centerline{\epsffile{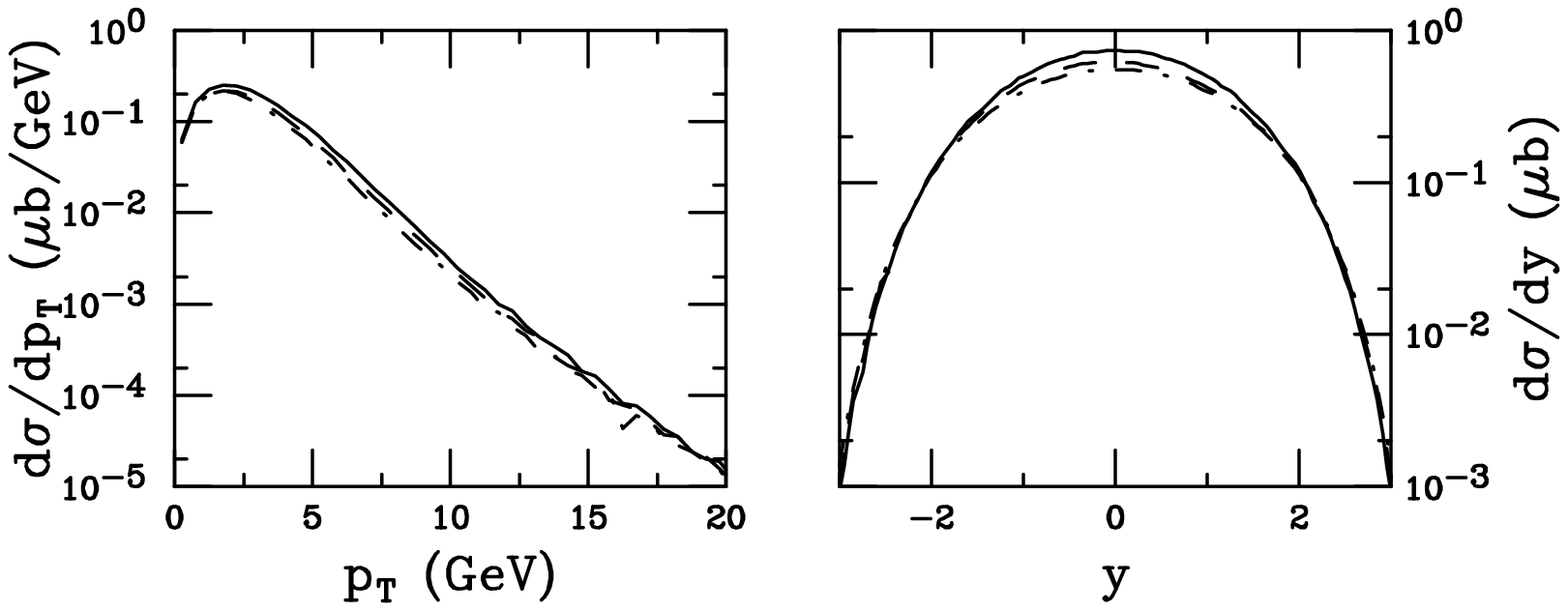}}
\caption[]{Inclusive NLO $b$ quark production in $pp$, $pA$ and $AA$
interactions at $\sqrt{s} = 200$ GeV 
as a function of $p_T$ and $y$.
The $p_T$ distributions are integrated over $x_F > 0$ only.
The per nucleon cross section is given for $pA$ and $AA$ interactions.
The dot-dashed curves are the $pp$ results, the dashed curves are the $pA$
results, and the solid curves are the $AA$ 
results.} 
\label{b200singles}
\end{figure}

\begin{figure}[p]
\setlength{\epsfxsize=\textwidth}
\setlength{\epsfysize=0.6\textheight}
\centerline{\epsffile{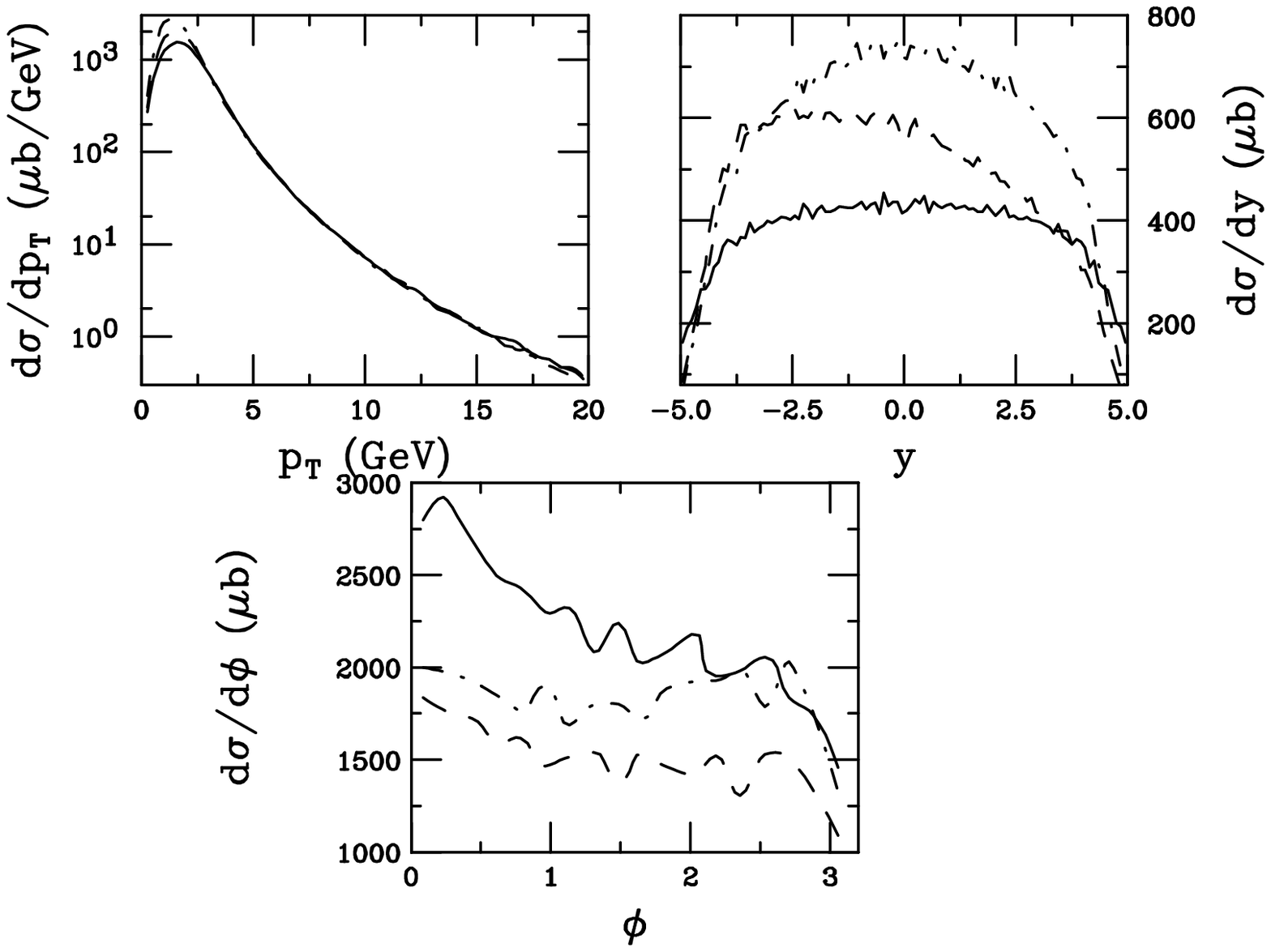}}
\caption[]{Exclusive NLO $c \overline c$ pair production in $pp$, $pA$ and $AA$
interactions at $\sqrt{s} = 5.5$ TeV 
as a function of $p_T$, $y$, and $\phi$.
The per nucleon cross section is given for $pA$ and $AA$ interactions.
The dot-dashed curves are the $pp$ results, the dashed curves are the $pA$
results, and the solid curves are the $AA$ 
results.} 
\label{c5500pairs}
\end{figure}

\begin{figure}[p]
\setlength{\epsfxsize=\textwidth}
\setlength{\epsfysize=0.3\textheight}
\centerline{\epsffile{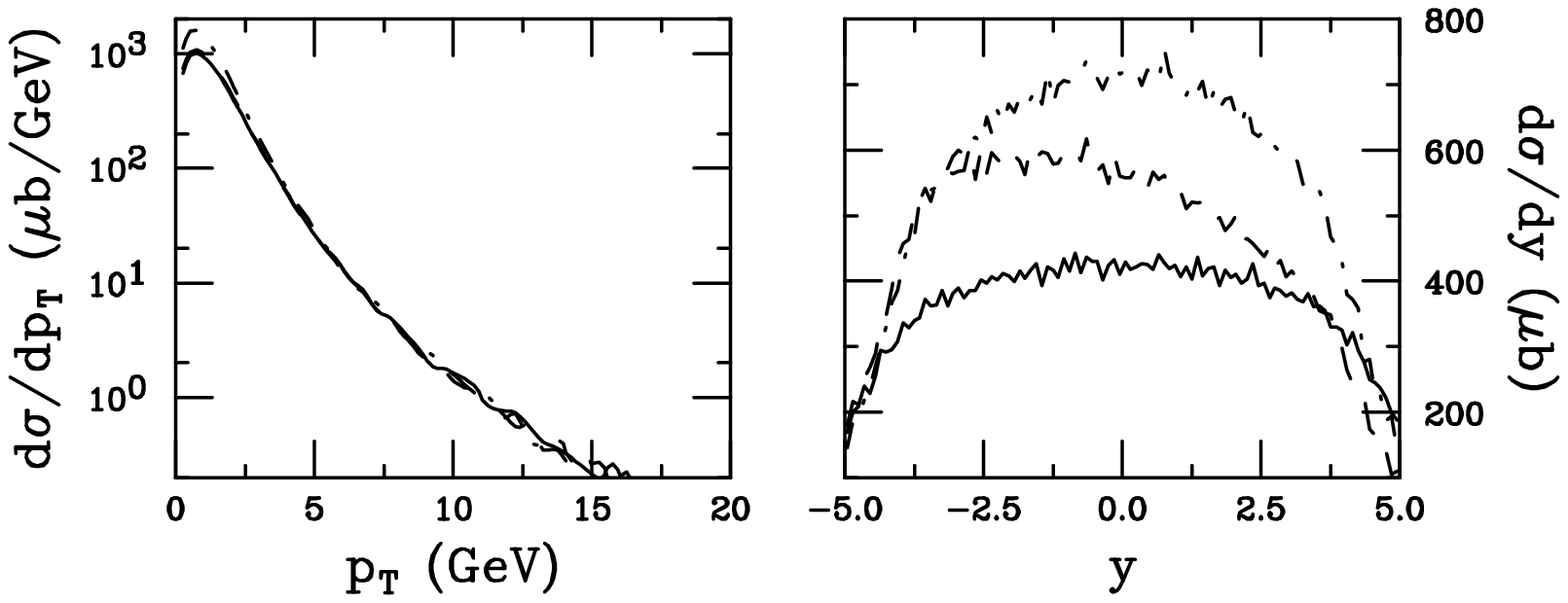}}
\caption[]{Inclusive NLO $c$ quark production in $pp$, $pA$ and $AA$
interactions at $\sqrt{s} = 5.5$ TeV 
as a function of $p_T$ and $y$.
The $p_T$ distributions are integrated over $x_F > 0$ only.
The per nucleon cross section is given for $pA$ and $AA$ interactions.
The dot-dashed curves are the $pp$ results, the dashed curves are the $pA$
results, and the solid curves are the $AA$ 
results.} 
\label{c5500singles}
\end{figure}

\begin{figure}[p]
\setlength{\epsfxsize=\textwidth}
\setlength{\epsfysize=0.6\textheight}
\centerline{\epsffile{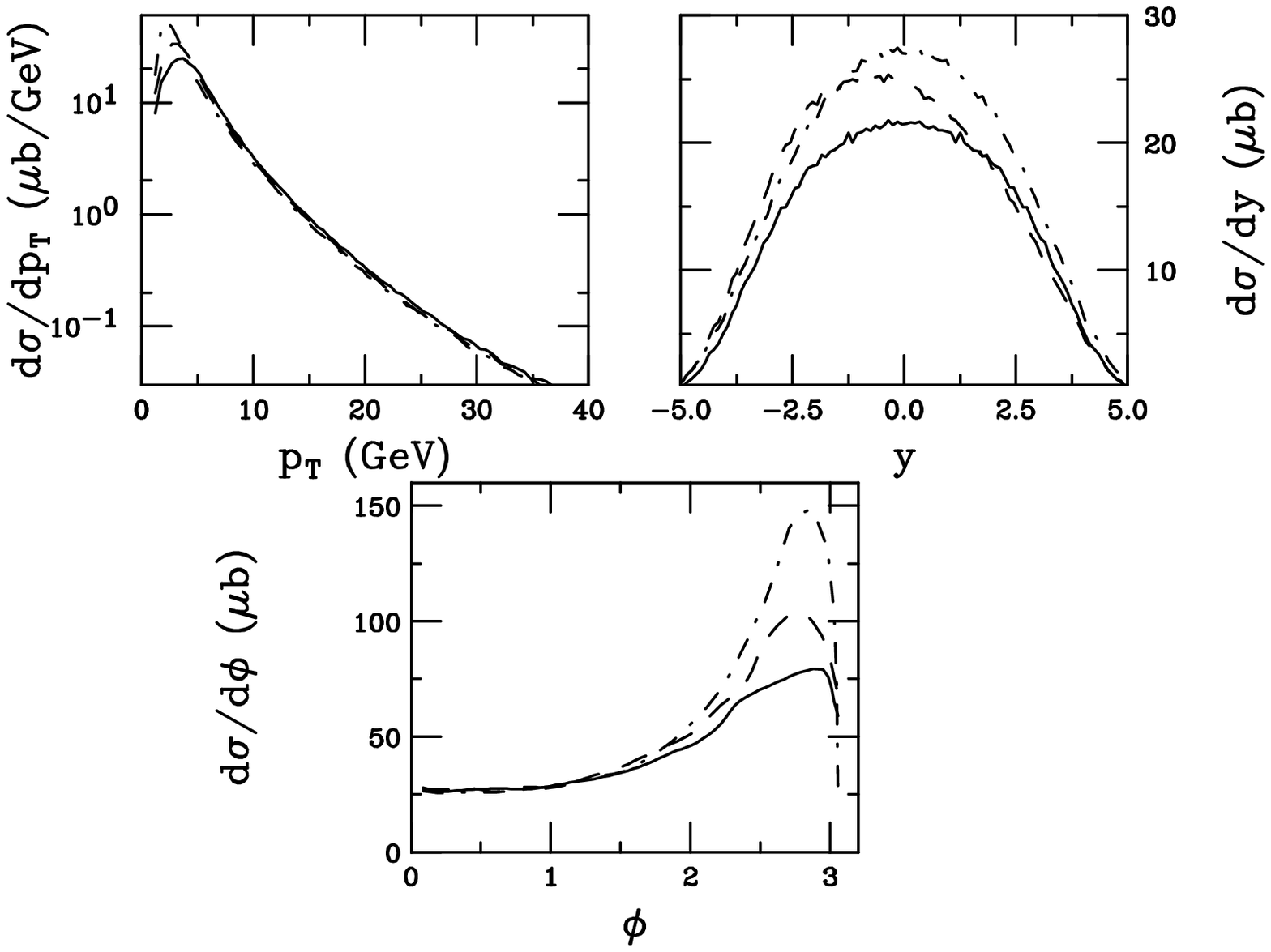}}
\caption[]{Exclusive NLO $b \overline b$ pair production in $pp$, $pA$ and $AA$
interactions at $\sqrt{s} = 5.5$ TeV 
as a function of $p_T$, $y$, and $\phi$.
The per nucleon cross section is given for $pA$ and $AA$ interactions.
The dot-dashed curves are the $pp$ results, the dashed curves are the $pA$
results, and the solid curves are the $AA$ 
results.} 
\label{b5500pairs}
\end{figure}

\begin{figure}[p]
\setlength{\epsfxsize=\textwidth}
\setlength{\epsfysize=0.3\textheight}
\centerline{\epsffile{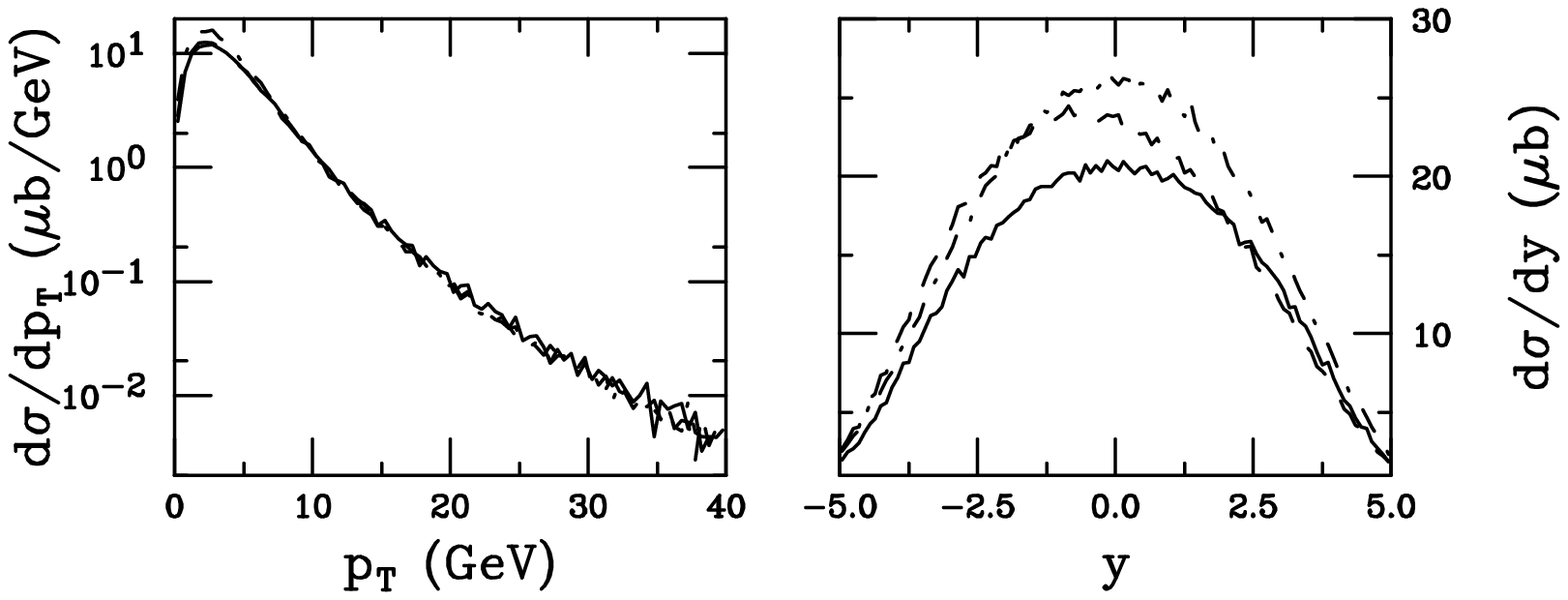}}
\caption[]{Inclusive NLO $b$ quark production in $pp$, $pA$ and $AA$
interactions at $\sqrt{s} = 5.5$ TeV 
as a function of $p_T$ and $y$.
The $p_T$ distributions are integrated over $x_F > 0$ only.
The per nucleon cross section is given for $pA$ and $AA$ interactions.
The dot-dashed curves are the $pp$ results, the dashed curves are the $pA$
results, and the solid curves are the $AA$ 
results.} 
\label{b5500singles}
\end{figure}

The exclusive and inclusive cross sections are shown for all energies
considered in Figs.~\ref{158pairs}-\ref{b5500singles}.  The exclusive results
are given for the pair $p_T$, rapidity, and $\phi$ distributions while the
inclusive results are given as a function of quark $p_T$, $x_F$, and rapidity.
Note that at collider energies, we do not show the $x_F$ results since the
distributions are very narrow at high energies and the rapidity distributions 
are thus more useful.  
The pair $p_T$ and $\phi$ distributions and the quark $p_T$
distribution all show the effect of $k_T$ broadening in $pA$ and $AA$ over $pp$
interactions.  The broadening is most apparent at fixed target energies and
at low $p_T$ for the collider energies.  The effect is always stronger on the
pair quantities.  On the other hand, information on nuclear shadowing is more 
likely to be obtained from the rapidity distributions.  The most useful
variable for studying shadowing effects is the pair rapidity distribution since
Fig.~\ref{158pairskt} showed that neither $k_T$ broadening nor fragmentation 
affects the pair rapidity. While the quark rapidity distribution in
Fig.~\ref{158singleskt} did show a difference between the bare quark rapidity
distribution and the distributions with $k_T$ broadening included, 
the change is due to
fragmentation rather than the $k_T$ effect.  Since shadowing is not symmetric
around $y=0$, we present results over all rapidity and $x_F$.  Tables of the
pair and single quark $p_T$ and pair and single quark rapidity for the
$pp$, $pA$ and $AA$ results are also given.

As the center of mass energy is increased, the effect of $k_T$ 
broadening on the
$c \overline c$ pair $p_T$ and $\phi$ distributions is seen to diminish in
Figs.~\ref{158pairs}, \ref{450pairs}, \ref{800pairs}, \ref{c200pairs}, and
\ref{c5500pairs}.  At the same time, the $pA$ pair rapidity distributions
exhibit a depletion at forward $x_F$ relative to the $pp$ distribution.  This
effect, most obvious on the linear scale of Fig.~\ref{c5500pairs}, is due to
shadowing and is biased toward forward rapidities where the smallest target $x$
values are probed, $x_2 \sim 3 \times 10^{-6}$ for pairs with $p_T \sim 0$ and
$y=5$.  The $b \overline b$ pair results in Figs.~\ref{b200pairs} and
\ref{b5500pairs} exhibit the same basic trends.  Note however that the $b
\overline b$ pairs tend to have much larger $p_T$ than the charm quarks.
In $AA$ collisions, the rapidity distributions are again symmetric about $y=0$
because now negative rapidity corresponds to low $x_1$ and the 
shadowing results
are thus essentially the product of the target shadowing ($x_2$) and the 
projectile shadowing ($x_1$).  

The pair $p_T$ is generally higher than for single quarks at any given energy.
At NLO, large $p_T$ $Q \overline Q$ pairs come from {\it e.g.}\ $gg \rightarrow
gg$ jets where one final state gluon splits into $Q \overline Q$, $g
\rightarrow Q \overline Q$.  In this way, the pair gets the entire $p_T$ of the
gluon while the single quark gets only a fraction of the gluon $p_T$.
Conversely the single quark rapidity is generally greater than that of the
pair.  That is because $y_{Q \overline Q} = 1/2 \ln ((E_Q +E_{\overline Q} +
p_{z Q} + p_{z \overline Q})/(E_Q +E_{\overline Q} -
p_{z Q} - p_{z \overline Q}))$ so that if $p_{z Q}$ is large and positive, it
may be balanced by a large negative $p_{z \overline Q}$.  The single
quark nuclear effects show the same trends as those of the pair.  Note that the
single quark $p_T$ distributions are all for $x_F > 0$ only.

\begin{figure}[p]
\setlength{\epsfxsize=\textwidth}
\setlength{\epsfysize=0.6\textheight}
\centerline{\epsffile{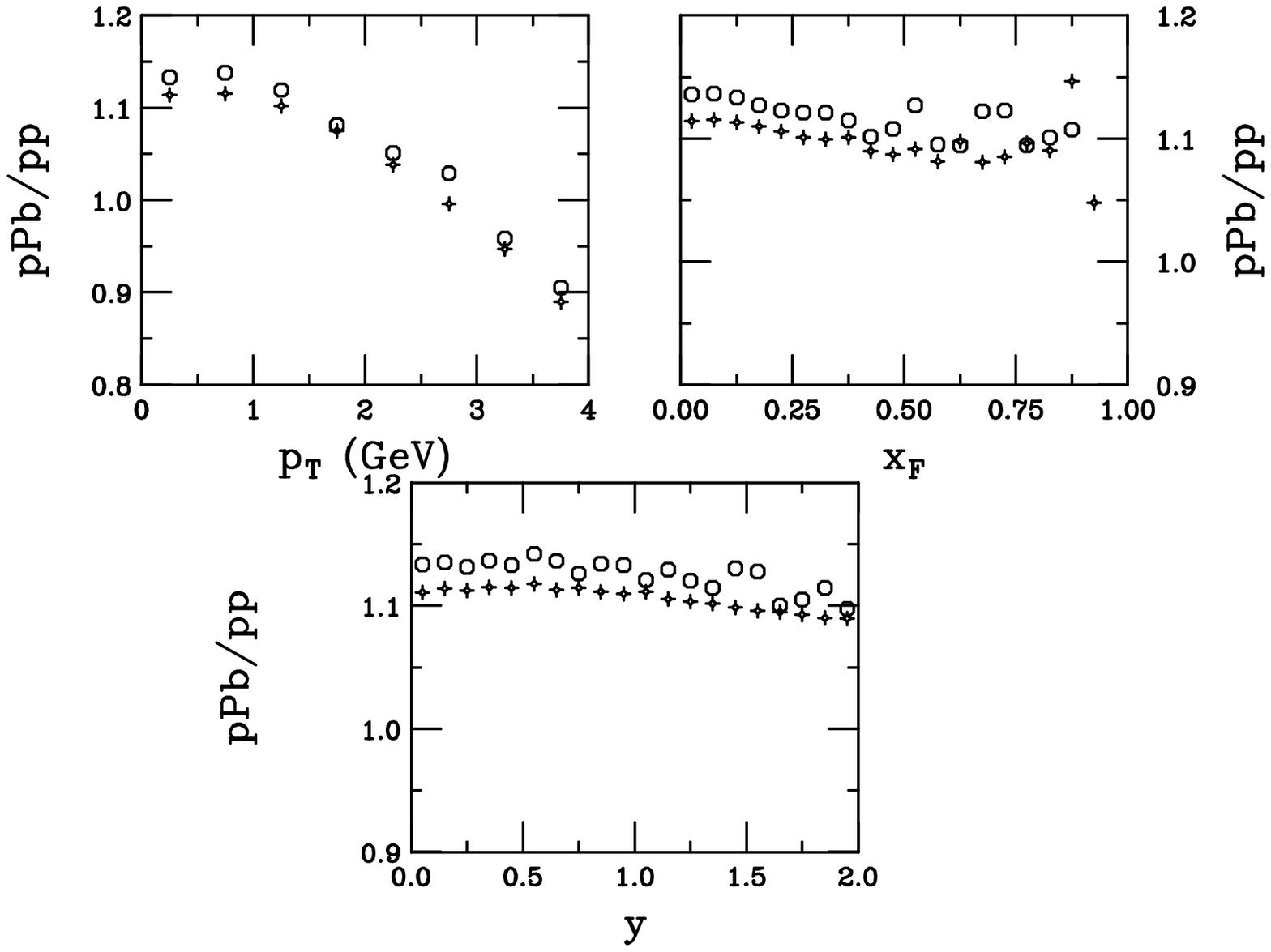}}
\caption[]{Ratio of $c$ quark production in $pA$ to $pp$
interactions at 158 GeV as a function of $p_T$, $x_F$, and $y$.
The crosses are the LO results while the circles are the NLO
results.} 
\label{s158nopt}
\end{figure}

Figure~\ref{s158nopt} shows the effect of shadowing alone on the $pA/pp$ ratios
of the bare charm quark distributions at 158 GeV.  Without $k_T$ broadening,
the ratio decreases with $p_T$ since $x_2 \sim 0.5$ at $y=0$, well into the EMC
region where the shadowing ratio decreases with $x$.  The effect at low $p_T$
can be observed in the $x_F$ and $y$ ratios where the ratios tend to be greater
than unity since $x_2$ is in the antishadowing region, $x_2 \sim 0.14$ at $y=0$
and $\sim 0.05$ at $y=1$, still in the antishadowing region for gluons with low
$\mu^2$.  The LO and NLO ratios, both calculated with MRST HO, are quite
similar so that the order of the calculation is not strongly influenced by
shadowing. 

\begin{figure}[p]
\setlength{\epsfxsize=\textwidth}
\setlength{\epsfysize=0.6\textheight}
\centerline{\epsffile{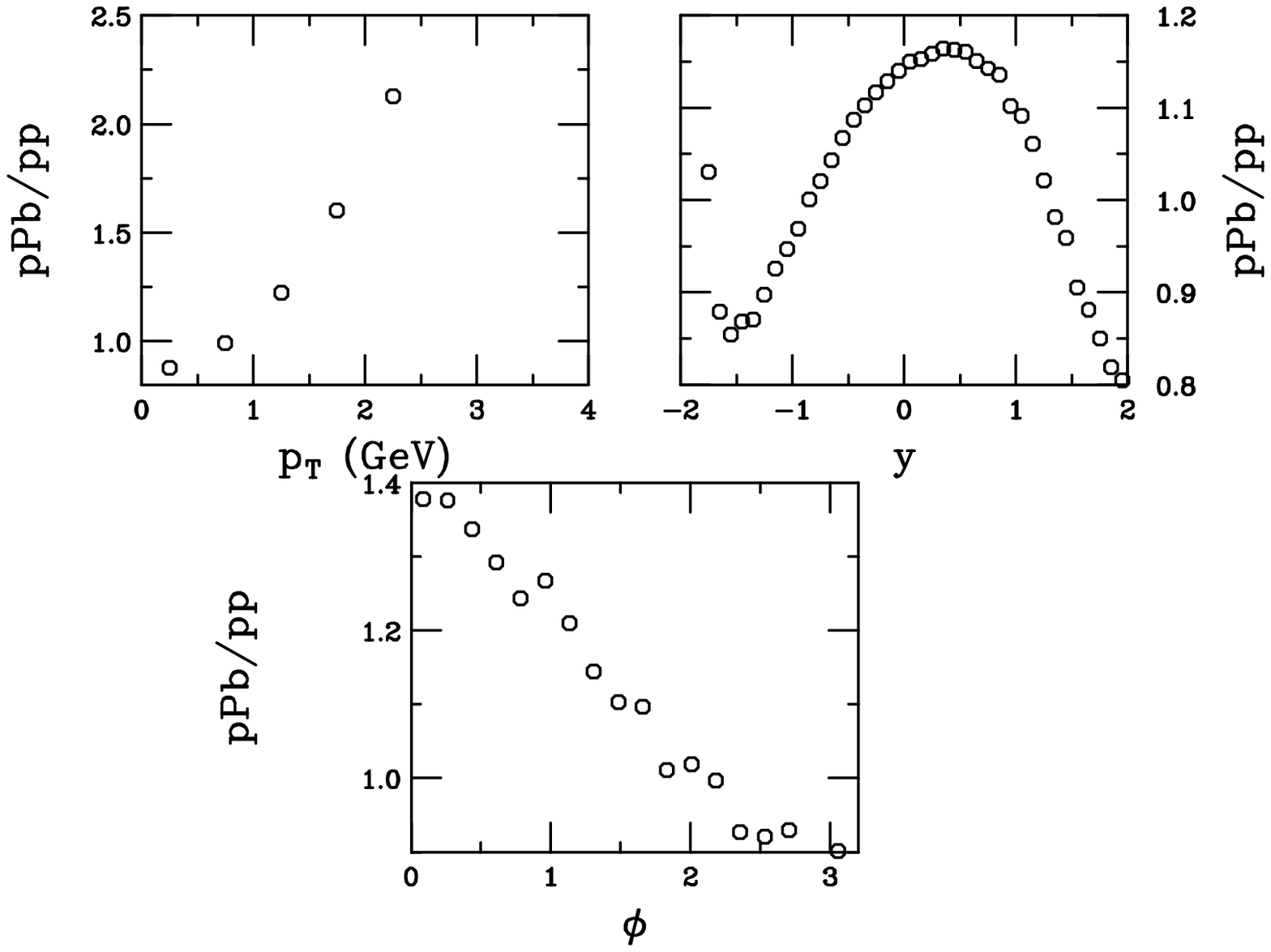}}
\caption[]{The ratio of $pA$ to $pp$ exclusive NLO $c \overline c$ pair 
production at 158 GeV as a function of $p_T$, $y$, and $\phi$.}
\label{158rat_p}
\end{figure}

\begin{figure}[p]
\setlength{\epsfxsize=\textwidth}
\setlength{\epsfysize=0.6\textheight}
\centerline{\epsffile{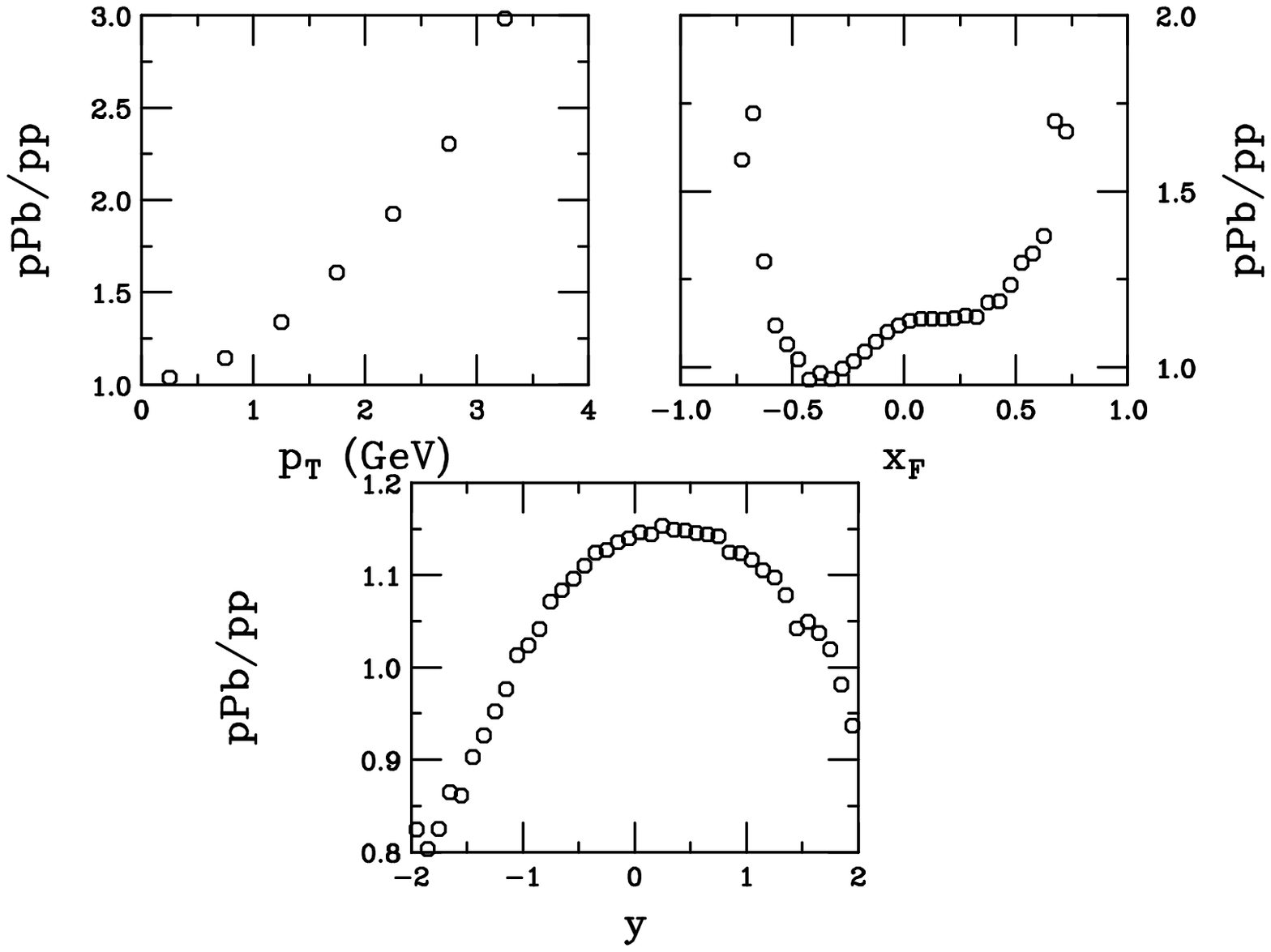}}
\caption[]{The ratio of $pA$ to $pp$ inclusive NLO $c$ quark production 
at 158 GeV as a function of $p_T$, $x_F$, and $y$.}
\label{158rat_s}
\end{figure}

\begin{figure}[p]
\setlength{\epsfxsize=\textwidth}
\setlength{\epsfysize=0.6\textheight}
\centerline{\epsffile{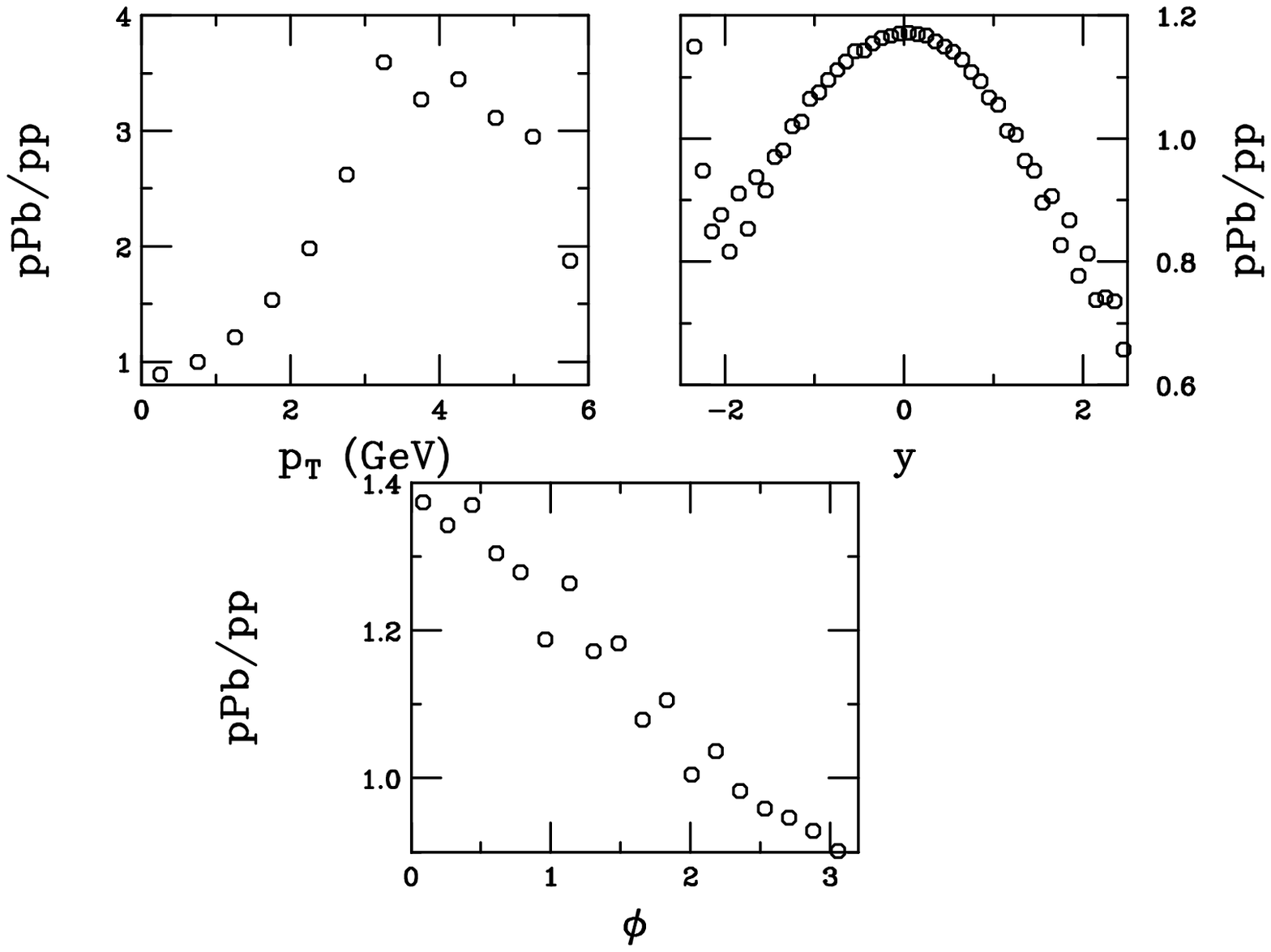}}
\caption[]{The ratio of $pA$ to $pp$ exclusive NLO $c \overline c$ pair 
production at 450 GeV as a function of $p_T$, $y$, and $\phi$.}
\label{450rat_p}
\end{figure}

\begin{figure}[p]
\setlength{\epsfxsize=\textwidth}
\setlength{\epsfysize=0.6\textheight}
\centerline{\epsffile{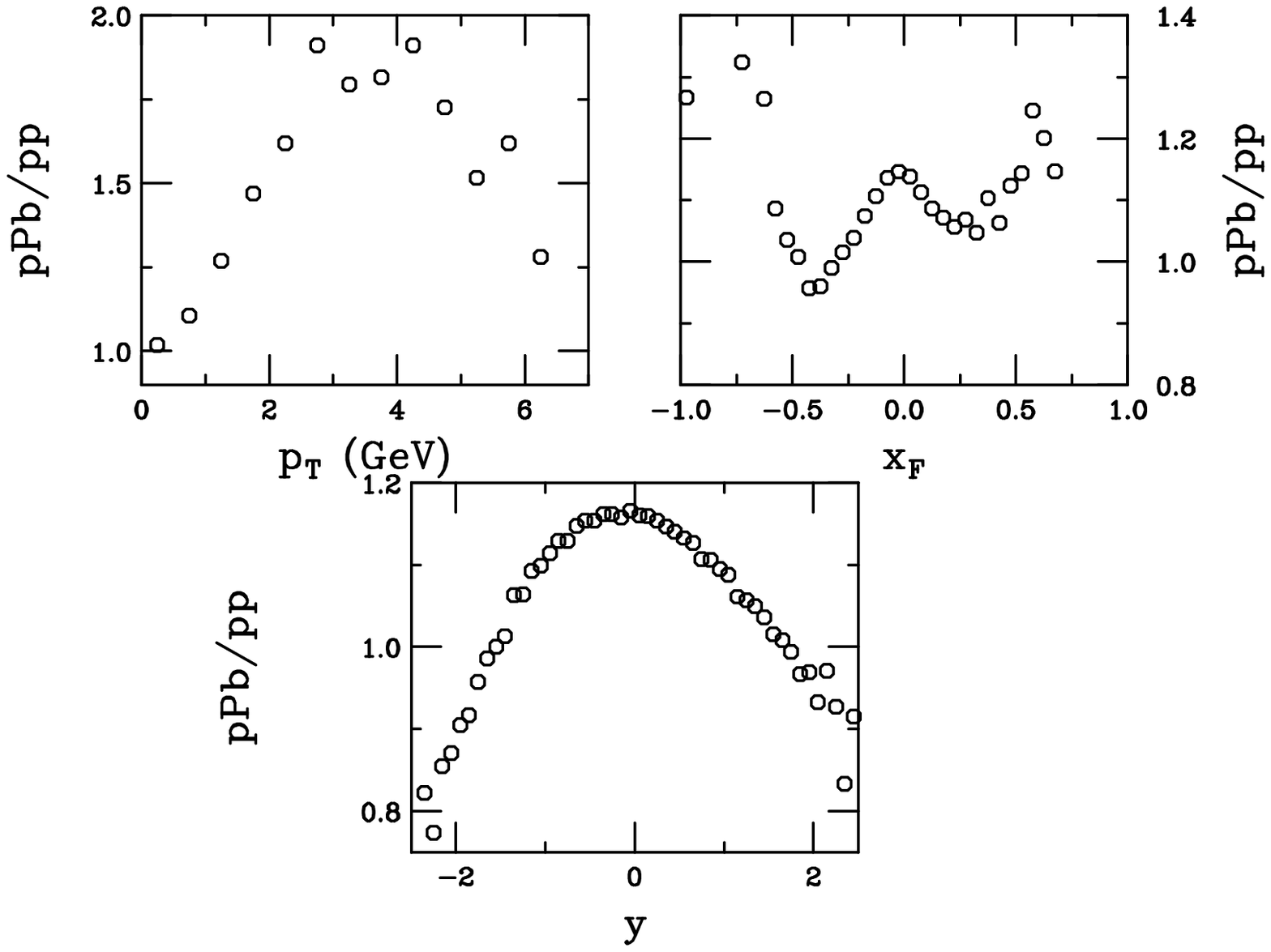}}
\caption[]{The ratio of $pA$ to $pp$ inclusive NLO $c$ quark production  
at 450 GeV as a function of $p_T$, $x_F$, and $y$.}
\label{450rat_s}
\end{figure}

\begin{figure}[p]
\setlength{\epsfxsize=\textwidth}
\setlength{\epsfysize=0.6\textheight}
\centerline{\epsffile{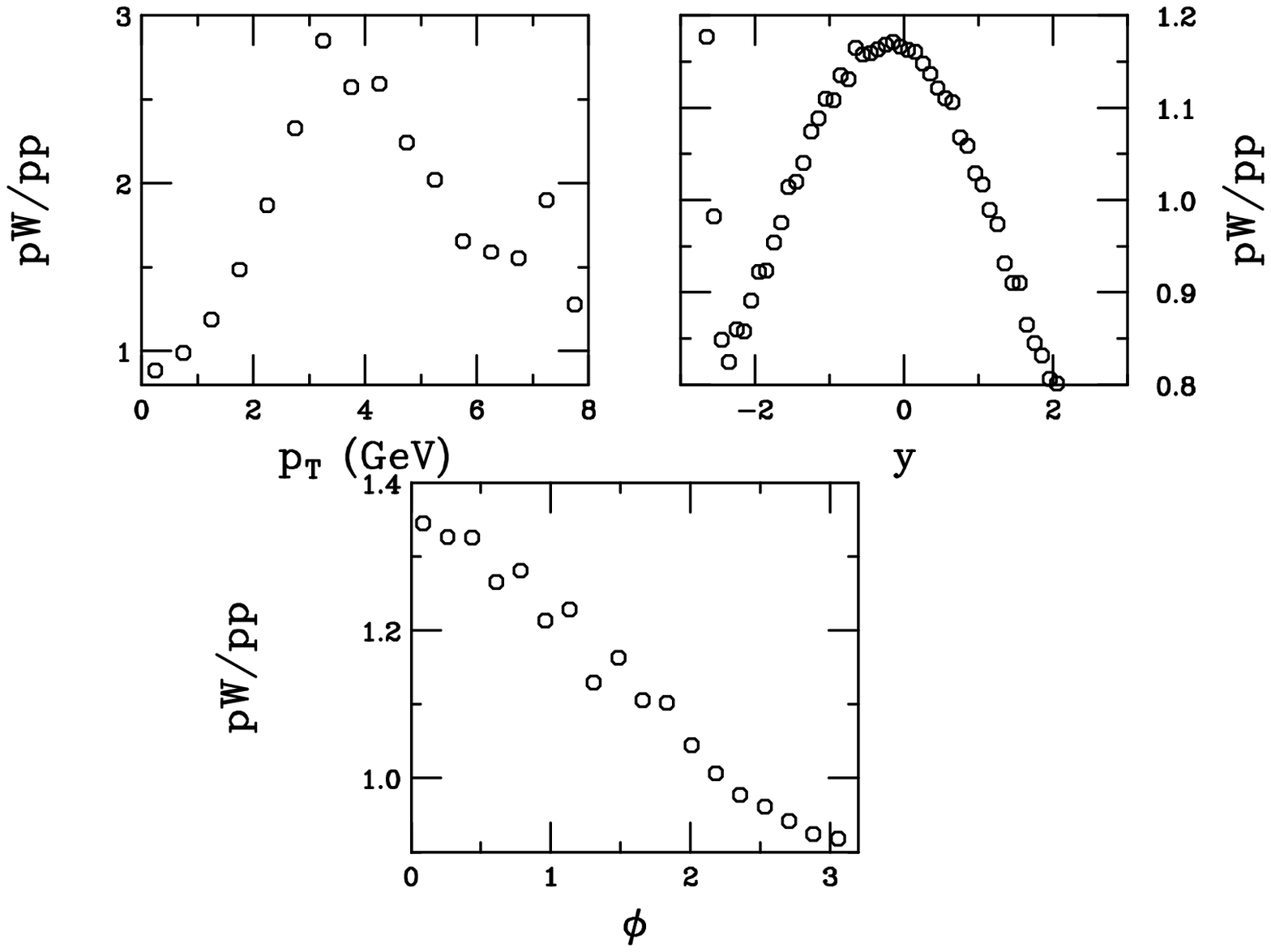}}
\caption[]{The ratio of $pA$ to $pp$ exclusive NLO $c \overline c$ pair 
production at 800 GeV as a function of $p_T$, $y$, and $\phi$.}
\label{800rat_p}
\end{figure}

\begin{figure}[p]
\setlength{\epsfxsize=\textwidth}
\setlength{\epsfysize=0.6\textheight}
\centerline{\epsffile{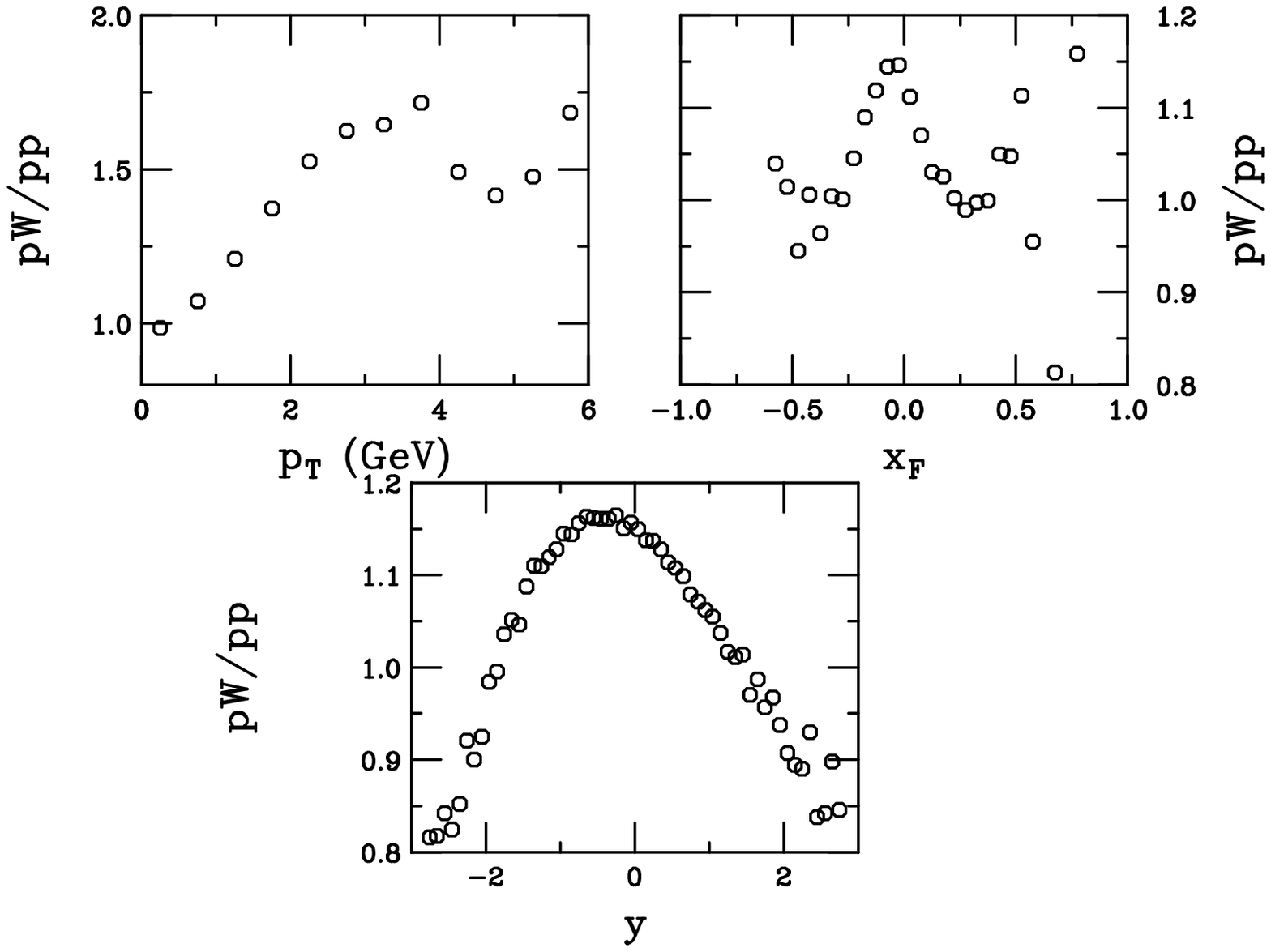}}
\caption[]{The ratio of $pA$ to $pp$ inclusive NLO $c$ quark production 
at 800 GeV as a function of $p_T$, $x_F$, and $y$.}
\label{800rat_s}
\end{figure}

\begin{figure}[p]
\setlength{\epsfxsize=\textwidth}
\setlength{\epsfysize=0.6\textheight}
\centerline{\epsffile{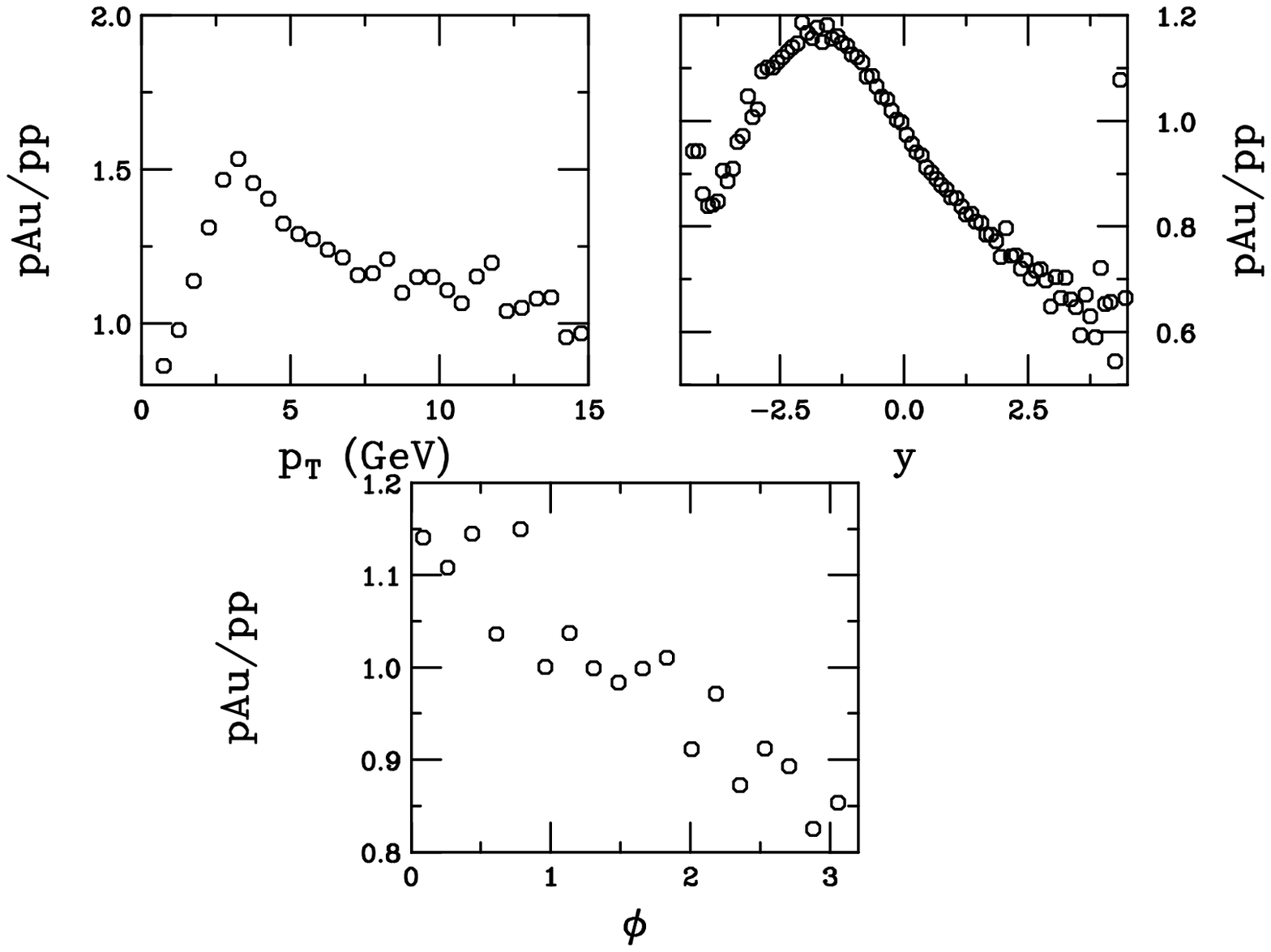}}
\caption[]{The ratio of $pA$ to $pp$ exclusive NLO $c \overline c$ pair 
production at $\sqrt{s} = 200$ GeV as a function of 
$p_T$, $y$, and $\phi$.}
\label{c200rat_p}
\end{figure}

\begin{figure}[p]
\setlength{\epsfxsize=\textwidth}
\setlength{\epsfysize=0.3\textheight}
\centerline{\epsffile{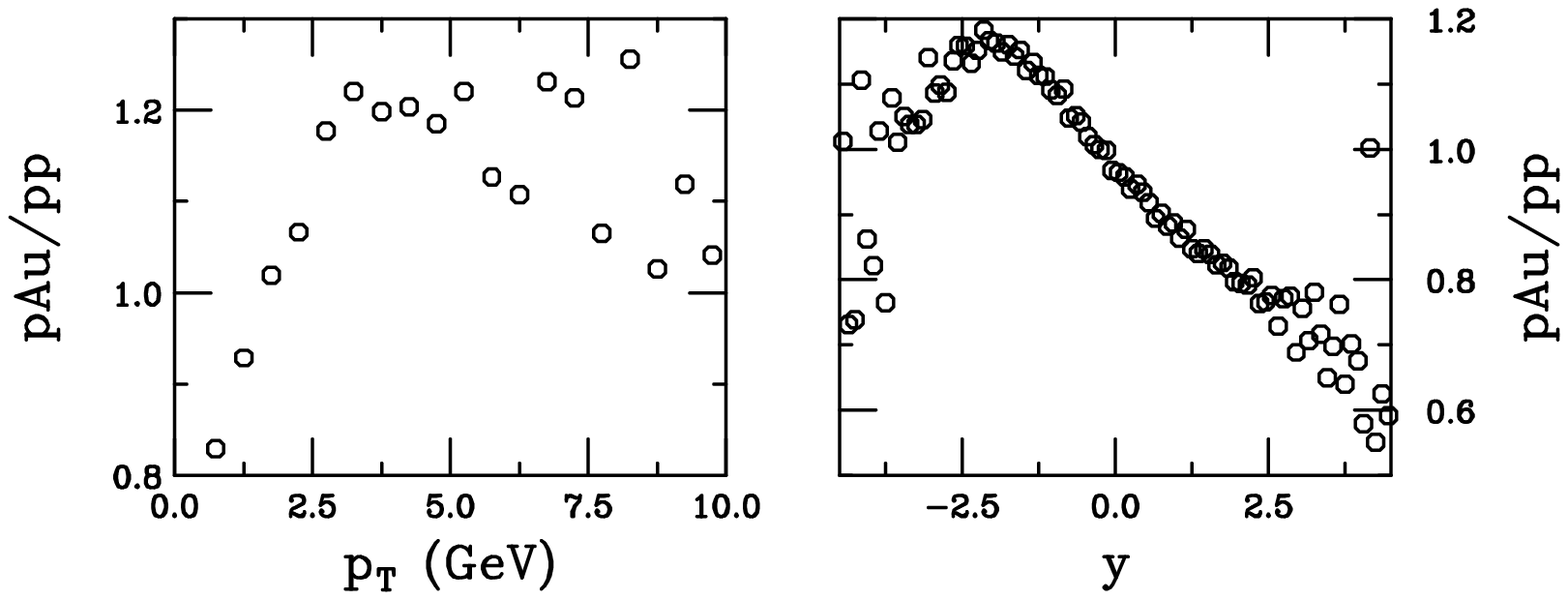}}
\caption[]{The ratio of $pA$ to $pp$ inclusive NLO $c$ quark production 
at $\sqrt{s} = 200$ GeV 
as a function of $p_T$ and $y$.}
\label{c200rat_s}
\end{figure}

\begin{figure}[p]
\setlength{\epsfxsize=\textwidth}
\setlength{\epsfysize=0.6\textheight}
\centerline{\epsffile{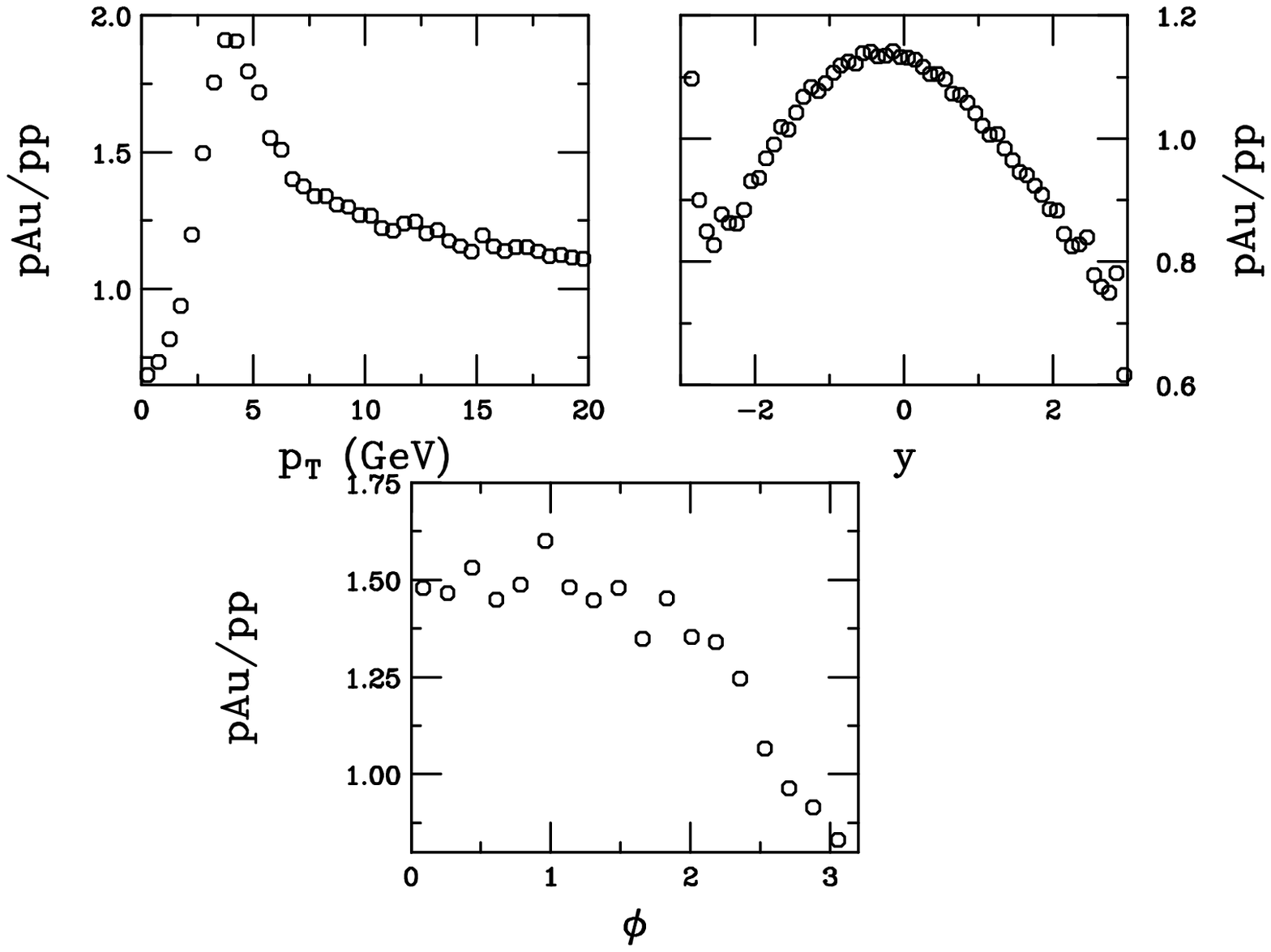}}
\caption[]{The ratio of $pA$ to $pp$ exclusive NLO $b \overline b$ pair 
production at $\sqrt{s} = 200$ GeV 
as a function of $p_T$, $y$, and $\phi$.}
\label{b200rat_p}
\end{figure}

\begin{figure}[p]
\setlength{\epsfxsize=\textwidth}
\setlength{\epsfysize=0.3\textheight}
\centerline{\epsffile{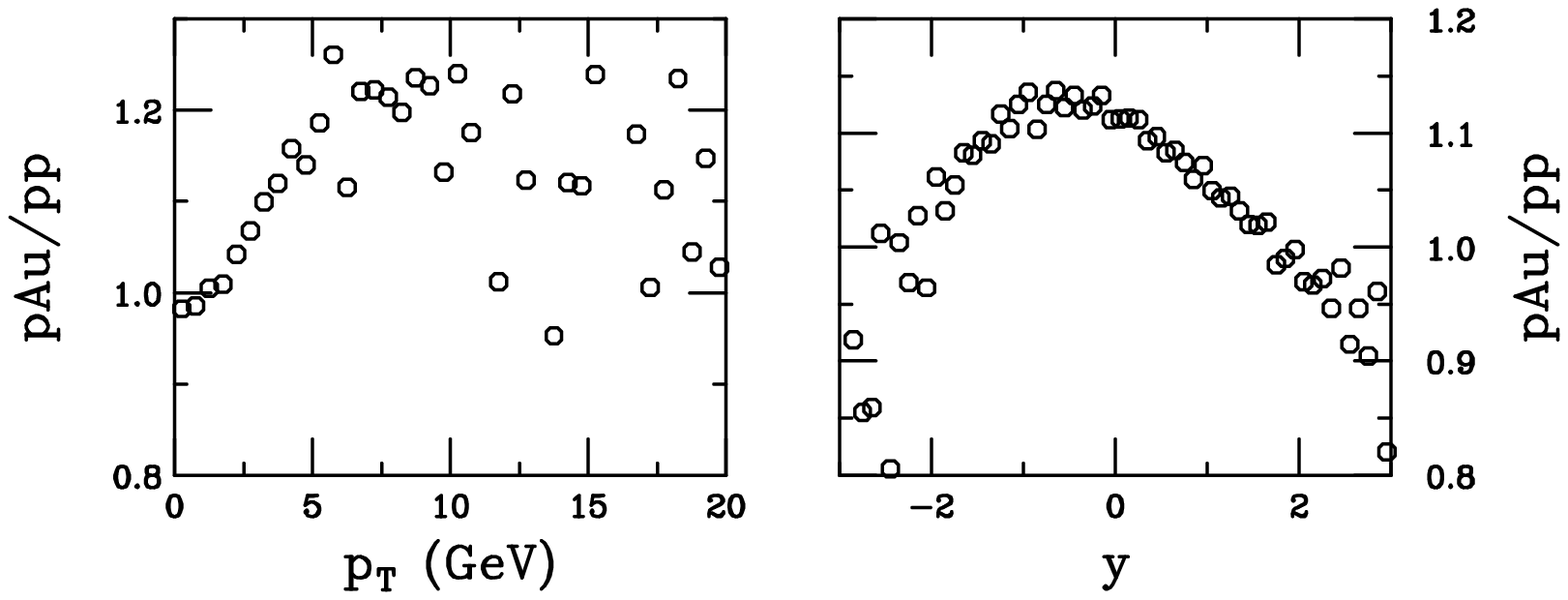}}
\caption[]{The ratio of $pA$ to $pp$ inclusive NLO $b$ quark production 
at $\sqrt{s} = 200$ GeV 
as a function of $p_T$ and $y$.}
\label{b200rat_s}
\end{figure}

\begin{figure}[p]
\setlength{\epsfxsize=\textwidth}
\setlength{\epsfysize=0.6\textheight}
\centerline{\epsffile{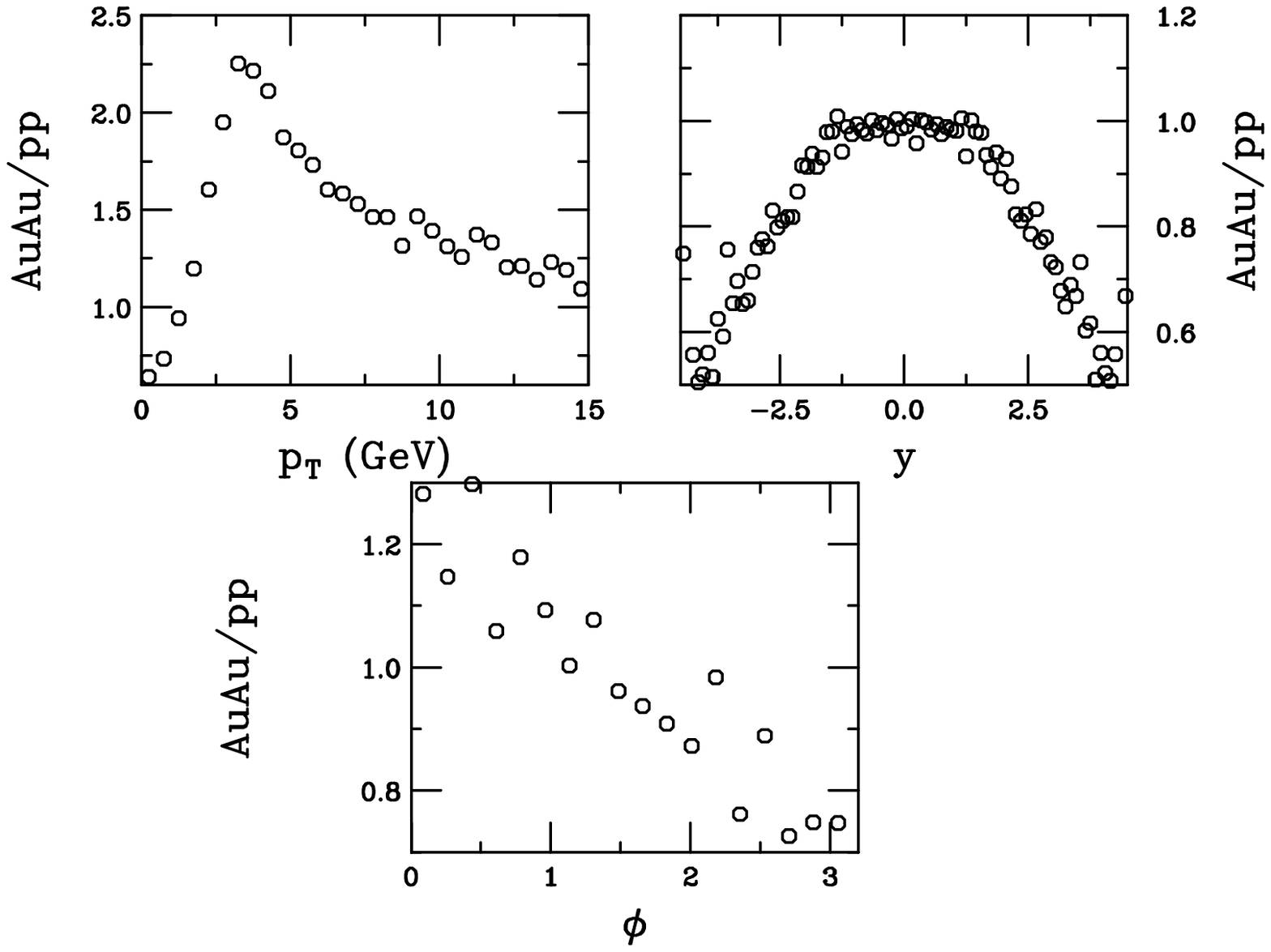}}
\caption[]{The ratio of $AA$ to $pp$ exclusive NLO $c \overline c$ pair 
production at $\sqrt{s} = 200$ GeV as a function of 
$p_T$, $y$, and $\phi$.}
\label{c200rat_p_a}
\end{figure}

\begin{figure}[p]
\setlength{\epsfxsize=\textwidth}
\setlength{\epsfysize=0.3\textheight}
\centerline{\epsffile{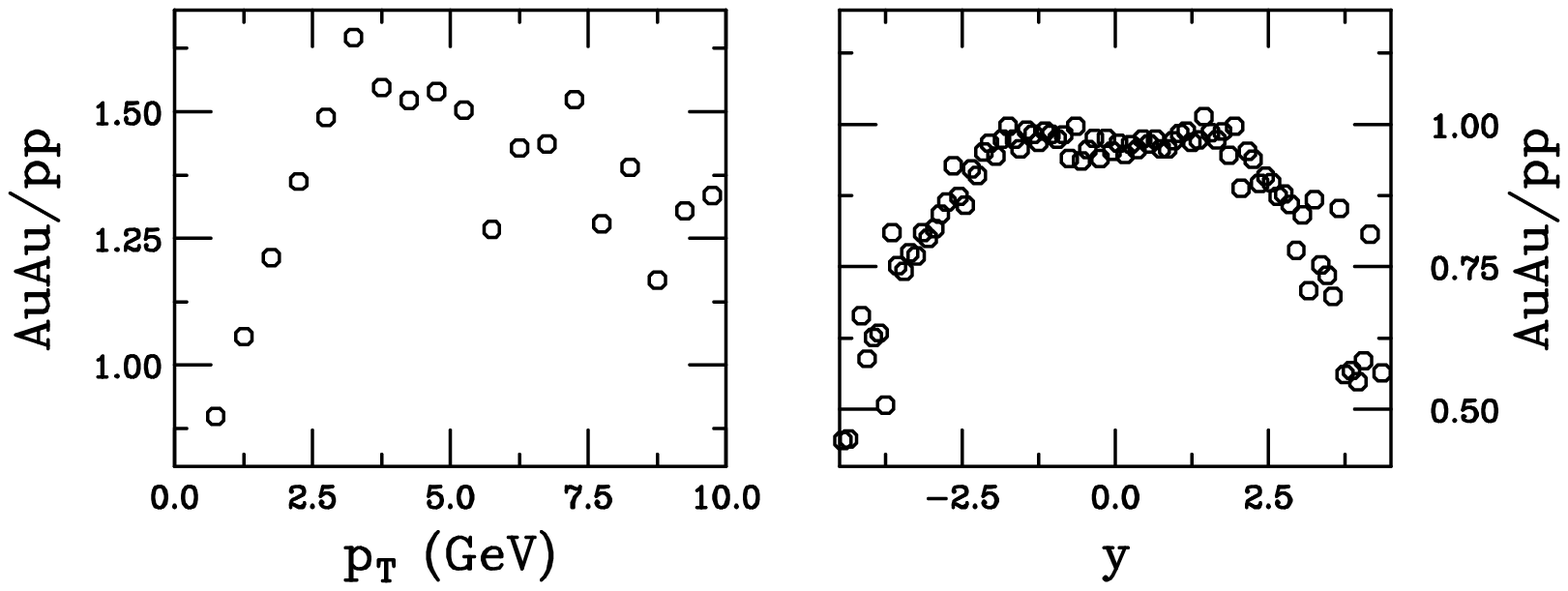}}
\caption[]{The ratio of $AA$ to $pp$ inclusive NLO $c$ quark production 
at $\sqrt{s} = 200$ GeV 
as a function of $p_T$ and $y$.}
\label{c200rat_s_a}
\end{figure}

\begin{figure}[p]
\setlength{\epsfxsize=\textwidth}
\setlength{\epsfysize=0.6\textheight}
\centerline{\epsffile{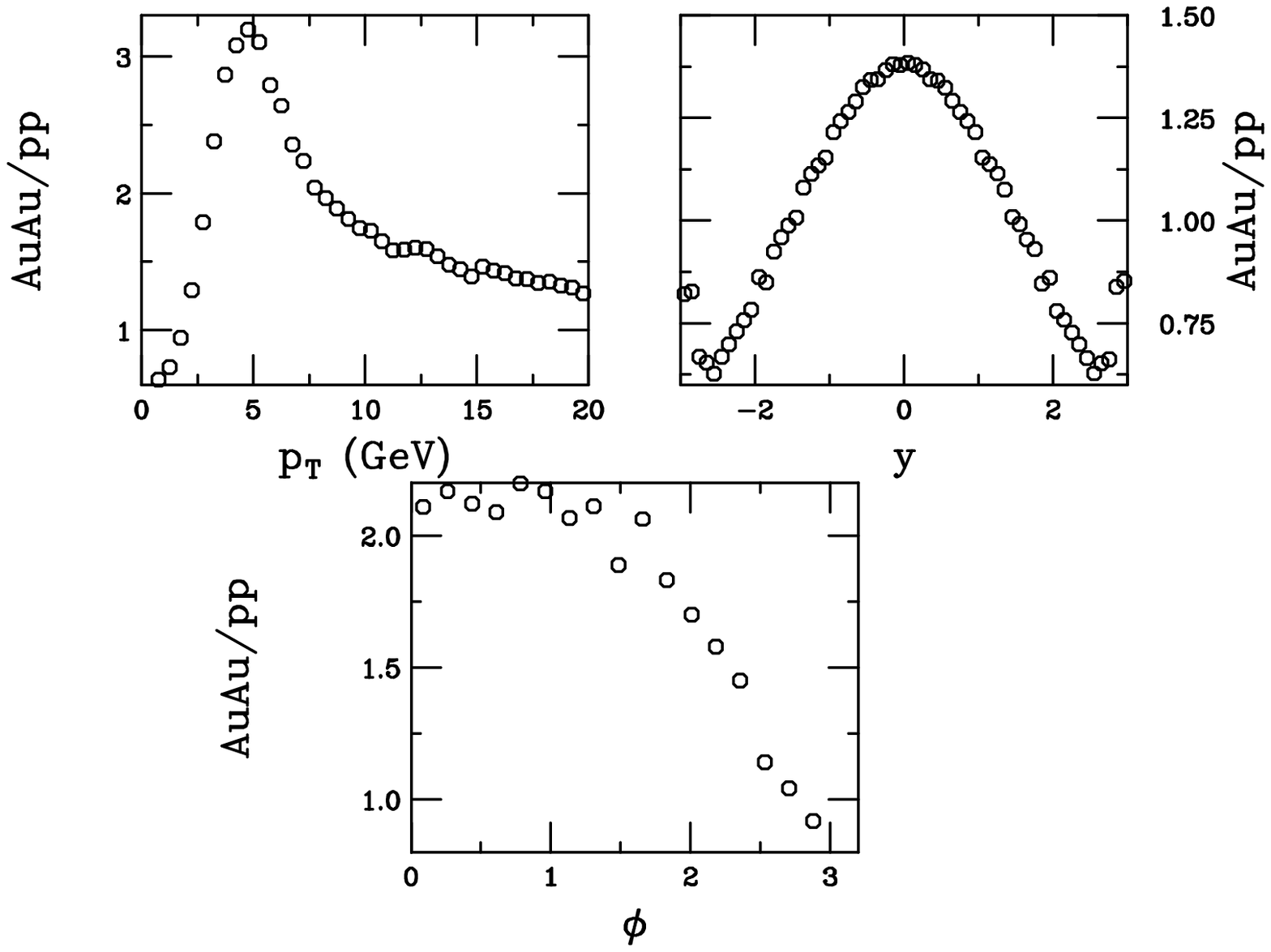}}
\caption[]{The ratio of $AA$ to $pp$ exclusive NLO $b \overline b$ pair 
production at $\sqrt{s} = 200$ GeV 
as a function of $p_T$, $y$, and $\phi$.}
\label{b200rat_p_a}
\end{figure}

\begin{figure}[p]
\setlength{\epsfxsize=\textwidth}
\setlength{\epsfysize=0.3\textheight}
\centerline{\epsffile{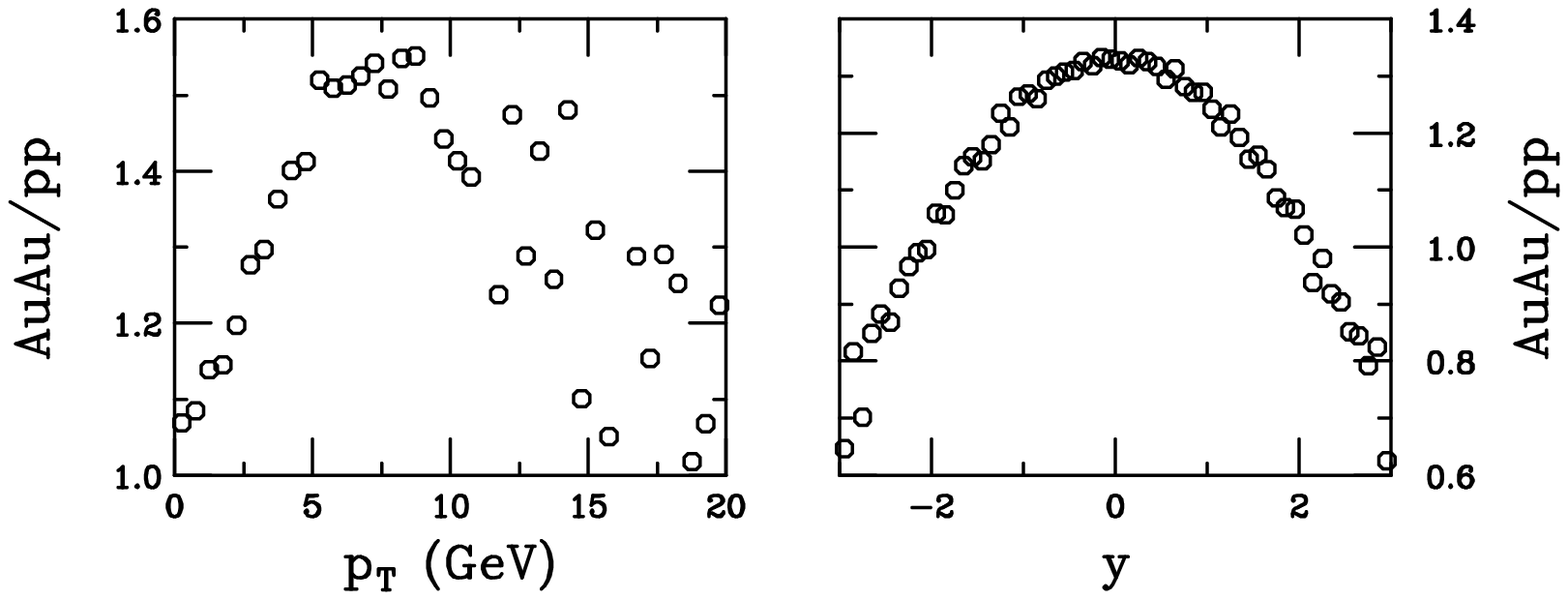}}
\caption[]{The ratio of $AA$ to $pp$ inclusive NLO $b$ quark production 
at $\sqrt{s} = 200$ GeV 
as a function of $p_T$ and $y$.}
\label{b200rat_s_a}
\end{figure}

\begin{figure}[p]
\setlength{\epsfxsize=\textwidth}
\setlength{\epsfysize=0.6\textheight}
\centerline{\epsffile{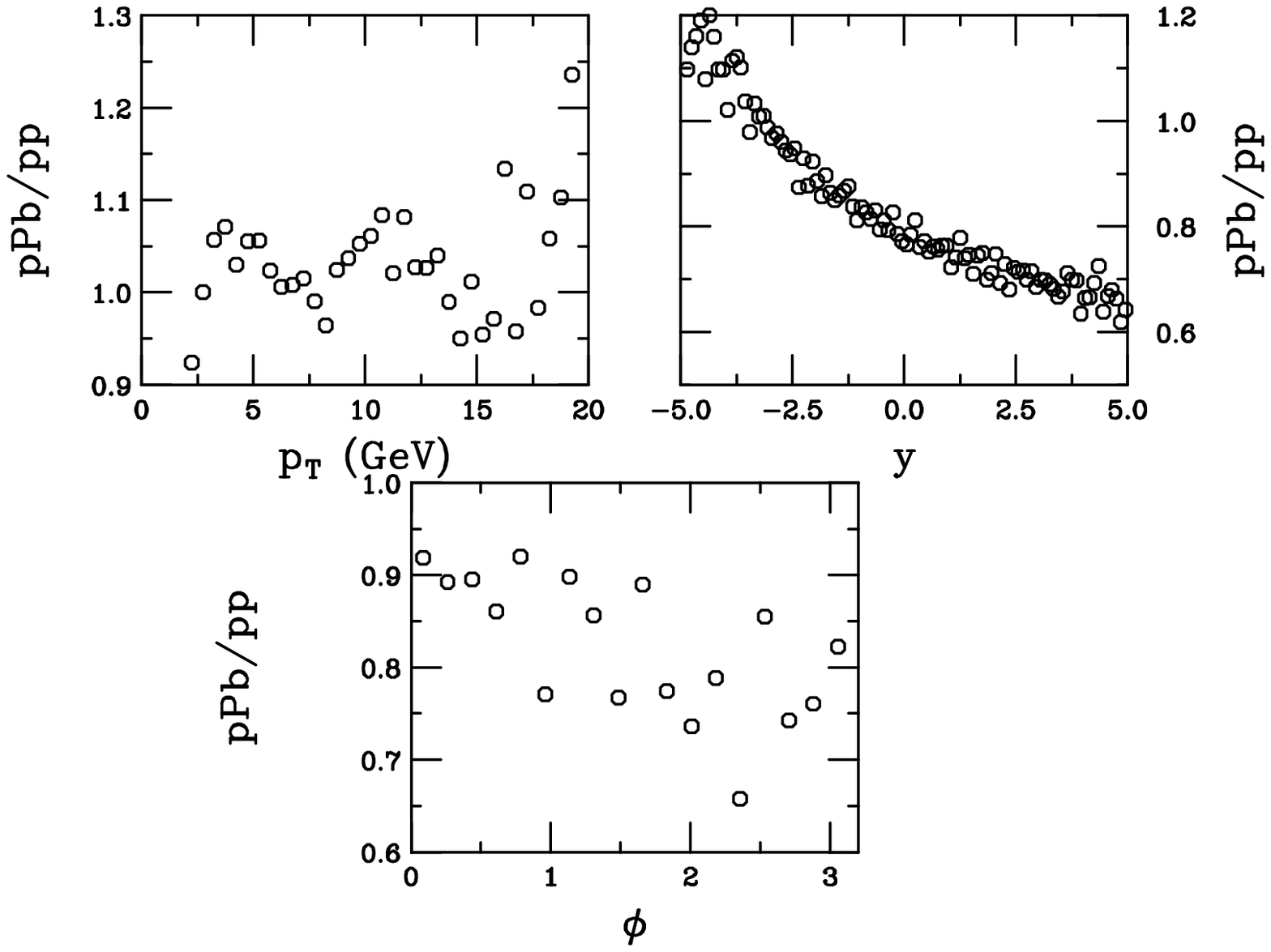}}
\caption[]{The ratio of $pA$ to $pp$ exclusive NLO $c \overline c$ pair 
production at $\sqrt{s} = 5.5$ TeV 
as a function of $p_T$, $y$, and $\phi$.}
\label{c5500rat_p}
\end{figure}

\begin{figure}[p]
\setlength{\epsfxsize=\textwidth}
\setlength{\epsfysize=0.3\textheight}
\centerline{\epsffile{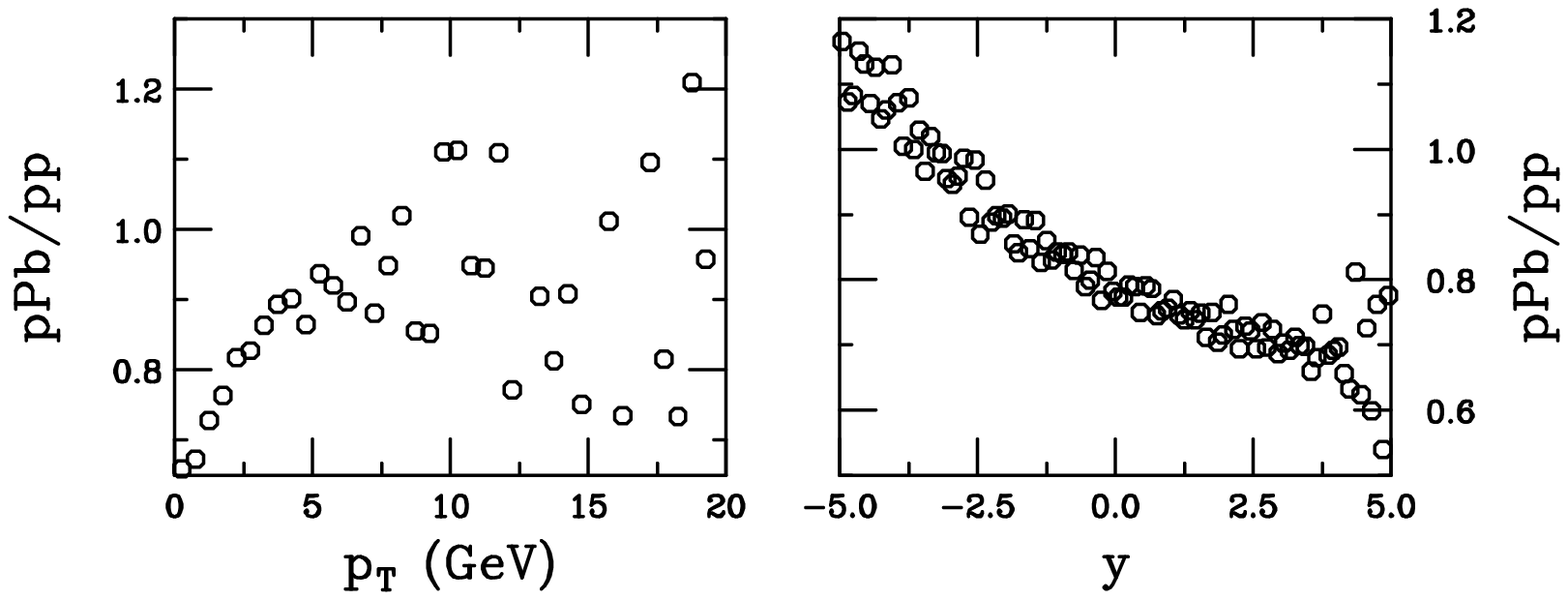}}
\caption[]{The ratio of $pA$ to $pp$ inclusive NLO $c$ quark production 
at $\sqrt{s} = 5.5$ TeV 
as a function of $p_T$ and $y$.}
\label{c5500rat_s}
\end{figure}

\begin{figure}[p]
\setlength{\epsfxsize=\textwidth}
\setlength{\epsfysize=0.6\textheight}
\centerline{\epsffile{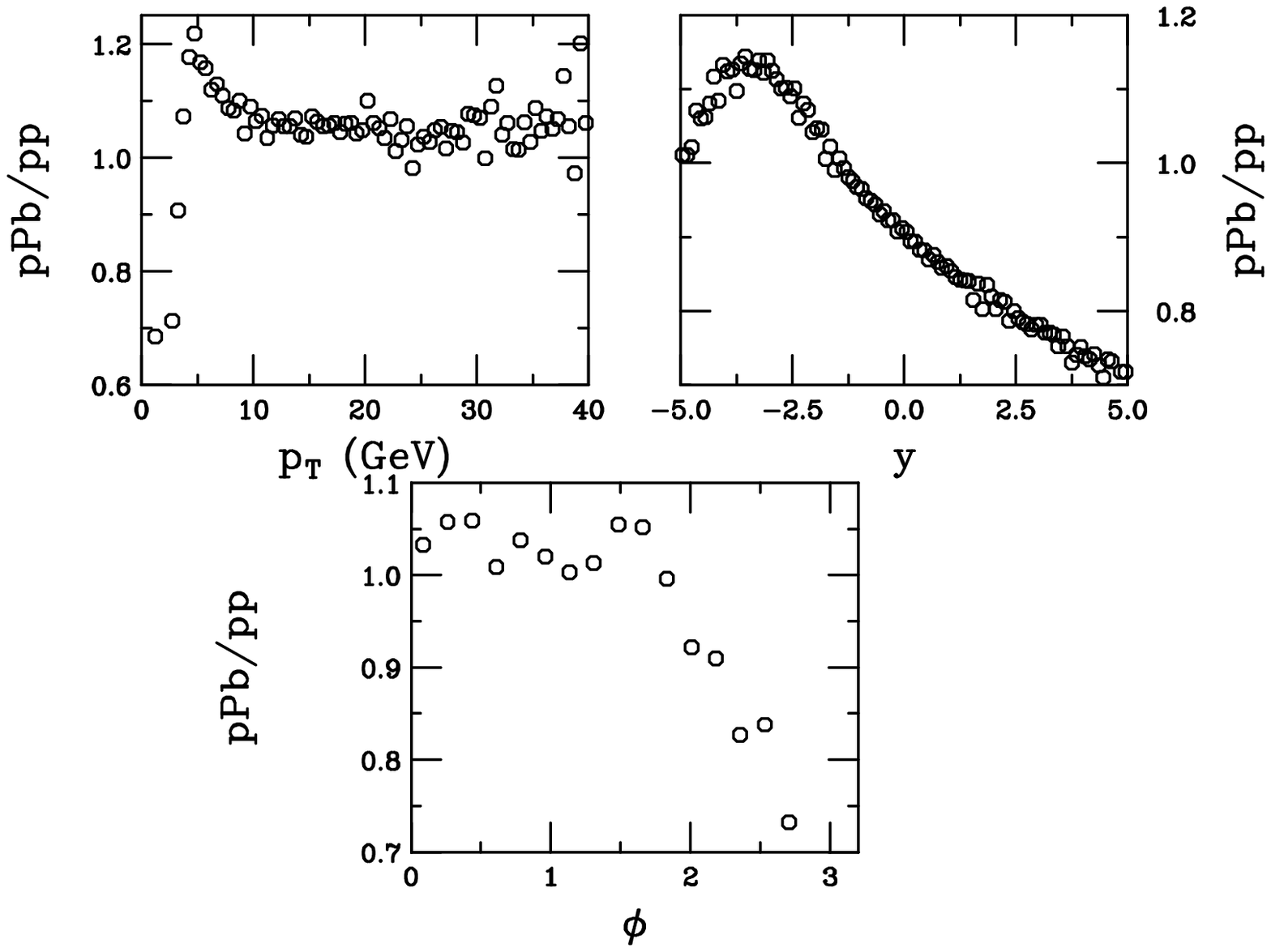}}
\caption[]{The ratio of $pA$ to $pp$ exclusive NLO $b \overline b$ pair 
production at $\sqrt{s} = 5.5$ TeV 
as a function of $p_T$, $y$, and $\phi$.}
\label{b5500rat_p}
\end{figure}

\begin{figure}[p]
\setlength{\epsfxsize=\textwidth}
\setlength{\epsfysize=0.3\textheight}
\centerline{\epsffile{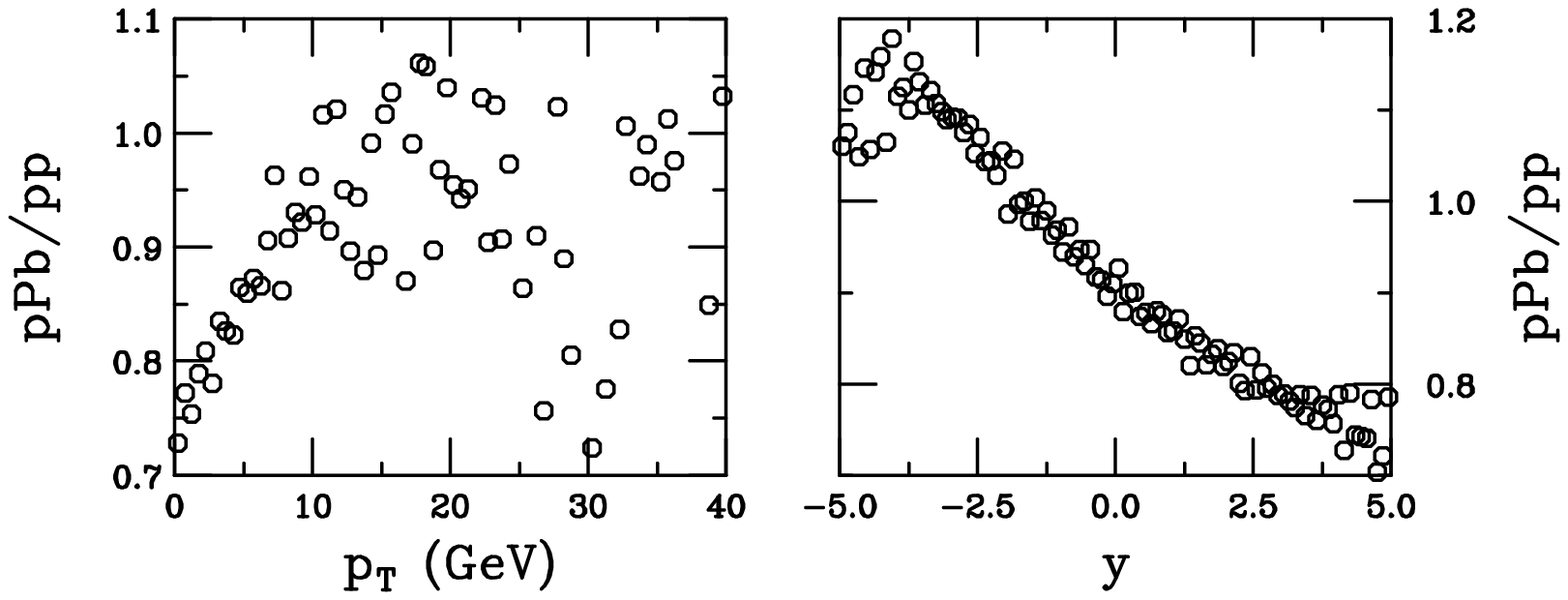}}
\caption[]{The ratio of $pA$ to $pp$ inclusive NLO $b$ quark production 
at $\sqrt{s} = 5.5$ TeV 
as a function of $p_T$ and $y$.}
\label{b5500rat_s}
\end{figure}

\begin{figure}[p]
\setlength{\epsfxsize=\textwidth}
\setlength{\epsfysize=0.6\textheight}
\centerline{\epsffile{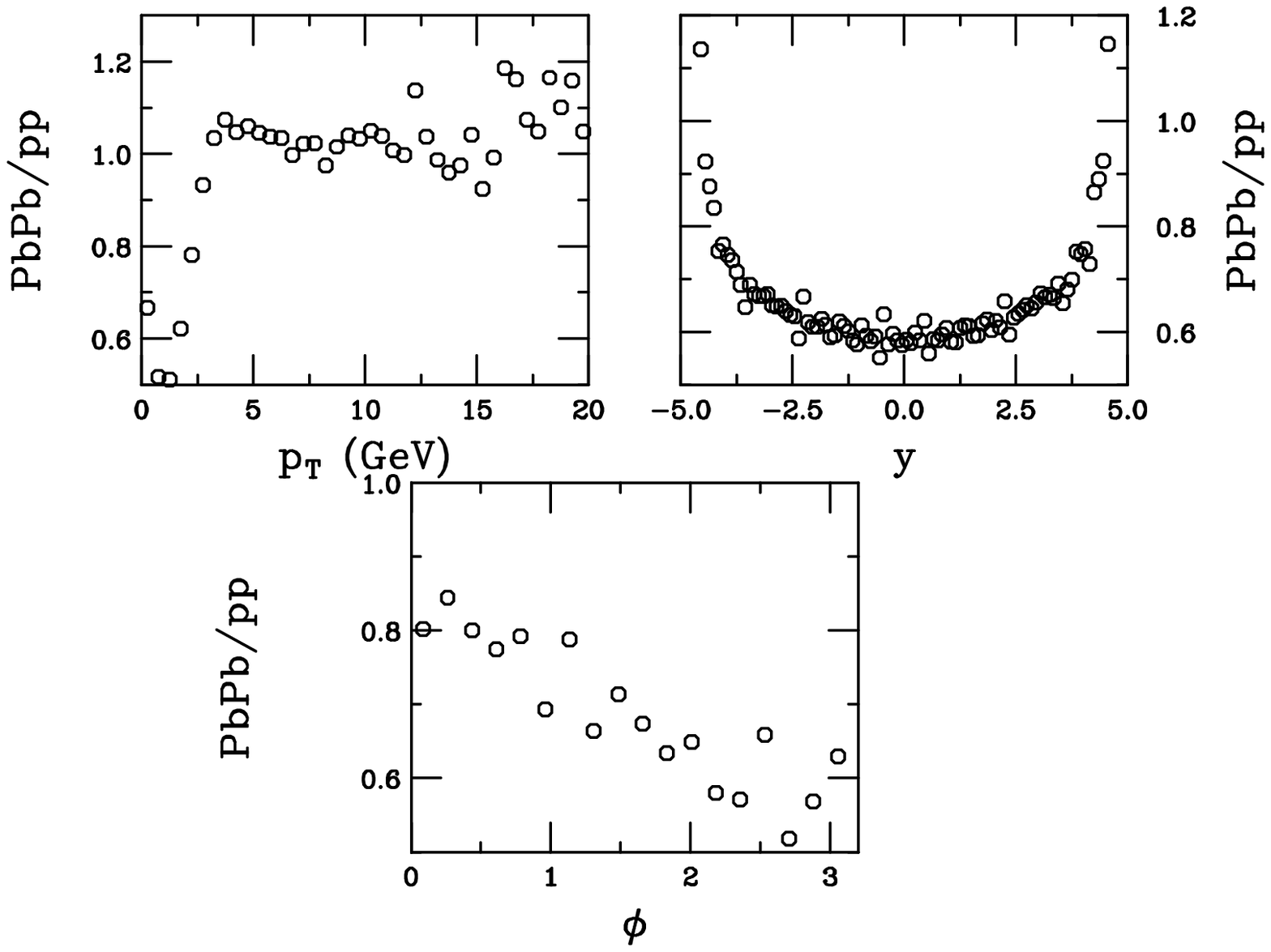}}
\caption[]{The ratio of $AA$ to $pp$ exclusive NLO $c \overline c$ pair 
production at $\sqrt{s} = 5.5$ TeV 
as a function of $p_T$, $y$, and $\phi$.}
\label{c5500rat_p_a}
\end{figure}

\begin{figure}[p]
\setlength{\epsfxsize=\textwidth}
\setlength{\epsfysize=0.3\textheight}
\centerline{\epsffile{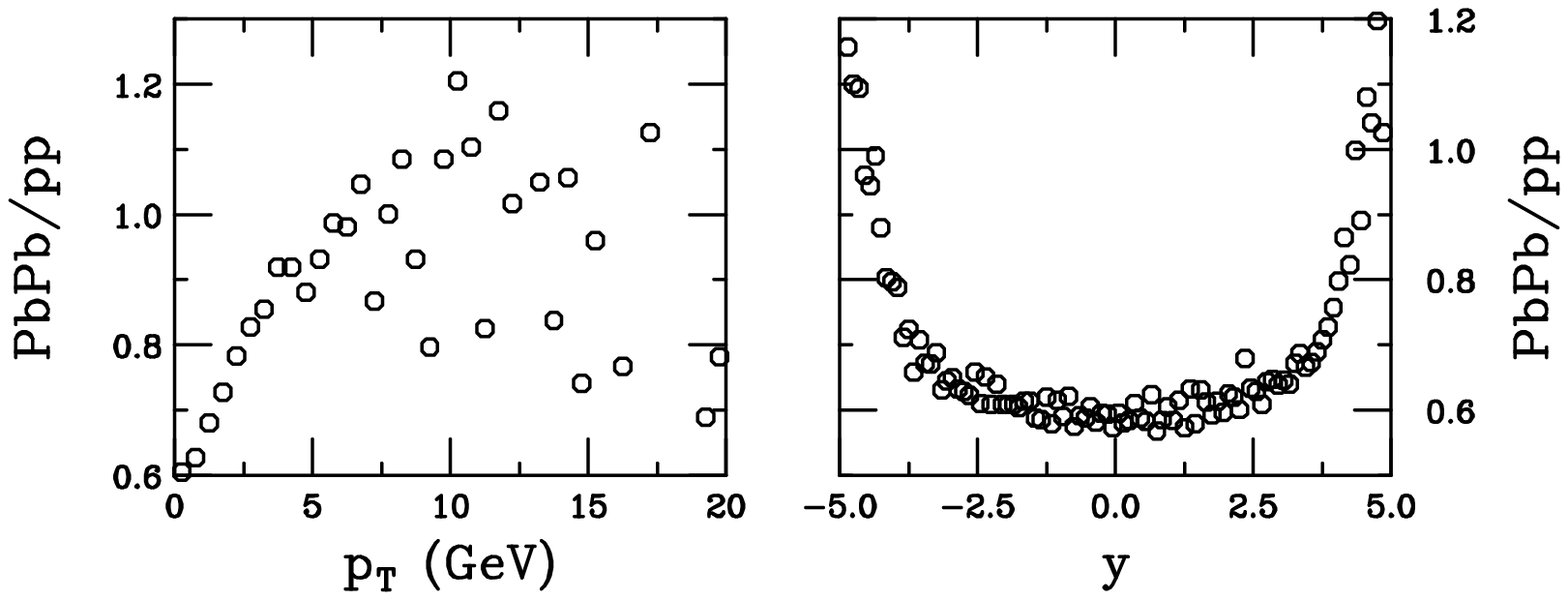}}
\caption[]{The ratio of $AA$ to $pp$ inclusive NLO $c$ quark production 
at $\sqrt{s} = 5.5$ TeV 
as a function of $p_T$ and $y$.}
\label{c5500rat_s_a}
\end{figure}

\begin{figure}[p]
\setlength{\epsfxsize=\textwidth}
\setlength{\epsfysize=0.6\textheight}
\centerline{\epsffile{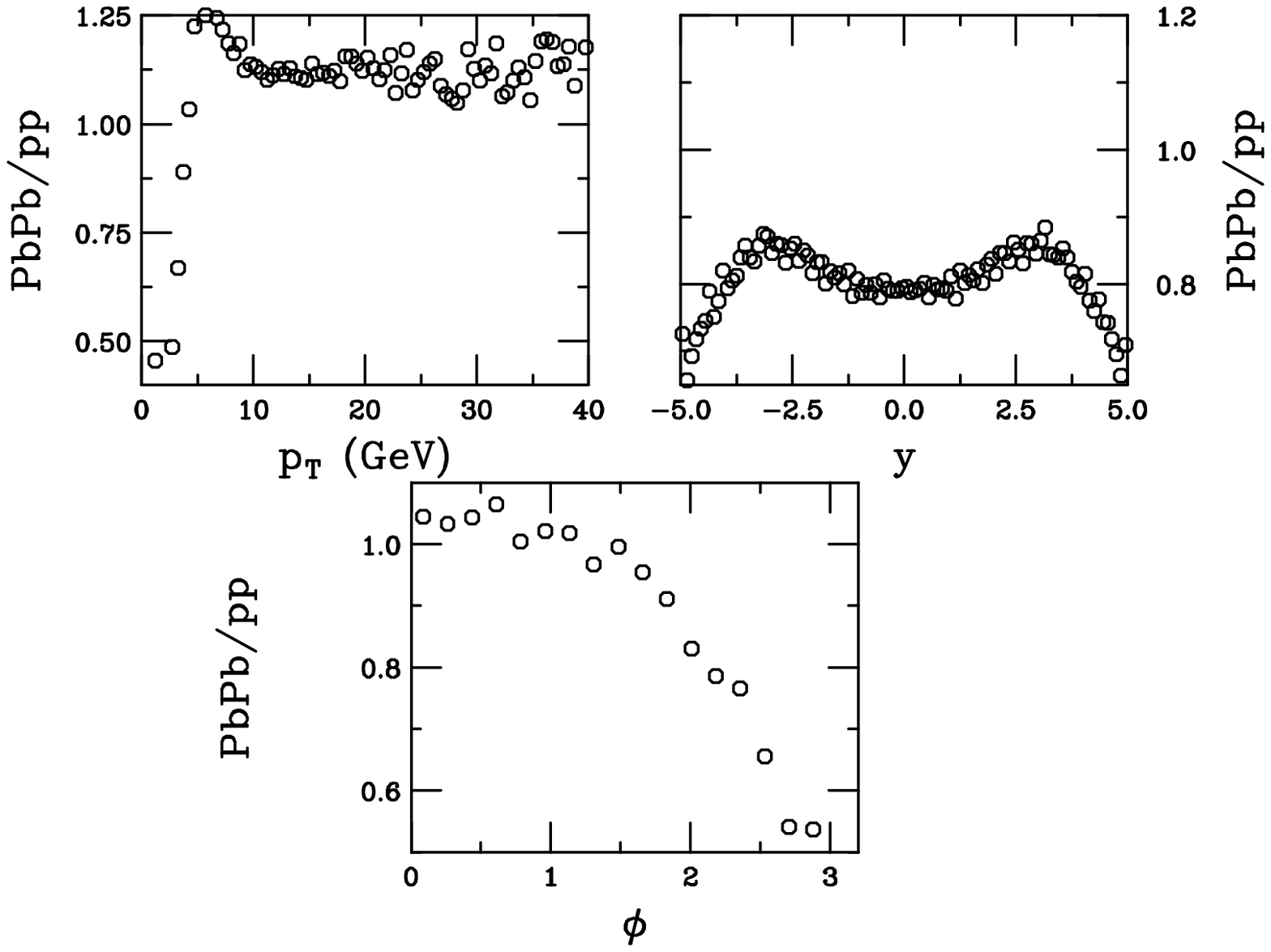}}
\caption[]{The ratio of $AA$ to $pp$ exclusive NLO $b \overline b$ pair 
production at $\sqrt{s} = 5.5$ TeV 
as a function of $p_T$, $y$, and $\phi$.}
\label{b5500rat_p_a}
\end{figure}

\begin{figure}[p]
\setlength{\epsfxsize=\textwidth}
\setlength{\epsfysize=0.3\textheight}
\centerline{\epsffile{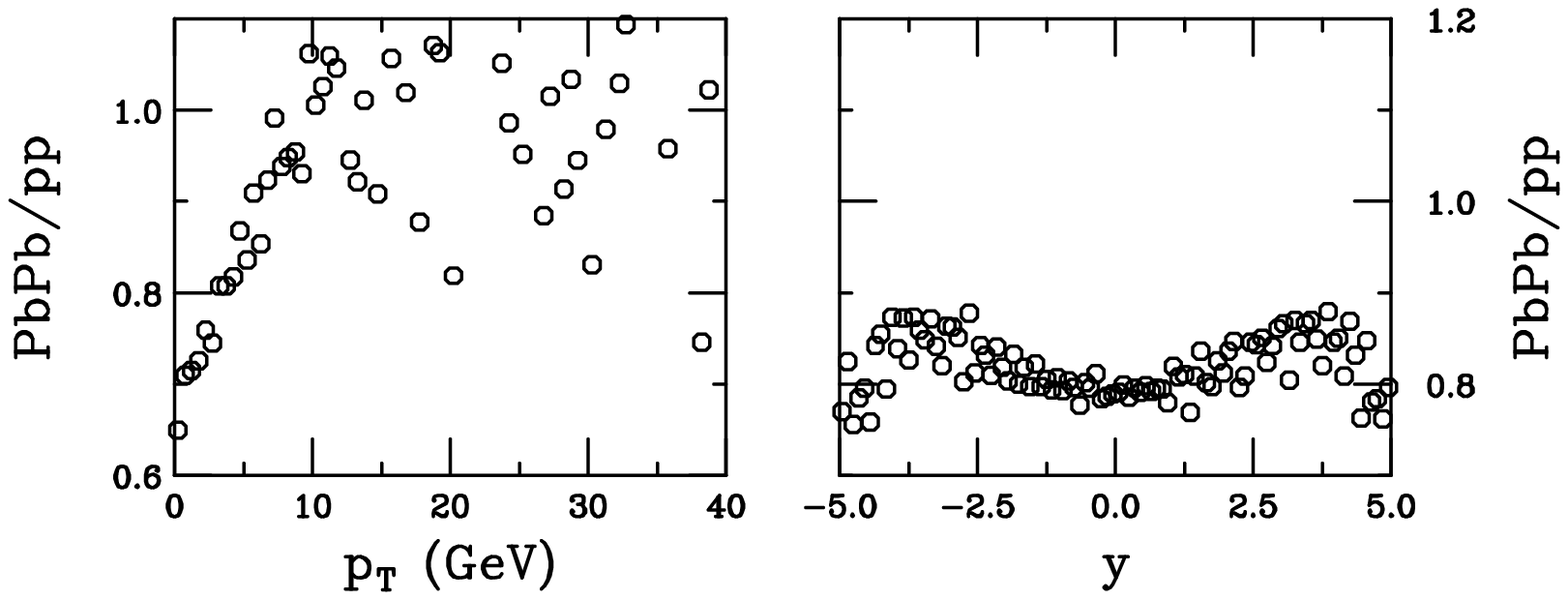}}
\caption[]{The ratio of $AA$ to $pp$ inclusive NLO $b$ quark production 
at $\sqrt{s} = 5.5$ TeV 
as a function of $p_T$ and $y$.}
\label{b5500rat_s_a}
\end{figure}

Since the differences between the distributions are not always obvious in
Figs.~\ref{158pairs}-\ref{b5500singles}, we have also calculated the $pA$ to
$pp$ ratios in Figs.~\ref{158rat_p}-\ref{b200rat_s} and 
\ref{c5500rat_p}-\ref{b5500rat_s} and the $AA$ to $pp$ ratios at the same 
energies in Figs.~\ref{c200rat_p_a}-\ref{b200rat_s_a} and 
\ref{c5500rat_p_a}-\ref{b5500rat_s_a}.  
The pair and single quark $p_T$ ratios reflect the effects
of the $k_T$ broadening, as do the $x_F$ distributions for $|x_F| > 0.5$.  The
$k_T$ broadening follows the energy dependent trend predicted in
Ref.~\cite{XNWrat}.  The $pA/pp$ ratio increases with $p_T$ at low energies
and, as the energy increases, begins to turn over at pair $p_T \sim 3.5 - 4$
GeV for charm, decreasing to unity at large pair $p_T$.  This turnover is at
$\sim 5$ GeV for bottom.  As the energy increases, shadowing becomes more
important at low $p_T$, decreasing the ratio below unity in this region.  The
magnitude of this decrease is a good indication of the strength of gluon
shadowing, particularly for the pair.  While the single quark ratios follow the
trends of the pair ratios at low $p_T$, the lower statistics of the Monte Carlo
at high $p_T$ lead to
large fluctuations.  We have checked that taking the single $p_T$ ratio for all
$x_F$ rather than $x_F > 0$, as is shown in the figures, neither strongly
influences the final results nor significantly reduces the fluctuations.
These fluctuations are particularly strong in $c \overline c$ production at
$\sqrt{s} = 5.5$ TeV where both the pair and single charm $p_T$ distributions
decrease 4-5 orders of magnitude over the $p_T$ interval shown.
The $\phi$ ratios decrease as $\phi$ goes from 0 to $\pi$
because the larger $k_T$ kick in $pA$ flattens this distribution relative to
$pp$ so that $pA/pp > 1$ at $\phi \sim 0$ and less than unity at $\phi \sim
\pi$, compare Fig.~\ref{158pairs} to Fig.~\ref{158rat_p}.  The general trend is
similar for all energies because variables more sensitive to shadowing are
integrated over.  The $AA/pp$ ratios exhibit the same trends except that the
depletion at low $p_T$ is stronger and the enhancement is larger before the
turnover than in the $pA/pp$ ratios.  The position of the maximum enhancement
is not significantly changed. The stronger depletion at low $p_T$ is, in part,
due to nuclear shadowing while the change in the relative enhancement between
$pA/pp$ and $AA/pp$ is due to the larger $\langle k_T^2 \rangle$ in $AA$
collisions.

On the other hand, the $pA/pp$
rapidity ratios follow the gluon ratios shown in
Fig.~\ref{fshadow}.  The peak at $x \rightarrow 1$ from Fermi motion appears at
large negative rapidity, followed by the dip in the EMC region, a broad peak
tracing out the antishadowing region, and finally decreasing again as the
shadowing region is reached at large, positive rapidity.  The shape is inverted
from that in Fig.~\ref{fshadow} because large $x$ corresponds to large negative
rapidity and low $x$ to large positive rapidity.  The strength of the shadowing
effect increases as energy increases while the antishadowing peak, broad at 158
GeV, narrows with increasing energy.  At collider energies, the full rapidity
distribution is not shown in the plots so that the $x$ distribution of
Fig.~\ref{fshadow} is somewhat truncated.  Both the single and pair rapidity
distributions follow the shape of the gluon shadowing ratio, suggesting that
either can help elucidate gluon modifications in the nucleus without
significant distortion from $k_T$ broadening which affects the $x_F$ ratios.
The $AA/pp$ rapidity ratios show the shadowing in both the target and 
projectile.  Now, in addition to the shadowing function for the target nucleus,
there is a shadowing function for the projectile nucleus that is essentially
that of Fig.~\ref{fshadow} since here low $x$ corresponds to large negative
rapidity.  Thus the $AA/pp$ ratios are symmetric around $y=0$.

\yoursection{Conclusions}

We have studied the effects of $k_T$ broadening and nuclear shadowing on charm
and bottom production over a wide range of energies.  
We have found that the $Q \overline Q$ $p_T$, the
single quark $p_T$, and the pair $\phi$ distribution are most sensitive to the
effects of $k_T$ broadening while the pair and single quark rapidities are
sensitive to shadowing.  There is significantly less distortion of the nuclear
effects in the pair distributions which are the most difficult to measure.  The
same results may be obtained through measurements of the single heavy quarks
event though the definitions of $x$ and $k_T$ are smeared relative to those of
the pair.

\clearpage

\clearpage

\clearpage

\clearpage

\clearpage

\clearpage

\begin{table}[p]
\begin{center}
\begin{tabular}{|c|c|c||c|c|} \hline
& \multicolumn{2}{c||}{ $d\sigma_{c \overline c}/dp_T$ (nb/GeV)} & 
\multicolumn{2}{c|}{$d\sigma_c/dp_T$ (nb/GeV)} \\ 
$p_T$ (GeV) & $pp$ & $pA$ & $pp$ & $pA$ \\ \hline
 0.25 &  2104.000 &  1848.000 &  1970.400 &  2046.000 \\
 0.75 &  3470.000 &  3440.000 &  1669.400 &  1909.400 \\
 1.25 &  2048.000 &  2504.000 &   527.000 &   705.200 \\
 1.75 &   747.600 &  1197.000 &   115.180 &   185.060 \\
 2.25 &   198.080 &   421.600 &    20.100 &    38.720 \\
 2.75 &    38.140 &   116.940 &     3.040 &     7.006 \\
 3.25 &     5.188 &    27.220 &     0.391 &     1.166 \\
 3.75 &     0.788 &     3.852 &     0.044 &     0.135 \\
 4.25 &     0.107 &     0.867 &     0.008 &     0.017 \\ \hline
\end{tabular}
\end{center}
\caption[]{The NLO exclusive $c \overline c$ and inclusive single charm $p_T$
distributions for a proton beam of 158 GeV.  
The exclusive $c \overline c$ distributions are integrated over
all rapidity while the inclusive single charm $p_T$ distributions are
integrated over $x_F > 0$ only.  The distributions for $pp$ and $pA$
interactions are both given.  
The per nucleon cross section is given for $pA$ interactions.
Recall that the intrinsic $\langle k_T^2 \rangle$
is 1 GeV$^2$ for $pp$ interactions and 1.35 GeV$^2$ for $pA$ interactions.}
\end{table}

\begin{table}[p]
\begin{center}
\begin{tabular}{|c|c|c||c|c|} \hline
& \multicolumn{2}{c||}{ $d\sigma_{c \overline c}/dp_T$ (nb/GeV)} & 
\multicolumn{2}{c|}{$d\sigma_c/dp_T$ (nb/GeV)} \\ 
$p_T$ (GeV) & $pp$ & $pA$ & $pp$ & $pA$ \\ \hline
 0.25 &  8044.000 &  7156.000 &  7214.000 &  7342.000 \\
 0.75 & 13546.000 & 13544.000 &  6704.000 &  7416.000 \\
 1.25 &  8352.000 & 10130.000 &  2456.000 &  3118.000 \\
 1.75 &  3268.000 &  5020.000 &   642.200 &   943.800 \\
 2.25 &   939.200 &  1862.600 &   152.060 &   246.200 \\
 2.75 &   210.600 &   552.400 &    29.800 &    56.960 \\
 3.25 &    40.220 &   144.480 &     6.476 &    11.616 \\
 3.75 &     8.492 &    27.780 &     1.405 &     2.552 \\
 4.25 &     1.934 &     6.662 &     0.313 &     0.598 \\
 4.75 &     0.476 &     1.481 &     0.078 &     0.135 \\
 5.25 &     0.128 &     0.378 &     0.020 &     0.030 \\
 5.75 &     0.042 &     0.078 &     0.004 &     0.007 \\
 6.25 &     0.011 &     0.024 &     0.001 &     0.001 \\ \hline
\end{tabular}
\end{center}
\caption[]{The NLO exclusive $c \overline c$ and inclusive single charm $p_T$
distributions for a proton beam of 450 GeV.  
The exclusive $c \overline c$ distributions are integrated over
all rapidity while the inclusive single charm $p_T$ distributions are
integrated over $x_F > 0$ only.  The distributions for $pp$ and $pA$
interactions are both given.  
The per nucleon cross section is given for $pA$ interactions.
Recall that the intrinsic $\langle k_T^2 \rangle$
is 1 GeV$^2$ for $pp$ interactions and 1.35 GeV$^2$ for $pA$ interactions.}
\end{table}

\begin{table}[p]
\begin{center}
\begin{tabular}{|c|c|c||c|c|} \hline
& \multicolumn{2}{c||}{ $d\sigma_{c \overline c}/dp_T$ (nb/GeV)} & 
\multicolumn{2}{c|}{$d\sigma_c/dp_T$ (nb/GeV)} \\ 
$p_T$ (GeV) & $pp$ & $pA$ & $pp$ & $pA$ \\ \hline
 0.25 & 14168.000 & 12530.000 & 12486.000 & 12300.000 \\
 0.75 & 24160.000 & 23880.000 & 12100.000 & 12952.000 \\
 1.25 & 15346.000 & 18216.000 &  4788.000 &  5790.000 \\
 1.75 &  6266.000 &  9312.000 &  1383.600 &  1899.800 \\
 2.25 &  1902.400 &  3558.000 &   356.200 &   543.000 \\
 2.75 &   480.600 &  1119.000 &    86.240 &   140.140 \\
 3.25 &   109.220 &   311.200 &    21.080 &    34.680 \\
 3.75 &    27.720 &    71.280 &     5.578 &     9.570 \\
 4.25 &     7.820 &    20.280 &     1.661 &     2.478 \\
 4.75 &     2.456 &     5.510 &     0.493 &     0.698 \\
 5.25 &     0.787 &     1.591 &     0.166 &     0.245 \\
 5.75 &     0.297 &     0.491 &     0.048 &     0.081 \\
 6.25 &     0.111 &     0.176 &     0.018 &     0.021 \\
 6.75 &     0.044 &     0.068 &     0.005 &     0.007 \\
 7.25 &     0.014 &     0.027 &     0.002 &     0.003 \\ \hline
\end{tabular}
\end{center}
\caption[]{The NLO exclusive $c \overline c$ and inclusive single charm $p_T$
distributions for a proton beam of 800 GeV.  
The exclusive $c \overline c$ distributions are integrated over
all rapidity while the inclusive single charm $p_T$ distributions are
integrated over $x_F > 0$ only.  The distributions for $pp$ and $pA$
interactions are both given.  
The per nucleon cross section is given for $pA$ interactions.
Recall that the intrinsic $\langle k_T^2 \rangle$
is 1 GeV$^2$ for $pp$ interactions and 1.35 GeV$^2$ for $pA$ interactions.}
\end{table}

\begin{table}[p]
\begin{center}
\begin{tabular}{|c|c|c|c||c|c|c|} \hline
& \multicolumn{3}{c||}{ $d\sigma_{c \overline c}/dp_T$ ($\mu$b/GeV)} & 
\multicolumn{3}{c|}{$d\sigma_c/dp_T$ ($\mu$b/GeV)} \\ 
$p_T$ (GeV) & $pp$ & $pA$ & $AA$ & $pp$ & $pA$ & $AA$ \\ \hline
 0.25 & 106.640 &  84.080 &  68.300 &  98.400 &  75.740 &  77.900 \\
 0.75 & 204.800 & 176.440 & 150.640 & 113.660 &  94.220 & 102.140 \\
 1.25 & 162.540 & 158.960 & 153.020 &  60.160 &  55.860 &  63.540 \\
 1.75 &  89.340 & 101.580 & 106.940 &  25.480 &  25.960 &  30.880 \\
 2.25 &  38.820 &  50.860 &  62.220 &  10.330 &  11.016 &  14.720 \\
 2.75 &  15.744 &  23.080 &  30.700 &   4.082 &   4.806 &   6.076 \\
 3.25 &   6.390 &   9.796 &  14.398 &   1.612 &   1.967 &   2.652 \\
 3.75 &   2.918 &   4.246 &   7.614 &   0.752 &   0.902 &   1.164 \\
 4.25 &   1.433 &   2.014 &   3.024 &   0.377 &   0.454 &   0.574 \\
 4.75 &   0.765 &   1.013 &   1.433 &   0.185 &   0.220 &   0.285 \\
 5.25 &   0.421 &   0.543 &   0.761 &   0.101 &   0.123 &   0.152 \\
 5.75 &   0.243 &   0.310 &   0.422 &   0.058 &   0.066 &   0.074 \\
 6.25 &   0.145 &   0.180 &   0.233 &   0.034 &   0.038 &   0.048 \\
 6.75 &   0.091 &   0.110 &   0.143 &   0.021 &   0.026 &   0.030 \\
 7.25 &   0.058 &   0.067 &   0.089 &   0.013 &   0.015 &   0.019 \\
 7.75 &   0.038 &   0.044 &   0.055 &   0.008 &   0.009 &   0.010 \\
 8.25 &   0.024 &   0.029 &   0.035 &   0.005 &   0.006 &   0.007 \\
 8.75 &   0.017 &   0.019 &   0.023 &   0.004 &   0.004 &   0.005 \\
 9.25 &   0.011 &   0.013 &   0.016 &   0.002 &   0.003 &   0.003 \\
 9.75 &   0.008 &   0.009 &   0.011 &   0.002 &   0.002 &   0.002 \\ \hline
\end{tabular}
\end{center}
\caption[]{The NLO exclusive $c \overline c$ and inclusive single charm $p_T$
distributions for collisions at $\sqrt{s} = 200$ GeV.
The exclusive $c \overline c$ distributions are integrated over
all rapidity while the inclusive single charm $p_T$ distributions are
integrated over $x_F > 0$ only.  The distributions for $pp$, $pA$ and $AA$
interactions are all given.  
The per nucleon cross section is given for $pA$ and $AA$ interactions.
Recall that the intrinsic $\langle k_T^2 \rangle$
is 1 GeV$^2$ for $pp$ interactions, 1.35 GeV$^2$ for $pA$ interactions,
and 1.7 GeV$^2$ for $AA$.}
\end{table}

\begin{table}[p]
\begin{center}
\begin{tabular}{|c|c|c|c|||c|c|c|} \hline
& \multicolumn{3}{c||}{ $d\sigma_{c \overline c}/dp_T$ ($\mu$b/GeV)} & 
\multicolumn{3}{c|}{$d\sigma_c/dp_T$ ($\mu$b/GeV)} \\ 
$p_T$ (GeV) & $pp$ & $pA$ & $AA$ & $pp$ & $pA$ & $AA$ \\ \hline
 0.25 &  407.800 &  301.400 &  271.800 & 1123.800 &  740.400 &  679.600 \\
 0.75 & 1791.000 & 1221.600 &  924.400 & 1604.400 & 1078.800 & 1005.800 \\
 1.25 & 2732.000 & 1921.200 & 1395.200 & 1202.200 &  875.000 &  817.400 \\
 1.75 & 2464.000 & 1932.400 & 1531.600 &  755.600 &  577.000 &  549.600 \\
 2.25 & 1615.200 & 1492.800 & 1262.000 &  428.800 &  350.800 &  335.600 \\
 2.75 &  938.400 &  939.000 &  875.600 &  242.200 &  200.400 &  200.400 \\
 3.25 &  540.200 &  570.800 &  559.000 &  143.700 &  124.140 &  122.820 \\
 3.75 &  324.200 &  347.200 &  348.200 &   85.380 &   76.260 &   78.460 \\
 4.25 &  207.800 &  214.000 &  217.600 &   53.500 &   48.260 &   49.200 \\
 4.75 &  133.820 &  141.280 &  141.760 &   36.640 &   31.680 &   32.280 \\
 5.25 &   92.220 &   97.380 &   96.480 &   23.160 &   21.700 &   21.580 \\
 5.75 &   65.940 &   67.480 &   68.420 &   16.696 &   15.368 &   16.480 \\
 6.25 &   47.900 &   48.180 &   49.540 &   11.192 &   10.042 &   10.974 \\
 6.75 &   35.580 &   35.880 &   35.480 &    8.354 &    8.282 &    8.742 \\
 7.25 &   26.420 &   26.820 &   27.000 &    6.568 &    5.788 &    5.694 \\
 7.75 &   20.400 &   20.200 &   20.860 &    4.900 &    4.650 &    4.902 \\
 8.25 &   16.306 &   15.728 &   15.904 &    3.378 &    3.446 &    3.664 \\
 8.75 &   12.646 &   12.958 &   12.840 &    2.886 &    2.468 &    2.688 \\
 9.25 &    9.750 &   10.112 &   11.420 &    2.316 &    1.973 &    1.844 \\
 9.75 &    7.740 &    8.150 &    7.996 &    1.631 &    1.812 &    1.769 \\
10.25 &    6.162 &    6.538 &    6.474 &    1.267 &    1.411 &    1.527 \\
10.75 &    5.046 &    5.468 &    5.238 &    1.159 &    1.099 &    1.279 \\
11.25 &    4.216 &    4.304 &    4.246 &    1.024 &    0.968 &    0.846 \\
11.75 &    3.512 &    3.798 &    3.504 &    0.660 &    0.733 &    0.765 \\
12.25 &    2.850 &    2.928 &    3.242 &    0.719 &    0.554 &    0.730 \\
12.75 &    2.474 &    2.540 &    2.566 &    0.406 &    0.580 &    0.554 \\
13.25 &    2.016 &    2.096 &    1.990 &    0.386 &    0.350 &    0.405 \\
13.75 &    1.881 &    1.862 &    1.804 &    0.430 &    0.349 &    0.360 \\
14.25 &    1.601 &    1.522 &    1.562 &    0.283 &    0.257 &    0.299 \\
14.75 &    1.280 &    1.296 &    1.334 &    0.322 &    0.241 &    0.238 \\
15.25 &    1.200 &    1.146 &    1.184 &    0.203 &    0.271 &    0.195 \\
15.75 &    1.011 &    0.982 &    1.022 &    0.176 &    0.178 &    0.261 \\
16.25 &    0.799 &    0.905 &    0.947 &    0.223 &    0.164 &    0.171 \\
16.75 &    0.752 &    0.720 &    0.873 &    0.117 &    0.166 &    0.174 \\
17.25 &    0.647 &    0.718 &    0.695 &    0.102 &    0.111 &    0.114 \\
17.75 &    0.576 &    0.566 &    0.604 &    0.128 &    0.104 &    0.171 \\ 
\hline
\end{tabular}
\end{center}
\caption[]{The NLO exclusive $c \overline c$ and inclusive single charm $p_T$
distributions for collisions at $\sqrt{s} = 5.5$ TeV.
The exclusive $c \overline c$ distributions are integrated over
all rapidity while the inclusive single charm $p_T$ distributions are
integrated over $x_F > 0$ only.  The distributions for $pp$, $pA$ and $AA$
interactions are all given.  
The per nucleon cross section is given for $pA$ and $AA$ interactions.
Recall that the intrinsic $\langle k_T^2 \rangle$
is 1 GeV$^2$ for $pp$ interactions, 1.35 GeV$^2$ for $pA$ interactions, and
1.7 GeV$^2$ for $AA$.}
\end{table}

\begin{table}[p]
\begin{center}
\begin{tabular}{|c|c|c|c||c|c|c|} \hline
& \multicolumn{3}{c||}{ $d\sigma_{b \overline b}/dp_T$ (nb/GeV)} & 
\multicolumn{3}{c|}{$d\sigma_b/dp_T$ (nb/GeV)} \\ 
$p_T$ (GeV) & $pp$ & $pA$ & $AA$ & $pp$ & $pA$ & $AA$ \\ \hline
 0.25 & 196.260 & 134.740 & 115.740 &  60.320 &  59.260 &  64.460 \\
 0.75 & 512.400 & 375.200 & 326.800 & 152.900 & 150.680 & 165.800 \\
 1.25 & 632.600 & 516.800 & 461.600 & 199.820 & 200.800 & 227.600 \\
 1.75 & 583.400 & 547.400 & 549.200 & 218.200 & 220.200 & 249.800 \\
 2.25 & 411.800 & 493.600 & 531.400 & 205.200 & 213.800 & 245.600 \\
 2.75 & 271.200 & 406.000 & 484.400 & 176.080 & 188.020 & 224.800 \\
 3.25 & 167.880 & 294.400 & 399.200 & 143.840 & 158.220 & 186.580 \\
 3.75 & 108.580 & 207.400 & 311.200 & 112.120 & 125.600 & 152.840 \\
 4.25 &  72.620 & 138.440 & 223.800 &  83.480 &  96.660 & 116.940 \\
 4.75 &  51.360 &  92.200 & 164.180 &  64.340 &  73.340 &  90.860 \\
 5.25 &  36.100 &  62.080 & 112.040 &  44.880 &  53.240 &  68.200 \\
 5.75 &  26.980 &  41.860 &  75.280 &  32.120 &  40.500 &  48.460 \\
 6.25 &  20.000 &  30.180 &  52.800 &  24.240 &  27.040 &  36.680 \\
 6.75 &  15.304 &  21.440 &  36.080 &  16.774 &  20.480 &  25.580 \\
 7.25 &  11.450 &  15.744 &  25.620 &  11.866 &  14.502 &  18.294 \\
 7.75 &   8.794 &  11.774 &  17.932 &   8.878 &  10.784 &  13.860 \\
 8.25 &   6.684 &   8.946 &  13.118 &   6.260 &   7.496 &   9.694 \\
 8.75 &   5.188 &   6.792 &   9.788 &   4.486 &   5.542 &   6.958 \\
 9.25 &   4.046 &   5.260 &   7.328 &   3.276 &   4.020 &   4.902 \\
 9.75 &   3.178 &   4.036 &   5.552 &   2.472 &   2.798 &   3.564 \\
10.25 &   2.464 &   3.124 &   4.256 &   1.741 &   2.158 &   2.460 \\
10.75 &   1.970 &   2.408 &   3.254 &   1.340 &   1.575 &   1.866 \\
11.25 &   1.572 &   1.906 &   2.488 &   0.891 &   1.168 &   1.434 \\
11.75 &   1.220 &   1.512 &   1.939 &   0.810 &   0.820 &   1.022 \\
12.25 &   0.969 &   1.209 &   1.554 &   0.567 &   0.691 &   0.835 \\
12.75 &   0.772 &   0.929 &   1.229 &   0.448 &   0.503 &   0.577 \\
13.25 &   0.625 &   0.759 &   0.963 &   0.301 &   0.432 &   0.430 \\
13.75 &   0.511 &   0.601 &   0.756 &   0.276 &   0.263 &   0.347 \\
14.25 &   0.413 &   0.478 &   0.596 &   0.187 &   0.210 &   0.277 \\
14.75 &   0.342 &   0.388 &   0.475 &   0.166 &   0.185 &   0.182 \\
15.25 &   0.266 &   0.318 &   0.389 &   0.123 &   0.152 &   0.162 \\
15.75 &   0.217 &   0.250 &   0.311 &   0.112 &   0.088 &   0.118 \\
16.25 &   0.176 &   0.201 &   0.250 &   0.043 &   0.079 &   0.081 \\
16.75 &   0.142 &   0.164 &   0.195 &   0.060 &   0.070 &   0.077 \\
17.25 &   0.116 &   0.134 &   0.160 &   0.051 &   0.051 &   0.059 \\
17.75 &   0.095 &   0.108 &   0.128 &   0.033 &   0.037 &   0.042 \\ \hline
\end{tabular}
\end{center}
\caption[]{The NLO exclusive $b \overline b$ and inclusive single bottom $p_T$
distributions for collisions at $\sqrt{s} = 200$ GeV.
The exclusive $b \overline b$ distributions are integrated over
all rapidity while the inclusive single bottom $p_T$ distributions are
integrated over $x_F > 0$ only.  The distributions for $pp$, $pA$ and $AA$
interactions are all given.  
The per nucleon cross section is given for $pA$ and $AA$ interactions.
Recall that the intrinsic $\langle k_T^2 \rangle$
is 1 GeV$^2$ for $pp$ interactions, 2.57 GeV$^2$ for $pA$ interactions, and
4.16 GeV$^2$ for $AA$.}
\end{table}

\begin{table}[p]
\begin{center}
\begin{tabular}{|c|c|c|c||c|c|c|} \hline
& \multicolumn{3}{c||}{ $d\sigma_{b \overline b}/dp_T$ ($\mu$b/GeV)} & 
\multicolumn{3}{c|}{$d\sigma_b/dp_T$ ($\mu$b/GeV)} \\ 
$p_T$ (GeV) & $pp$ & $pA$ & $AA$ & $pp$ & $pA$ & $AA$ \\ \hline
 1.25 & 17.738 & 12.142 &  8.068 & 13.646 & 10.286 &  9.754 \\
 1.75 & 45.340 & 20.480 & 15.058 & 15.568 & 12.280 & 11.296 \\
 2.25 & 53.480 & 29.200 & 18.176 & 15.502 & 12.538 & 11.764 \\
 2.75 & 46.660 & 33.260 & 22.700 & 15.890 & 12.408 & 11.838 \\
 3.25 & 36.540 & 33.120 & 24.460 & 13.544 & 11.308 & 10.934 \\
 3.75 & 27.760 & 29.760 & 24.680 & 12.574 & 10.390 & 10.154 \\
 4.25 & 21.680 & 25.520 & 22.420 & 10.944 &  9.006 &  8.946 \\
 4.75 & 16.842 & 20.520 & 20.620 &  8.946 &  7.738 &  7.762 \\
 5.25 & 13.938 & 16.272 & 17.608 &  7.910 &  6.802 &  6.610 \\
 5.75 & 11.326 & 13.108 & 14.156 &  6.424 &  5.606 &  5.840 \\
 6.25 &  9.462 & 10.594 & 11.854 &  5.540 &  4.798 &  4.730 \\
 6.75 &  7.872 &  8.888 &  9.796 &  4.524 &  4.096 &  4.176 \\
 7.25 &  6.600 &  7.318 &  8.036 &  3.656 &  3.520 &  3.624 \\
 7.75 &  5.664 &  6.156 &  6.720 &  3.128 &  2.696 &  2.936 \\
 8.25 &  4.798 &  5.192 &  5.582 &  2.616 &  2.374 &  2.480 \\
 8.75 &  4.112 &  4.520 &  4.868 &  2.154 &  2.004 &  2.056 \\
 9.25 &  3.614 &  3.768 &  4.062 &  1.829 &  1.686 &  1.701 \\
 9.75 &  3.096 &  3.372 &  3.520 &  1.502 &  1.444 &  1.596 \\
10.25 &  2.672 &  2.844 &  3.022 &  1.279 &  1.187 &  1.286 \\
10.75 &  2.334 &  2.504 &  2.612 &  1.061 &  1.078 &  1.088 \\
11.25 &  2.094 &  2.166 &  2.308 &  0.904 &  0.827 &  0.958 \\
11.75 &  1.821 &  1.923 &  2.026 &  0.741 &  0.756 &  0.775 \\
12.25 &  1.586 &  1.693 &  1.787 &  0.645 &  0.613 &  0.726 \\
12.75 &  1.430 &  1.509 &  1.594 &  0.608 &  0.545 &  0.575 \\
13.25 &  1.239 &  1.308 &  1.400 &  0.520 &  0.491 &  0.479 \\
13.75 &  1.107 &  1.184 &  1.229 &  0.478 &  0.420 &  0.483 \\
14.25 &  0.985 &  1.025 &  1.091 &  0.387 &  0.383 &  0.425 \\
14.75 &  0.903 &  0.935 &  0.994 &  0.344 &  0.307 &  0.313 \\
15.25 &  0.775 &  0.831 &  0.883 &  0.302 &  0.307 &  0.345 \\
15.75 &  0.698 &  0.741 &  0.778 &  0.261 &  0.270 &  0.276 \\
16.25 &  0.645 &  0.680 &  0.720 &  0.202 &  0.232 &  0.237 \\
16.75 &  0.572 &  0.605 &  0.635 &  0.207 &  0.180 &  0.211 \\
17.25 &  0.518 &  0.550 &  0.581 &  0.183 &  0.182 &  0.205 \\
17.75 &  0.464 &  0.485 &  0.510 &  0.156 &  0.166 &  0.137 \\ \hline
\end{tabular}
\end{center}
\caption[]{The NLO exclusive $b \overline b$ and inclusive single bottom $p_T$
distributions for collisions at $\sqrt{s} = 5.5$ TeV.  
The exclusive $b \overline b$ distributions are integrated over
all rapidity while the inclusive single bottom $p_T$ distributions are
integrated over $x_F > 0$ only.  The distributions for $pp$, $pA$ and $AA$
interactions are all given.  
The per nucleon cross section is given for $pA$ and $AA$ interactions.
Recall that the intrinsic $\langle k_T^2 \rangle$
is 1 GeV$^2$ for $pp$ interactions, 2.57 GeV$^2$ for $pA$ interactions, and
4.16 GeV$^2$ for $AA$.}
\end{table}

\begin{table}[p]
\begin{center}
\begin{tabular}{|c|c|c||c|c|} \hline
& \multicolumn{2}{c||}{ $d\sigma_{c \overline c}/dy$ ($\mu$b)} & 
\multicolumn{2}{c|}{$d\sigma_c/dy$ ($\mu$b)} \\ 
$y$ & $pp$ & $pA$ & $pp$ & $pA$ \\ \hline
-1.85 &     0.000 &     0.000 &     0.017 &     0.014 \\
-1.65 &     0.009 &     0.008 &     0.084 &     0.072 \\
-1.45 &     0.069 &     0.060 &     0.247 &     0.223 \\
-1.25 &     0.243 &     0.218 &     0.530 &     0.505 \\
-1.05 &     0.601 &     0.569 &     0.888 &     0.900 \\
-0.85 &     1.131 &     1.132 &     1.295 &     1.349 \\
-0.65 &     1.707 &     1.781 &     1.699 &     1.841 \\
-0.45 &     2.237 &     2.431 &     2.009 &     2.230 \\
-0.25 &     2.642 &     2.951 &     2.254 &     2.541 \\
-0.05 &     2.834 &     3.231 &     2.346 &     2.674 \\
 0.05 &     2.828 &     3.252 &     2.340 &     2.682 \\
 0.25 &     2.644 &     3.063 &     2.248 &     2.593 \\
 0.45 &     2.238 &     2.602 &     2.011 &     2.309 \\
 0.65 &     1.710 &     1.968 &     1.692 &     1.936 \\
 0.85 &     1.129 &     1.282 &     1.300 &     1.462 \\
 1.05 &     0.600 &     0.655 &     0.885 &     0.988 \\
 1.25 &     0.243 &     0.248 &     0.532 &     0.584 \\
 1.45 &     0.069 &     0.066 &     0.250 &     0.261 \\
 1.65 &     0.009 &     0.008 &     0.083 &     0.086 \\
 1.85 &     0.000 &     0.000 &     0.017 &     0.017 \\ \hline
\end{tabular}
\end{center}
\caption[]{The NLO exclusive $c \overline c$ and inclusive single charm $y$
distributions for a proton beam of 158 GeV.  
The distributions for $pp$ and $pA$
interactions are both given.  
The per nucleon cross section is given for $pA$ interactions.
Recall that the intrinsic $\langle k_T^2 \rangle$
is 1 GeV$^2$ for $pp$ interactions and 1.35 GeV$^2$ for $pA$ interactions.}
\end{table}

\begin{table}[p]
\begin{center}
\begin{tabular}{|c|c|c||c|c|} \hline
& \multicolumn{2}{c||}{ $d\sigma_{c \overline c}/dy$ ($\mu$b)} & 
\multicolumn{2}{c|}{$d\sigma_c/dy$ ($\mu$b)} \\ 
$y$ & $pp$ & $pA$ & $pp$ & $pA$ \\ \hline
-2.25 &     0.004 &     0.004 &     0.085 &     0.066 \\
-2.05 &     0.043 &     0.038 &     0.301 &     0.262 \\
-1.85 &     0.210 &     0.192 &     0.708 &     0.648 \\
-1.65 &     0.604 &     0.566 &     1.365 &     1.346 \\
-1.45 &     1.367 &     1.326 &     2.214 &     2.243 \\
-1.25 &     2.389 &     2.437 &     3.158 &     3.360 \\
-1.05 &     3.719 &     3.960 &     4.162 &     4.572 \\
-0.85 &     5.147 &     5.639 &     5.200 &     5.870 \\
-0.65 &     6.516 &     7.329 &     6.102 &     7.002 \\
-0.45 &     7.712 &     8.813 &     6.783 &     7.824 \\
-0.25 &     8.536 &     9.930 &     7.319 &     8.505 \\
-0.05 &     8.998 &    10.530 &     7.506 &     8.750 \\
 0.05 &     8.998 &    10.540 &     7.529 &     8.736 \\
 0.25 &     8.534 &     9.964 &     7.284 &     8.406 \\
 0.45 &     7.702 &     8.851 &     6.807 &     7.761 \\
 0.65 &     6.507 &     7.344 &     6.071 &     6.840 \\
 0.85 &     5.152 &     5.632 &     5.193 &     5.743 \\
 1.05 &     3.722 &     3.925 &     4.178 &     4.543 \\
 1.25 &     2.387 &     2.402 &     3.149 &     3.327 \\
 1.45 &     1.367 &     1.295 &     2.176 &     2.254 \\
 1.65 &     0.605 &     0.548 &     1.394 &     1.405 \\
 1.85 &     0.211 &     0.183 &     0.710 &     0.686 \\
 2.05 &     0.043 &     0.035 &     0.292 &     0.272 \\
 2.25 &     0.004 &     0.003 &     0.083 &     0.077 \\ \hline
\end{tabular}
\end{center}
\caption[]{The NLO exclusive $c \overline c$ and inclusive single charm $y$
distributions for a proton beam of 450 GeV.  
The distributions for $pp$ and $pA$
interactions are both given.  
The per nucleon cross section is given for $pA$ interactions.
Recall that the intrinsic $\langle k_T^2 \rangle$
is 1 GeV$^2$ for $pp$ interactions and 1.35 GeV$^2$ for $pA$ interactions.}
\end{table}

\begin{table}[p]
\begin{center}
\begin{tabular}{|c|c|c||c|c|} \hline
& \multicolumn{2}{c||}{ $d\sigma_{c \overline c}/dy$ ($\mu$b)} & 
\multicolumn{2}{c|}{$d\sigma_c/dy$ ($\mu$b)} \\ 
$y$ & $pp$ & $pA$ & $pp$ & $pA$ \\ \hline
-2.45 &     0.015 &     0.013 &     0.187 &     0.154 \\
-2.25 &     0.106 &     0.091 &     0.555 &     0.510 \\
-2.05 &     0.390 &     0.347 &     1.199 &     1.109 \\
-1.85 &     1.019 &     0.941 &     2.164 &     2.155 \\
-1.65 &     2.093 &     2.042 &     3.320 &     3.490 \\
-1.45 &     3.506 &     3.576 &     4.685 &     5.096 \\
-1.25 &     5.345 &     5.741 &     6.120 &     6.789 \\
-1.05 &     7.181 &     7.967 &     7.684 &     8.666 \\
-0.85 &     9.321 &    10.580 &     9.081 &    10.390 \\
-0.65 &    11.060 &    12.880 &    10.340 &    12.030 \\
-0.45 &    12.860 &    14.910 &    11.290 &    13.110 \\
-0.25 &    14.050 &    16.420 &    11.980 &    13.950 \\
-0.05 &    14.610 &    17.040 &    12.300 &    14.230 \\
 0.05 &    14.640 &    17.020 &    12.270 &    14.110 \\
 0.25 &    14.030 &    16.100 &    11.970 &    13.610 \\
 0.45 &    12.860 &    14.420 &    11.330 &    12.620 \\
 0.65 &    11.060 &    12.230 &    10.320 &    11.340 \\
 0.85 &     9.324 &     9.872 &     9.030 &     9.675 \\
 1.05 &     7.191 &     7.314 &     7.720 &     8.143 \\
 1.25 &     5.347 &     5.209 &     6.141 &     6.242 \\
 1.45 &     3.510 &     3.196 &     4.672 &     4.737 \\
 1.65 &     2.103 &     1.819 &     3.297 &     3.255 \\
 1.85 &     1.014 &     0.844 &     2.189 &     2.118 \\
 2.05 &     0.391 &     0.313 &     1.209 &     1.097 \\
 2.25 &     0.107 &     0.078 &     0.550 &     0.490 \\
 2.45 &     0.015 &     0.011 &     0.193 &     0.161 \\ \hline
\end{tabular}
\end{center}
\caption[]{The NLO exclusive $c \overline c$ and inclusive single charm $y$
distributions for a proton beam of 800 GeV.  
The distributions for $pp$ and $pA$
interactions are both given.  
The per nucleon cross section is given for $pA$ interactions.
Recall that the intrinsic $\langle k_T^2 \rangle$
is 1 GeV$^2$ for $pp$ interactions and 1.35 GeV$^2$ for $pA$ interactions.}
\end{table}

\begin{table}[p]
\begin{center}
\begin{tabular}{|c|c|c|c|||c|c|c|} \hline
& \multicolumn{3}{c||}{ $d\sigma_{c \overline c}/dy$ ($\mu$b)} & 
\multicolumn{3}{c|}{$d\sigma_c/dy$ ($\mu$b)} \\ 
$y$ & $pp$ & $pA$ & $AA$ & $pp$ & $pA$ & $AA$ \\ \hline
-4.25 &  0.003 &  0.003 &  0.002 &  0.148 &  0.109 &  0.056 \\
-3.95 &  0.125 &  0.105 &  0.070 &  0.939 &  0.772 &  0.588 \\
-3.65 &  1.022 &  0.925 &  0.604 &  2.988 &  3.223 &  2.422 \\
-3.35 &  3.398 &  3.263 &  2.367 &  7.240 &  7.515 &  5.610 \\
-3.05 &  8.264 &  8.325 &  5.900 & 13.420 & 15.310 & 10.730 \\
-2.75 & 15.050 & 16.570 & 11.480 & 20.550 & 22.350 & 17.740 \\
-2.45 & 23.850 & 26.750 & 19.320 & 28.580 & 33.120 & 24.520 \\
-2.15 & 33.690 & 38.640 & 29.170 & 36.600 & 43.300 & 34.830 \\
-1.85 & 43.780 & 50.660 & 41.070 & 45.650 & 52.460 & 44.450 \\
-1.55 & 54.080 & 63.880 & 52.950 & 53.330 & 61.470 & 51.050 \\
-1.25 & 63.830 & 73.310 & 60.100 & 60.650 & 67.530 & 58.730 \\
-0.95 & 69.960 & 78.450 & 69.490 & 65.320 & 70.720 & 63.590 \\
-0.65 & 75.700 & 82.120 & 75.790 & 68.690 & 72.230 & 68.490 \\
-0.35 & 79.700 & 82.990 & 79.100 & 71.950 & 72.470 & 70.220 \\
-0.05 & 80.720 & 80.510 & 79.630 & 72.860 & 70.470 & 69.430 \\
 0.05 & 80.840 & 78.680 & 79.990 & 72.730 & 70.150 & 70.350 \\
 0.35 & 79.520 & 74.340 & 79.610 & 72.890 & 68.970 & 69.670 \\
 0.65 & 75.930 & 67.520 & 75.420 & 69.340 & 61.990 & 67.540 \\
 0.95 & 70.440 & 60.230 & 69.290 & 65.230 & 57.870 & 63.320 \\
 1.25 & 64.000 & 52.670 & 59.680 & 61.030 & 51.720 & 59.100 \\
 1.55 & 54.240 & 43.760 & 53.000 & 53.110 & 44.530 & 52.360 \\
 1.85 & 43.610 & 33.690 & 40.990 & 46.350 & 37.890 & 43.830 \\
 2.15 & 33.430 & 24.880 & 29.270 & 37.400 & 29.630 & 35.640 \\
 2.45 & 23.680 & 17.420 & 19.480 & 27.760 & 21.260 & 25.230 \\
 2.75 & 15.050 & 10.820 & 11.600 & 20.230 & 15.600 & 17.760 \\
 3.05 &  8.150 &  5.742 &  5.894 & 12.770 &  9.649 & 10.740 \\
 3.35 &  3.403 &  2.252 &  2.345 &  7.394 &  5.295 &  5.571 \\
 3.65 &  1.006 &  0.674 &  0.606 &  2.844 &  2.169 &  2.426 \\
 3.95 &  0.125 &  0.090 &  0.070 &  0.980 &  0.662 &  0.537 \\
 4.25 &  0.003 &  0.002 &  0.002 &  0.180 &  0.099 &  0.050 \\ \hline
\end{tabular}
\end{center}
\caption[]{The NLO exclusive $c \overline c$ and inclusive single charm $y$
distributions for collisions at $\sqrt{s} = 200$ GeV.
The distributions for $pp$, $pA$ and $AA$
interactions are all given.  
The per nucleon cross section is given for $pA$ and $AA$ interactions.
Recall that the intrinsic $\langle k_T^2 \rangle$
is 1 GeV$^2$ for $pp$ interactions, 1.35 GeV$^2$ for $pA$ interactions, and
1.7 GeV$^2$ for $AA$.}
\end{table}

\begin{table}[p]
\begin{center}
\begin{tabular}{|c|c|c|c||c|c|c|} \hline
& \multicolumn{3}{c||}{ $d\sigma_{c \overline c}/dy$ ($\mu$b)} & 
\multicolumn{3}{c|}{$d\sigma_c/dy$ ($\mu$b)} \\ 
$y$ & $pp$ & $pA$ & $AA$ & $pp$ & $pA$ & $AA$ \\ \hline
-4.85 & 117.0 & 128.4 & 190.7 & 186.8 & 200.4 & 216.2 \\
-4.55 & 234.1 & 278.6 & 265.9 & 237.5 & 268.7 & 228.0 \\
-4.25 & 369.4 & 428.5 & 308.5 & 331.2 & 346.9 & 291.3 \\
-3.95 & 485.5 & 495.9 & 362.1 & 426.5 & 457.1 & 336.4 \\
-3.65 & 532.9 & 586.9 & 367.1 & 522.0 & 522.1 & 343.8 \\
-3.35 & 561.1 & 580.0 & 377.3 & 542.4 & 553.1 & 364.0 \\
-3.05 & 594.1 & 586.8 & 399.1 & 590.5 & 564.2 & 380.4 \\
-2.75 & 627.1 & 602.2 & 407.0 & 614.5 & 606.1 & 386.3 \\
-2.45 & 650.1 & 616.5 & 409.4 & 650.6 & 565.6 & 396.7 \\
-2.15 & 683.6 & 600.0 & 422.8 & 642.2 & 577.1 & 411.1 \\
-1.85 & 685.5 & 587.8 & 428.7 & 681.9 & 582.8 & 415.0 \\
-1.55 & 710.3 & 603.6 & 421.4 & 694.6 & 588.5 & 426.7 \\
-1.25 & 705.9 & 618.8 & 424.9 & 672.0 & 578.2 & 416.3 \\
-0.95 & 711.8 & 595.5 & 435.7 & 703.2 & 589.7 & 414.9 \\
-0.65 & 735.9 & 611.1 & 435.5 & 736.7 & 617.2 & 435.9 \\
-0.35 & 743.6 & 590.1 & 428.8 & 692.5 & 577.4 & 402.3 \\
-0.05 & 720.6 & 556.8 & 414.6 & 716.7 & 560.5 & 410.3 \\
 0.05 & 715.6 & 548.0 & 419.2 & 721.1 & 557.8 & 429.4 \\
 0.35 & 741.4 & 564.0 & 433.3 & 699.3 & 552.3 & 427.0 \\
 0.65 & 744.0 & 567.4 & 436.3 & 706.2 & 555.3 & 440.3 \\
 0.95 & 708.6 & 541.4 & 430.2 & 701.5 & 530.4 & 424.2 \\
 1.25 & 702.3 & 546.4 & 426.7 & 704.2 & 520.6 & 403.4 \\
 1.55 & 713.9 & 506.9 & 423.0 & 664.2 & 497.5 & 419.4 \\
 1.85 & 689.1 & 482.1 & 429.9 & 678.2 & 477.1 & 416.1 \\
 2.15 & 687.6 & 476.7 & 418.2 & 639.9 & 462.9 & 396.7 \\
 2.45 & 651.9 & 470.5 & 409.3 & 620.5 & 447.6 & 393.0 \\
 2.75 & 625.8 & 437.6 & 407.2 & 606.6 & 422.2 & 390.0 \\
 3.05 & 589.7 & 412.5 & 397.3 & 591.6 & 415.5 & 382.4 \\
 3.35 & 564.0 & 384.7 & 374.7 & 555.1 & 388.1 & 380.9 \\
 3.65 & 532.9 & 379.3 & 362.5 & 516.0 & 350.9 & 355.4 \\
 3.95 & 479.2 & 304.0 & 358.4 & 430.2 & 297.4 & 325.7 \\
 4.25 & 371.3 & 257.4 & 321.1 & 357.9 & 226.2 & 294.6 \\
 4.55 & 230.3 & 153.8 & 263.9 & 226.4 & 164.3 & 244.7 \\
 4.85 & 112.1 &  69.4 & 192.8 & 193.4 & 104.4 & 198.5 \\ \hline
\end{tabular}
\end{center}
\caption[]{The NLO exclusive $c \overline c$ and inclusive single charm $y$
distributions for collisions at $\sqrt{s} = 5.5$ TeV.
The distributions for $pp$, $pA$ and $AA$
interactions are all given.  
The per nucleon cross section is given for $pA$ and $AA$ interactions.
Recall that the intrinsic $\langle k_T^2 \rangle$
is 1 GeV$^2$ for $pp$ interactions, 1.35 GeV$^2$ for $pA$ interactions, and
1.7 GeV$^2$ for $AA$.}
\end{table}

\begin{table}[p]
\begin{center}
\begin{tabular}{|c|c|c|c||c|c|c|} \hline
& \multicolumn{3}{c||}{ $d\sigma_{b \overline b}/dy$ (nb)} & 
\multicolumn{3}{c|}{$d\sigma_b/dy$ (nb)} \\ 
$y$ & $pp$ & $pA$ & $AA$ & $pp$ & $pA$ & $AA$ \\ \hline
-2.65 &   0.8 &   0.7 &   0.5 &  14.1 &  12.1 &  12.0 \\
-2.45 &   4.8 &   4.2 &   3.2 &  33.0 &  26.6 &  28.7 \\
-2.25 &  16.6 &  14.3 &  12.1 &  61.1 &  59.2 &  59.1 \\
-2.05 &  41.1 &  38.3 &  32.2 & 103.7 & 100.0 & 103.2 \\
-1.85 &  82.8 &  80.2 &  70.4 & 150.0 & 154.8 & 158.6 \\
-1.65 & 138.2 & 140.8 & 132.5 & 203.4 & 220.3 & 232.5 \\
-1.45 & 213.3 & 222.4 & 214.9 & 262.8 & 287.3 & 302.5 \\
-1.25 & 289.1 & 313.4 & 321.9 & 323.3 & 361.1 & 399.4 \\
-1.05 & 382.7 & 417.3 & 441.6 & 383.9 & 432.0 & 485.2 \\
-0.85 & 463.7 & 518.7 & 576.3 & 443.2 & 489.1 & 558.3 \\
-0.65 & 543.8 & 609.8 & 702.0 & 482.7 & 548.9 & 627.4 \\
-0.45 & 606.8 & 692.5 & 814.8 & 520.3 & 589.4 & 681.2 \\
-0.25 & 651.4 & 739.2 & 890.1 & 547.5 & 615.2 & 722.1 \\
-0.05 & 674.7 & 764.2 & 930.9 & 555.3 & 617.2 & 738.3 \\
 0.05 & 674.1 & 762.7 & 933.5 & 553.5 & 615.6 & 734.7 \\
 0.25 & 650.8 & 726.9 & 890.2 & 545.0 & 606.1 & 725.8 \\
 0.45 & 607.3 & 671.0 & 814.3 & 520.1 & 570.5 & 684.9 \\
 0.65 & 543.3 & 582.9 & 701.9 & 481.7 & 522.5 & 632.2 \\
 0.85 & 463.9 & 491.0 & 576.4 & 440.1 & 466.3 & 559.4 \\
 1.05 & 383.1 & 391.0 & 442.2 & 387.5 & 406.7 & 481.3 \\
 1.25 & 289.0 & 291.1 & 322.1 & 323.5 & 338.0 & 399.0 \\
 1.45 & 213.3 & 205.7 & 215.1 & 264.0 & 269.3 & 304.8 \\
 1.65 & 138.7 & 130.4 & 132.3 & 205.7 & 210.2 & 233.8 \\
 1.85 &  82.9 &  75.3 &  70.2 & 148.6 & 147.1 & 159.0 \\
 2.05 &  41.2 &  36.4 &  32.1 & 102.1 &  99.0 & 104.3 \\
 2.25 &  16.6 &  13.7 &  12.1 &  59.9 &  58.3 &  58.7 \\
 2.45 &   4.8 &   4.0 &   3.2 &  31.3 &  30.7 &  28.2 \\
 2.65 &   0.8 &   0.6 &   0.5 &  13.5 &  12.8 &  11.4 \\ \hline
\end{tabular}
\end{center}
\caption[]{The NLO exclusive $b \overline b$ and inclusive single bottom $y$
distributions for collisions at $\sqrt{s} = 200$ GeV.
The distributions for $pp$, $pA$ and $AA$
interactions are all given.  
The per nucleon cross section is given for $pA$ and $AA$ interactions.
Recall that the intrinsic $\langle k_T^2 \rangle$
is 1 GeV$^2$ for $pp$ interactions, 2.57 GeV$^2$ for $pA$ interactions, and
4.16 GeV$^2$ for $AA$.}
\end{table}

\begin{table}[p]
\begin{center}
\begin{tabular}{|c|c|c|c||c|c|c|} \hline
& \multicolumn{3}{c||}{ $d\sigma_{b \overline b}/dy$ ($\mu$b)} & 
\multicolumn{3}{c|}{$d\sigma_b/dy$ ($\mu$b)} \\ 
$y$ & $pp$ & $pA$ & $AA$ & $pp$ & $pA$ & $AA$ \\ \hline
-4.85 &  1.811 &  1.831 &  1.189 &  3.012 &  3.239 &  2.484 \\
-4.55 &  2.940 &  3.117 &  2.157 &  4.652 &  5.331 &  3.699 \\
-4.25 &  5.035 &  5.625 &  3.782 &  6.328 &  7.331 &  5.407 \\
-3.95 &  7.070 &  7.946 &  5.609 &  8.606 &  9.596 &  7.215 \\
-3.65 &  9.872 & 11.200 &  8.283 & 10.500 & 12.110 &  9.164 \\
-3.35 & 12.260 & 13.800 & 10.220 & 12.780 & 14.330 & 11.140 \\
-3.05 & 14.330 & 16.330 & 12.480 & 14.770 & 16.100 & 12.750 \\
-2.75 & 17.350 & 19.110 & 14.890 & 16.950 & 18.230 & 13.600 \\
-2.45 & 19.100 & 21.030 & 16.430 & 18.730 & 20.040 & 15.780 \\
-2.15 & 21.350 & 22.880 & 17.990 & 20.070 & 20.650 & 16.870 \\
-1.85 & 22.890 & 23.940 & 19.050 & 21.300 & 22.290 & 17.730 \\
-1.55 & 24.010 & 23.790 & 19.440 & 23.250 & 22.740 & 18.530 \\
-1.25 & 25.320 & 24.830 & 20.760 & 24.040 & 23.790 & 19.360 \\
-0.95 & 26.500 & 25.570 & 20.850 & 25.360 & 23.960 & 20.110 \\
-0.65 & 26.660 & 25.160 & 21.330 & 25.380 & 24.050 & 19.720 \\
-0.35 & 27.460 & 25.330 & 21.770 & 25.480 & 23.380 & 20.670 \\
-0.05 & 27.050 & 24.680 & 21.470 & 26.170 & 23.810 & 20.650 \\
 0.05 & 27.010 & 24.500 & 21.500 & 25.790 & 23.920 & 20.390 \\
 0.35 & 27.330 & 24.130 & 21.680 & 25.320 & 22.810 & 20.150 \\
 0.65 & 26.660 & 23.340 & 21.300 & 25.380 & 21.990 & 20.100 \\
 0.95 & 26.460 & 22.770 & 20.900 & 25.480 & 21.810 & 19.870 \\
 1.25 & 25.310 & 21.310 & 20.760 & 23.830 & 20.230 & 19.310 \\
 1.55 & 24.080 & 19.630 & 19.390 & 22.540 & 19.050 & 18.840 \\
 1.85 & 22.830 & 19.060 & 18.930 & 21.370 & 17.920 & 17.640 \\
 2.15 & 21.320 & 17.380 & 18.040 & 20.000 & 16.690 & 16.930 \\
 2.45 & 19.100 & 15.280 & 16.470 & 18.770 & 15.570 & 15.870 \\
 2.75 & 17.360 & 13.580 & 14.940 & 17.320 & 13.780 & 14.260 \\
 3.05 & 14.410 & 11.260 & 12.460 & 14.670 & 11.580 & 12.710 \\
 3.35 & 12.140 &  9.320 & 10.240 & 12.740 & 10.040 & 10.770 \\
 3.65 &  9.831 &  7.393 &  8.249 & 10.430 &  7.929 &  8.860 \\
 3.95 &  7.049 &  5.294 &  5.607 &  8.431 &  6.380 &  7.131 \\
 4.25 &  5.017 &  3.721 &  3.810 &  6.190 &  4.889 &  5.378 \\
 4.55 &  2.925 &  2.147 &  2.171 &  4.660 &  3.450 &  3.950 \\
 4.85 &  1.797 &  1.290 &  1.193 &  3.080 &  2.222 &  2.346 \\ \hline
\end{tabular}
\end{center}
\caption[]{The NLO exclusive $b \overline b$ and inclusive single bottom $y$
distributions for collisions at $\sqrt{s} = 5.5$ TeV.  
The distributions for $pp$, $pA$ and $AA$
interactions are all given.  
The per nucleon cross section is given for $pA$ and $AA$ interactions.
Recall that the intrinsic $\langle k_T^2 \rangle$
is 1 GeV$^2$ for $pp$ interactions, 2.57 GeV$^2$ for $pA$ interactions, and
4.16 GeV$^2$ for $AA$.}
\end{table}

\end{document}